\documentclass{ar-1col-S2O}

\usepackage[comma]{natbib}
\usepackage{url}
\usepackage[colorlinks=true,allcolors=blue]{hyperref}
\usepackage{subcaption}
\usepackage{amssymb}
\usepackage{longtable}
\usepackage{multibib}
\usepackage{IEEEtrantools} 
\usepackage{bm} 
\usepackage[labelfont=bf]{caption} 
\usepackage{verbatim}

\jname{Annu. Rev. Astron. Astrophys}
\jvol{AA}
\jyear{2024}
\doi{10.1146/((please add article doi))}

\newcommand{\eqn}[1]{Eq.~\ref{#1}}
\newcommand{\sect}[1]{Sect.~\ref{#1}}
\newcommand{\fig}[1]{Fig.~\ref{#1}}
\newcommand{\tab}[1]{Table~\ref{#1}}

\newcommand{\stagger}{\texttt{Stagger}}
\newcommand{\cifist}{\texttt{CIFIST}}
\newcommand{\cobold}{\texttt{CO$^5$BOLD}}
\newcommand{\muram}{\texttt{MURaM}}
\newcommand{\antares}{\texttt{Antares}}

\newcommand{\hlinop}{\texttt{HLINOP}}
\newcommand{\multitd}{\texttt{Multi3D}}
\newcommand{\balder}{\texttt{Balder}}
\newcommand{\nltetd}{\texttt{NLTE3D}}

\newcommand{\nicole}{\texttt{NICOLE}}
\newcommand{\nataja}{\texttt{NATAJA}}
\newcommand{\dispatch}{\texttt{DISPATCH}}
\newcommand{\tlusty}{\texttt{TLUSTY}}
\newcommand{\porta}{\texttt{PORTA}}
\newcommand{\linfortd}{\texttt{LinFor3D}}
\newcommand{\scate}{\texttt{Scate}}
\newcommand{\freeeos}{\texttt{FreeEOS}}
\newcommand{\marcs}{\texttt{MARCS}}
\newcommand{\optimtd}{\texttt{OPTIM3D}}
\newcommand{\asset}{\texttt{ASSET}}
\newcommand{\mtd}{\rm\textlangle3D\textrangle}
\newcommand{\teff}{T_{\mathrm{eff}}}
\newcommand{\lgg}{\log{g}}
\newcommand{\feh}{\mathrm{[Fe/H]}}
\newcommand\ion[2]{#1\,\textsc{\lowercase{#2}}}	

\begin{document}
\markboth{Lind \& Amarsi}{3D non-LTE abundance analyses of late-type stars}

\title{3D non-LTE abundance analyses of late-type stars}

\author{Karin Lind$^1$ and Anish M. Amarsi$^2$
\affil{$^1$Department of Astronomy, Stockholm University, AlbaNova Research Centre, SE-106 91, Stockholm, Sweden; email: karin.lind@astro.su.se}
\affil{$^2$Theroretical Astrophysics, Department of Physics and Astronomy, Uppsala University, Box 516, SE-751 20, Uppsala, Sweden; email: anish.amarsi@physics.uu.se}}

\begin{abstract}
The chemical compositions of stars encode the history of the universe 
and are thus fundamental for advancing our knowledge of astrophysics and cosmology.
However, measurements of elemental abundances ratios, and our interpretations of them, strongly depend on the physical assumptions that dictate the generation of synthetic stellar spectra. Three-dimensional radiation-hydrodynamic (3D RHD) ``box-in-a-star'' simulations of stellar atmospheres offer a more realistic representation of surface convection occurring in late-type stars compared to traditional one-dimensional (1D) hydrostatic models. As evident from a multitude of observational tests, the coupling of 3D RHD models with line-formation in non-local thermodynamic equilibrium (non-LTE) today provides a solid foundation for abundance analysis for many elements. This review describes the ongoing and transformational work to advance the state-of-the-art and replace 1D LTE spectrum synthesis with its 3D non-LTE counterpart. In summary: \\

\begin{minipage}{10cm}
\begin{itemize}
\item 3D and non-LTE effects are intricately coupled and consistent modelling thereof is necessary for high-precision abundances, which is currently feasible for individual elements in large surveys. Mean 3D ($\mtd{}$) models are not adequate as substitutes. 
\item The solar abundance debate is presently dominated by choices and systematic uncertainties that are not specific to 3D non-LTE modelling.  
\item 3D non-LTE abundance corrections have a profound impact on our understanding of FGK-type stars, exoplanets, and the nucleosynthetic origins of the elements.  
\end{itemize}
\end{minipage}
\end{abstract}

\maketitle

\tableofcontents

\section{INTRODUCTION}

In 1925, the ground-breaking work of Cecilia Payne-Gaposchkin revealed that stars are primarily composed of hydrogen and helium, with traces of heavier elements. Half a century later saw the dawn of quantitative spectroscopy using grids of model atmospheres for FGK-type dwarfs and giants of varying composition. Synthetic spectra based on one-dimensional (1D) hydrostatic models have become more extensive and sophisticated over time, by inclusion of a more accurate equation of state (EOS) and opacities, and monochromatic radiative transfer, and are still heavily used today.  However, their applicability and physical realism are intrinsically limited by the unavoidably simplified treatment of convection, as well as by the commonly-employed assumption of local thermodynamic equilibrium (LTE).

A fundamental improvement to the modelling of spectra of FGK-type stars was introduced by lifting the assumption that all atomic level populations abide strictly to LTE. Pioneering 1D non-LTE studies targetted spectral lines of e.g., Li, O, and Fe \citep{1972ApJ...176..809A,1984A&A...130..319S,1991A&A...245L...9K}. Equipped with model atoms with approximately a dozen energy levels, the statistical equilibrium equations could be solved with the computational power available at the time, and sometimes yielded significant deviations from LTE. Importantly, spectral lines were found to react differently to non-LTE modelling depending on atomic properties, abundance, and atmospheric conditions, a conclusion that has been reinforced many times since. Because the necessary atomic data were often lacking, astrophysical calibration of model atoms to minimize abundance discrepancies between lines was common and the custom still prevails, but to a much lesser extent today.

This century has seen a dramatic improvement in the availability and quality of atomic and molecular data for neutrals and singly ionized species, such as energy levels, transition rates or oscillator strengths, photoionisation cross-sections and collisional cross-sections. In particular, the development of ab initio quantum chemistry calculations and asymptotic model approximations of inelastic hydrogen collisions has removed a crucial free parameter in the modelling of late-type stars \citep{2016A&ARv..24....9B}. Such data now exist for more than twenty metals of astrophysical importance.

Largely in parallel with the development of 1D non-LTE abundance analysis, three-dimensional radiation-hydrodynamic (RHD) simulations of stellar surface convection were introduced, allowing us to predict the inhomogeneous and dynamic state of the atmosphere from first principles \citep{1982A&A...107....1N,1990A&A...228..155N,1990CoPhC..59..119N,1998ApJ...499..914S}.
Observational validation of such models have revealed, for example, that they can adequately reproduce spectral line asymmetries and blueshifts, the statistical properties of granulation,
and the observed centre-to-limb (CLV) variation of solar continuum light
\citep[e.g.][]{1981A&A....96..345D,2000A&A...359..729A,2009A&A...503..225W,2011ApJ...736...69U,2013A&A...554A.118P}.

Despite the many successes of 3D RHD models, it soon became apparent that their use for abundance analyses must proceed with great care. The temperature and density fluctuations associated with granulation can lead to significant strengthening of lines under LTE, but the steeper temperature stratification of metal-poor stars also often lead to enhanced non-LTE effects in the opposite direction \citep{2005ARA&A..43..481A}. At the turn of the century, 3D non-LTE modelling that simultaneously accounted for departures from LTE and homogeneity was still in its infancy, with trailblazing studies for a handful of stars using simple atoms for elements with great astrophysical significance \citep{1995A&A...302..578K,2003A&A...399L..31A,2003A&A...409L...1B,2005ApJ...618..939S,2007A&A...473L..37C}. In contrast, leading 3D non-LTE studies are now performed with model atoms including hundreds of energy levels \citep{2016MNRAS.463.1518A,2017A&A...597A...6N,2017MNRAS.468.4311L,2019A&A...631A..80B}, and for tens or hundreds of model atmospheres \citep[e.g.][]{2022A&A...668A..68A}. This progress owes much to the exponential growth in computational power and the development of frequency- and domain-decomposed MPI-parallelised codes that solve the restricted non-LTE problem in 3D \citep{2009ASPC..415...87L,2018A&A...615A.139A}.   

Alongside advances on the theory and modelling side, spectroscopic stellar surveys have become a cornerstone of modern astrophysics, now routinely collecting medium and high-resolution data for hundreds of thousands of stars \citep{2019ARA&A..57..571J}. Accurate determination of chemical abundances and fundamental stellar parameters such as effective temperature, surface gravity, mass and age, is a primary goal of current and upcoming surveys and instrumental to the diverse endeavours of Galactic archaeology. Recently, the GALAH survey presented the largest existing 1D non-LTE abundance data set for 13 elements in 650,000 stars \citep{2020A&A...642A..62A,2021MNRAS.506..150B}, removing any lingering doubt of the feasibility of such analysis for million-star spectroscopic surveys. In fact, 3D non-LTE analysis is also possible for individual elements, such as Li \citep{Wang_li} and O \citep{2016PhDT.......279A}, but, as discussed here, more work is needed on the grids of 3D RHD simulations, grids of 3D non-LTE calculations, and spectrum interpolation techniques, before 3D analysis will be truly competitive with 1D analysis in this respect.    

In this review, we describe the methodology behind 3D non-LTE spectral line formation, its observational validation and practical implementation in spectroscopic analyses, and the impact seen for the Sun and other stars. We review the abundances of elements for which high-quality model atoms can be constructed and include a description of common data sources. We compare 3D non-LTE abundances with 1D LTE, 1D non-LTE and 3D LTE ones and discuss the usefulness of alternative approaches, such as horizontally- and temporally-averaged 3D models, so called $\mtd{}$ models.

\section{THEORY AND METHODS}
\label{sect:method}

In this section we summarise the theory and methods relevant to deriving
3D non-LTE abundances in late-type stars.  
\sect{sect:method-atmospheres} gives an overview of the
construction of three-dimensional RHDs
model atmospheres, 
and \sect{sect:method-atoms} describes the
construction of model atoms, as well as how their
complexities are reduced.  \sect{sect:method-spectra}
outlines the generation of 3D non-LTE stellar spectra
given these key inputs.
This section closes with a synopsis of the validation of the
3D model atmospheres and the 3D non-LTE stellar spectra,
in \sect{sect:method-validation}.

\subsection{Model atmospheres} 
\label{sect:method-atmospheres}

\begin{figure*}
    \includegraphics[width=5in]{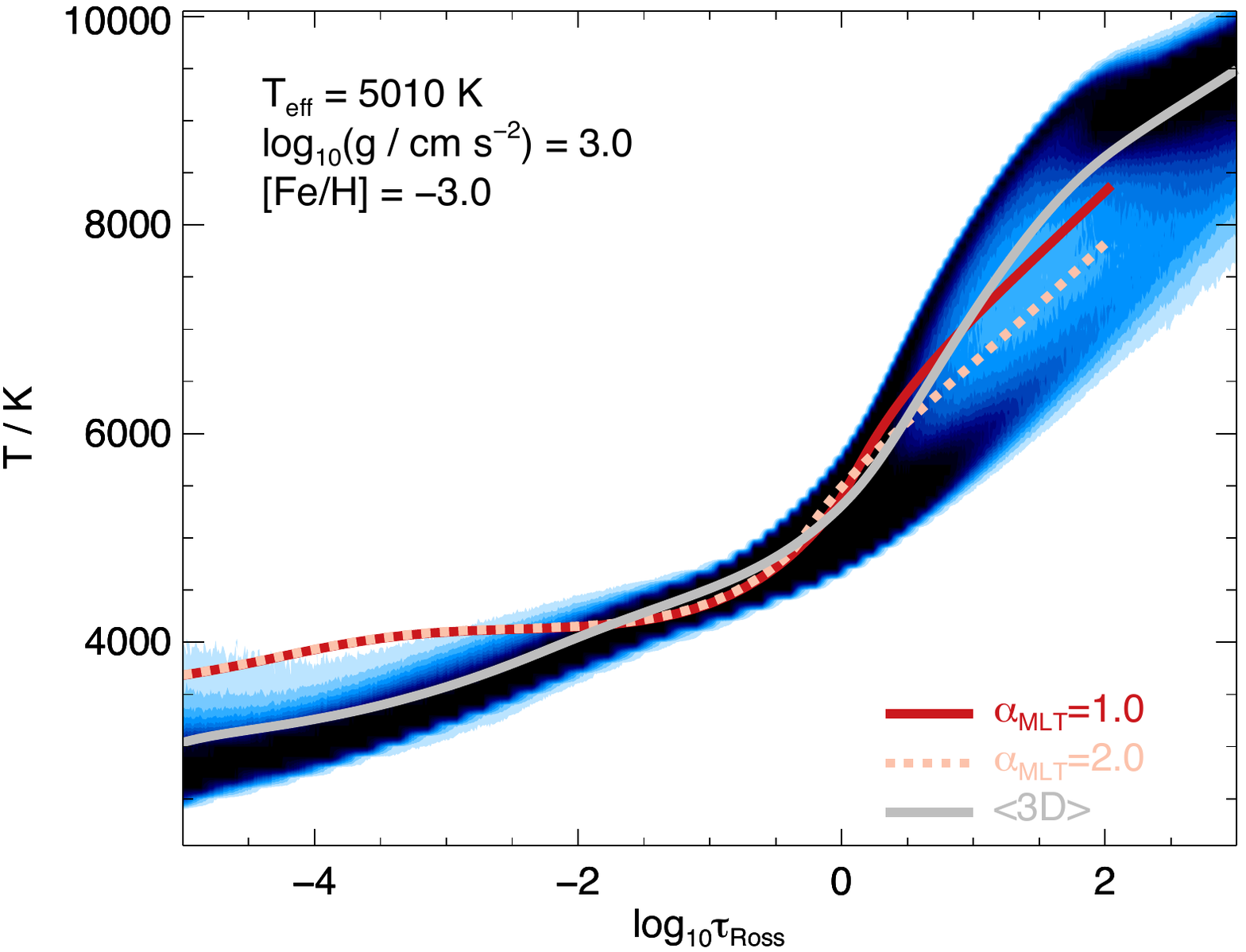}
    \caption{Distribution of gas temperature with logarithmic
    vertical optical depth in a
    3D RHD model atmosphere from the \stagger{}-grid.
    Also shown are theoretical 1D models computed
    with different mixing lengths, and a $\mtd{}$ model.
    From \citet{2018A&A...615A.139A}.}
\label{fig:atmos-structure}
\end{figure*}

The first key input to 3D non-LTE spectrum synthesis codes is the model atmosphere.
In the context of ``3D non-LTE abundance analyses of late-type stars'', we
usually refer to 3D RHD
``box-in-a-star'' large-eddy simulations of stellar surface convection.  
In general, simulations including magnetic fields
(3D radiation magnetohydrodynamics; 3D RMHD) are beyond the scope
of this paper, although some results are briefly discussed in 
\sect{sect:method-effects-mhd}. The 3D RHD simulations
aim to provide a realistic ab initio description of the continuum- and
line-forming regions of the stellar photosphere via a fundamental treatment of
the mass, momentum, and energy transfer due to convection, combined
with a non-grey treatment of radiative transfer.  In contrast, 1D
models must employ some simplified picture of turbulent
convection that involves adjustable parameters,
such as the Mixing Length Theory
\citep[][]{1965ApJ...142..841H,
1977ApJ...214..196G,1990SoPh..128..161B}.
Although one can
attempt to calibrate mixing length parameters to reproduce the mean
structures of the deeper layers of 3D RHD models
\citep[e.g.][]{2014MNRAS.445.4366T,2019A&A...621A..84S},
it fails to properly describe the temperature stratification
close to the optical surface and in the convectively-stable line-forming 
regions above it (\fig{fig:atmos-structure});
and there can be particularly large discrepancies for metal-poor stars
\citep{2007A&A...469..687C}.  The 1D hydrostatic
models also cannot make predictions related
to stellar granulation that cause line shifts, asymmetries, and broadening
\citep[e.g.][]{2021A&A...649A..16D}, 
that can also act to desaturate and strengthen strong lines.

We also note that 3D RHD models should not be
confused with horizontally- and temporally-averaged 3D RHD 
models, hereafter $\mtd{}$
models.  These $\mtd{}$ models are sometimes referred to in the
literature as 3D models, but this should be avoided.  Such models suffer
from several disadvantages that theoretical 1D hydrostatic models are also prone
to, due to the the anisotropy of convective motions and the non-linearity of
spectral line formation in stellar atmospheres \citep{2011ApJ...736...69U}.

\subsubsection{Codes and grids}
\label{sect:method-atmospheres-codes}

\begin{figure*}
    \includegraphics[width=5in]{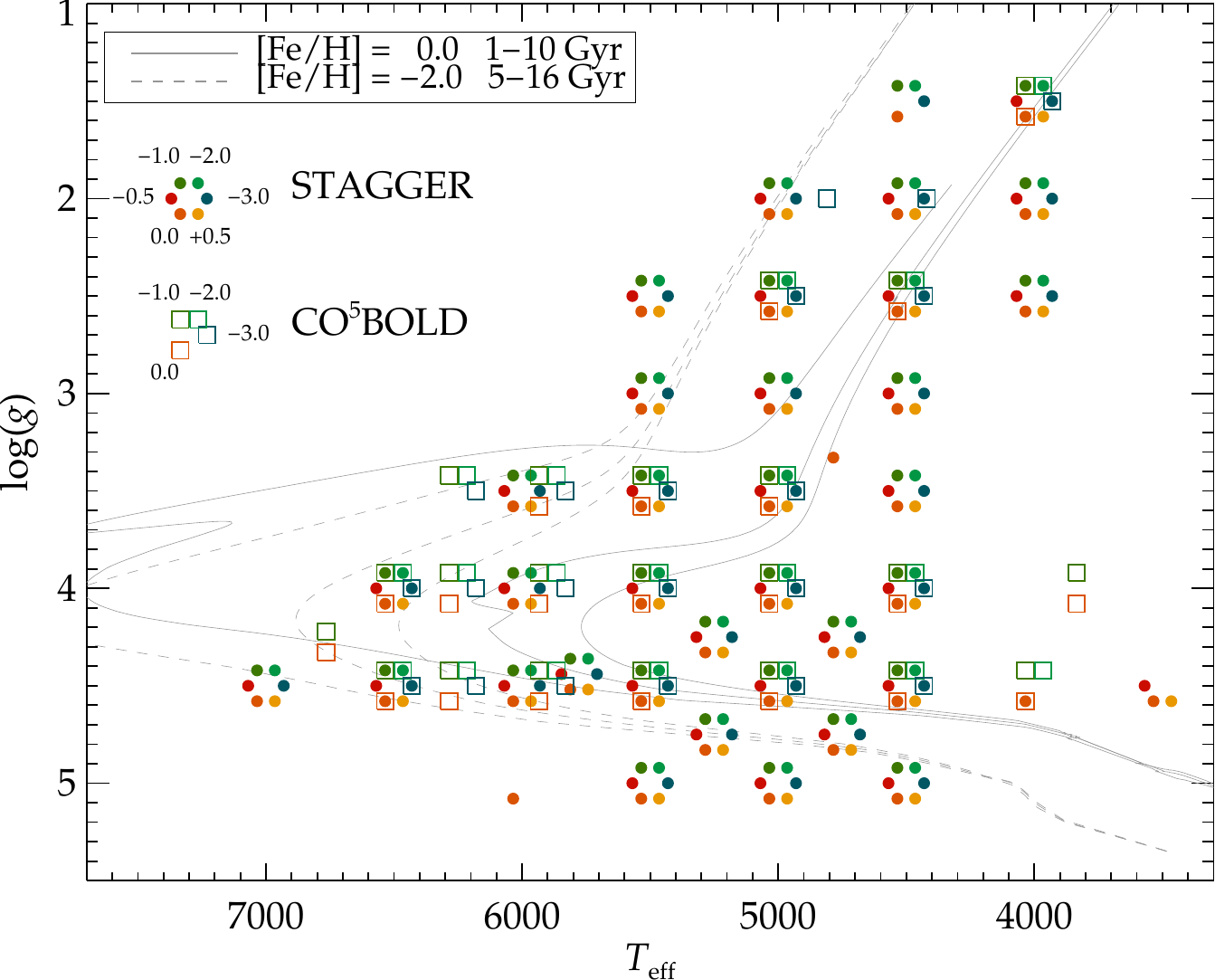}
    \caption{Extent of the \stagger{}-grid and 
    \cifist{}-grid of 3D RHD model atmospheres,
    in $\teff$, $\lgg$, and $\feh$.
    Grid data are shown for 214 \stagger{} models from \citet{Diaz_stagger}
    and 84 \cobold{} models from the \cifist{}-grid used by \citet{2022A&A...661A..76B}. The values of $\teff$ are centered on grid nodes for clarity and representative to within $\pm100$\,K of the true model. Metal-poor and solar-metallicity MIST isochrones are included for comparison, with ages as indicated by the legend. The $\feh=-4$ models of the \stagger{}-grid have been omitted from the figure (see \citet{Diaz_stagger}).}  
\label{fig:stagger-cifist-grids}
\end{figure*}

Most 3D non-LTE abundance analyses for stars other than the Sun have employed
simulations based on the parallelised 3D
R(M)HD codes \stagger{} \citep{Nordlund:1995, 2018MNRAS.475.3369C,stein:StaggerCode}
and \cobold{}
\citep{2004A&A...414.1121W,2012JCoPh.231..919F,2012A&A...547A.118L}.  Other  codes that are
commonly used for realistic simulations of
stellar surface convection include for example \muram{}
\citep{2005A&A...429..335V} and \antares{} \citep{2010NewA...15..460M};
further examples can be found within the 
review of \citet{2017LRCA....3....1K}.  
Reassuringly, comparisons of solar
models with independent codes tend to show rather
small differences in granulation and
mean structure \citep{2009MmSAI..80..701K,
2012A&A...539A.121B,2021ApJ...923..207C}.
These are likely to be negligible in the context of solar abundance
determinations.  For instance, the comparisons of 
\ion{Si}{I} lines based on \stagger{} and \cobold{} models 
(albeit also using different 3D LTE spectrum synthesis codes) 
presented in Appendix A of \citet{2022A&A...668A..48D}
suggests differences of at most $0.01\,\mathrm{dex}$ in abundance.
Similarly, only small differences are found in 3D LTE for lines of 
\ion{C}{I} (Sect.~4.2 of \citealt{2019A&A...624A.111A}) and 
\ion{N}{I} (Fig.~5 of \citealt{2020A&A...636A.120A}).

The reason for the dominance of \stagger{} and \cobold{} in this context is in
part due to them being used to construct grids of 3D RHD models; namely, 
the \stagger{}-grid \citep{2013A&A...557A..26M,2022MNRAS.514.1741R,Diaz_stagger} and the
\cifist{}-grid \citep{2009MmSAI..80..711L,2012A&A...547A.118L,2022A&A...661A..76B}.  These grids
span a wide enough range of three fundamental stellar parameters to be useful
for abundance analyses: effective temperature $\teff$, surface gravity $\lgg$,
and Fe abundance $\feh$, which serves as a proxy for the overall stellar
metallicity $Z$.  Other grids of models exist
\citep[e.g.][]{2013ApJ...767...78T,2013ApJ...778..117T,2013ApJ...769...18T}, but
are more limited in stellar parameter space. 
We illustrate the coverage of the 
\stagger{}-grid and \cifist{}-grid in \fig{fig:stagger-cifist-grids}. By comparing to MIST model isochrones \citep{2016ApJS..222....8D},
we see that
the grids have good coverage of the main sequence and subgiants with $4500<\teff<6500$\,K,
but would benefit from a more complete coverage 
of the metal-poor red giant branch. 
It may be important
to introduce more parameters into these grids to properly model stars with
peculiar abundances and recently a version
of \stagger{} was developed to help facilitate this \citep{2023A&A...677A..98Z}.
Selected snapshots of the original \stagger{}-grid models,
reprocessed and tailored for abundance analyses, will soon be publicly available
\citep{Diaz_stagger}.

\subsubsection{Overview of the simulation approach}
\label{sect:method-atmospheres-overview}


We refer the reader to other literature
\citep[e.g.][]{1998ApJ...499..914S,2012JCoPh.231..919F,
2017LRCA....3....1K,2018MNRAS.475.3369C} for a full discussion of performing the
3D RHD simulations; here we give just a brief overview with emphasis on details
pertinent to 3D non-LTE abundance analyses.  Fundamentally, the simulations
solve the coupled time-dependent equations describing the conservation of mass,
momentum, and energy.  The simulations in the \stagger{}-grid and \cifist{}-grid
are of similar extent and resolution. In the case of the former, they adopt a
Cartesian mesh with $240^{3}$ grid points that includes $10$ vertical ghost
zones distributed at the lower and upper boundaries, 
with finest resolution around the continuum forming regions.  They span a
vertical extent from $-5.0<\log\tau_{\mathrm{Ross.}}<+6.0$ or around 14 pressure
scale heights, thus capturing the outer part of the convective envelope as well
as the photosphere. They exclude the chromosphere, which is assumed to not
significantly impact the structure of the continuum- and line-forming regions of
the photosphere.  They cover a relatively small area of the stellar
disc, enclosing at least 10 granules at any time of the simulation; 
for the Sun this corresponds to around $0.004\%$ of the total disc area \citep{2022MNRAS.514.1741R}.  Given
the limited volume of the simulation box, the curvature of the stars can be
neglected.  Moreover a constant acceleration due to external forces is assumed
throughout the box that corresponds to the $\lgg$ of the model, and
centrifugal and Coriolis forces are neglected (the models are non-rotating).  
The horizontal boundaries are periodic and the vertical
boundaries are open.  The entropy and thermal pressure of the inflowing gas at
the bottom boundary are assumed to be constant.  The time-averaged $\teff$ of the simulation is a function of this entropy combined with the
other parameters of the simulation, and this is why the nodes of the
\stagger{}-grid and \cifist{}-grid are slightly irregular in $\teff$.  Since the
granulation as well as the $\teff$ of the simulations, calculated
via integration of the emergent flux, fluctuate with time, it is necessary to
carry out spectrum synthesis calculations on several snapshots; consequently the
\stagger{}-grid models span around two convective turnover times, corresponding
to several periods of the fundamental p-mode of the simulation domain.

The influence of radiation enters mainly through the energy equation, where it is necessary for
describing the radiative heating and cooling rates. The radiative
transfer equation is solved in the layers close to and above the optical surface
at each time step of the simulation, for a small number of opacity bins
(\sect{sect:method-atmospheres-micro}).  \stagger{} uses a modified version of the
long characteristics method of \citet{1964CR....258.3189F} as described in
\citet[][]{2003ASPC..288..519S}, and integrates the intensity for nine rays, or 
eighteen directions on the unit sphere. The latest version of \cobold{} has a short
characteristics solver that may give better overall time performance
\citep{2017MmSAI..88...22S}.  Radiation also enters as pressure in the momentum
equation, but it is generally negligible for the stars of interest here; in the
\stagger{}-grid it is estimated in an approximate way by assuming it to be
isotropic and carrying out radiative transfer calculations on $\mtd{}$ models.

\subsubsection{Microphysics}
\label{sect:method-atmospheres-micro}


The EOS relates state variables and gives the distribution
of nuclei in atomic, ionic, and molecular states.
These are used to calculate opacities under the assumption of LTE
(via the Boltzmann distribution; e.g.~\citealt{2022A&A...659A..87H}).
The code \cobold{} calculates this using the Saha equation and the analogous
equation for molecular equilibrium, whereas
\stagger{} adopts the free energy minimisation approach of
\citet{1988ApJ...331..815M}, often referred to as the MHD EOS, as implemented by
\citet{2013ApJ...769...18T}.  In the latter, the 17 most abundant elements as well as the
$\mathrm{H_{2}}$ and $\mathrm{H_{2}^{+}}$ molecules are considered, and the
occupation probability formalism is used to account for the dissolution of
Rydberg levels into the continuum \citep{1988ApJ...331..794H}.  
Recently \freeeos{} \citep{2012ascl.soft11002I}, also based on free energy
minimisation, was implemented into \stagger{}:  \citet{2023A&A...677A..98Z}
show, at least in the context of stellar spectroscopy, that this change in EOS
leads to negligible differences on the solar simulations, at least when the elemental
abundance ratios are kept fixed.  This is consistent with the findings of
\citet{2015AnGeo..33..703V}.  In all cases,
partition functions as well as ionisation energies and molecular equilibrium
constants for the species of interests \citep[e.g.][]{2016A&A...588A..96B} are
needed (in late-type stellar
atmospheres it is usually sufficient to consider up to doubly-ionised species), 
and instantaneous chemical equilibrium is assumed.

The larger uncertainty comes from the treatment of opacities.
This is a key aspect of 3D RHD simulations, as radiation has a
strong influence on the mean temperature stratification of the model via the
heating and cooling rates.  A full monochromatic solution of the radiative
transfer equation at every time step is computationally prohibitive, even in
LTE.  Consequently, 3D RHD models typically solve the radiative transfer for a
small number of representative opacity bins
\citep{1982A&A...107....1N,2000ApJ...536..465S}.  In the \stagger{}-grid, 12
such bins are used.
A careful selection of opacity bins is crucial to obtain the
correct temperature structure in the optically thin layers. 
\citet{2018MNRAS.475.3369C} performed simulations for the
metal-poor (adopted $\feh=-2.5$) red giant HD 122563 with varying number of
bins, and demonstrated that the 12-bin solution is in good agreement with the
48-bin one, provided that the opacity bins are constructed to take
into account both strength and frequency (see their Fig.~1 and Fig.~9). 
At higher metallicities, the simulations are expected to be less sensitive to the
detailed treatment of the radiation field.  Still, future studies might consider
improving the opacity binning scheme, for example by exploring
the use of clustering algorithms.

The source function is usually assumed to be Planckian as per LTE.  However, it
should be noted that the \stagger{}-grid models estimate the continuum
scattering contribution to the extinction (\sect{sect:method-spectra-opacity}) and
exclude it from the integrand in the optically thin layers.  This is necessary
to avoid errors of up to around $500\,\mathrm{K}$ in the upper layers of metal-poor
stars \citep{2011A&A...528A..32C}.  
The most recent models in the \cifist{}-grid adopt a similar approach.


\subsubsection{Cost}
\label{sect:method-atmospheres-cost}

The computational cost of 3D RHD models is determined by various timescales that are discussed in detail in for example \citet{2002A&A...395...99L}, \citet{2012JCoPh.231..919F}, and \citet{2017LRCA....3....1K}. For stars similar to the Sun, with a suitable choice of initial conditions such that the mean structure is already well described in the deeper layers, relaxation should occur after some number of convective turnover times, $\tau_{\mathrm{conv.}}\sim H_{\mathrm{p}}/v_{\mathrm{conv.}}$, where $H_{\mathrm{p}}\propto\teff/g$ is the pressure scale height and $v_{\mathrm{conv.}}$ is the convective velocity; here we assume the latter to scale as the sound speed $c_{\mathrm{s}}\propto\sqrt{\teff}$.  It is relevant to consider this time scale relative to the time step $\Delta t$ of the simulations, which is usually dictated by the minimum radiative timescale near the optical surface $\tau_{\mathrm{rad.}}\sim\teff^{-3}$. The ratio $\tau_{\mathrm{conv.}}/\tau_{\mathrm{rad.}}$ then scales as $\teff^{3.5}g^{-1}$. A typical red giant with $\teff=4500\,\mathrm{K}$ and $\lgg=1.5\,\mathrm{K}$ gives a cost around $400$ times more than that for the Sun, demonstrating how such models are often the most expensive to compute within the existing grids. If a solar model takes of the order $300$ CPU hours to run \citep{2012JCoPh.231..919F}, a red giant model takes $0.1$ MCPU hours. However, these estimates do not include the often time consuming steps to optimize the opacity bin limits, damp oscillations, and arrive at the target $\teff$ by adjusting the bottom boundary condition \citep{Diaz_stagger}; these can take significant time and may even dominate the overall costs.

\subsection{Model atoms}
\label{sect:method-atoms}

The second key input to 3D non-LTE spectrum synthesis codes is the so-called
``model atom''.  This represents all of the additional atomic (and ionic and
molecular) data needed to relax the assumption of LTE, via solving for the
statistical equilibrium (\sect{sect:method-spectra-nlte}), sometimes referred to
as collisional-radiative modelling:
a) the energies ($E_{i}$) and statistical weights
($g_{i}=2J_{i}+1$) of the atomic, ionic, and also (less commonly) molecular
levels; b) the Einstein coefficients or oscillator
strengths for the bound-bound radiative transitions, and
the cross-sections for the bound-free radiative transitions;
and c) the cross-sections or rate coefficients for the most important
inelastic collisional processes, which, in late-type stars,
are typically involving either electrons or neutral hydrogen atoms.

Our focus below is on the construction of tractable yet reliable 
model atoms for 3D non-LTE abundance analyses.
A full description of the methods for how to go about accurately
measuring or calculating the data themselves
is beyond the scope of this paper, and we recommend the review of
\citet{2016A&ARv..24....9B} as a starting point for more insight into this.
Moreover, we focus on neutral and singly-ionised atomic species, 
which have been the main interest of the vast
majority of non-LTE studies for late-type stars to date, although we briefly
discuss molecules later in \sect{sect:method-effects-nlte}. 
In this context, excellent resources are the National Institute of Standards and Technology Atomic Spectra Database (NIST ASD; \citealt{2020Atoms...8...56R}), the databases of the Opacity Project (TOPbase; \citealt{1993A&A...275L...5C}) and the Iron Project  (TIPbase; \citealt{2005A&A...437..345N}), the Kurucz Smithsonian Atomic and Molecular Database (Kurucz database; \citealt{2017CaJPh..95..825K}), and VALD (formerly the Vienna Atomic Line Database; \citealt{2015PhyS...90e4005R}).  Relevant and useful data can also be found in the Nahar OSU Radiative database (NORAD-Atomic-Data; \citealt{2020Atoms...8...68N}) the CHIANTI atomic database (CHIANTI; \citealt{2023ApJS..268...52D}), and the Atomic Data and Analysis Structure database (OPEN-ADAS; \citealt{2011AIPC.1344..179S}). These and other databases can also be accessed via the Virtual Atomic and Molecular Data Centre (VAMDC; \citealt{2020Atoms...8...76A}).
The databases mostly contain information on energies and radiative transitions, although some also include inelastic electron collision data.
However, they do no include results for inelastic
collisions with hydrogen atoms.  Such processes are crucially
important in non-LTE studies of late-type stars,
as we discuss below (\sect{sect:method-atoms-collisions}).



\subsubsection{Levels} 
\label{sect:method-atoms-levels}

\begin{figure*}
\includegraphics[width=5in]{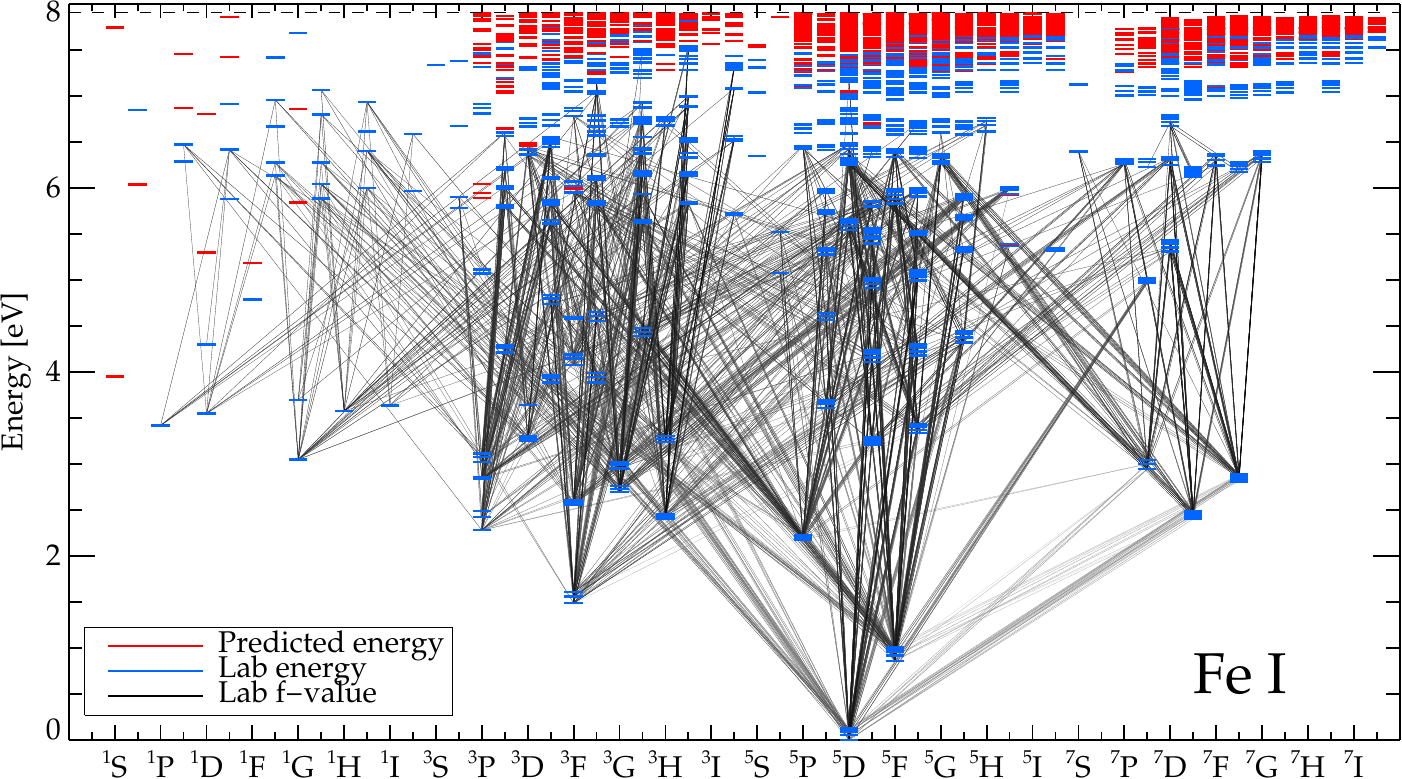}
\caption{Grotrian diagram illustrating part of a model
    atom of \ion{Fe}{I}. The energy levels are taken from the Kurucz database and include experimentally confirmed (blue) and predicted levels (red). Only transitions with experimentally measured transition probabilities are shown. Data sources are listed in \citet{2017MNRAS.468.4311L}.}
\label{fig:atom}
\end{figure*}

The energy levels establish the infrastructure for the rest of the model
atom. For species other than hydrogen (see \citealt{2010ADNDT..96..586K}), 
the accuracy of experimental measurements of energies, via
spectroscopic measurements of wavelengths and term analyses
\citep[e.g.][]{1994ApJS...94..221N}, almost always far exceeds that of
theoretical calculations.
As such, the usual starting point for energies
is NIST ASD, which contains a critical compilation of experimental energy
levels.  Often these data are not sufficiently complete for
non-LTE modelling, especially for heavier elements.
Model atoms must then make use of theoretical data.
For instance, for \ion{Fe}{I} \citet{2017MNRAS.468.4311L} use the Kurucz database, which complements experimental data with semi-empirical predictions;
in this case predicted fine-structure levels 
outnumber the experimental ones by a factor of three, as can be gauged from \fig{fig:atom}.

Model atoms usually must have a complete description of the levels
of the species under investigation (neutral or singly-ionised), as
well as the ground level(s) of the neighbouring ionisation stages.
Elements with intermediate excitation potentials 
may require a comprehensive description of 
both the neutral and singly-ionised species, as they have comparable populations in
late-type stellar atmospheres. A natural maximum
for the number of excited levels to include comes from considering the
dissolution of Rydberg levels due to perturbations by charged particles.  Where
this becomes significant can be roughly estimated from Eq.~8.86 of
\citet{2015tsaa.book.....H} for the reduction in binding energy, which gives
around $0.01\,\mathrm{eV}$ in the deepest layers of the atmospheres of warm
late-type stars, and smaller values in the line-forming regions.  
In reality, the levels this close to the ionisation threshold
have small Boltzmann factors and are not significantly populated in late-type
stars, and thus are often omitted.
Usually, for the neutral species for example, only 
levels up to some tenths of an electron volt
from the ionisation threshold are included explicitly.  This is
sufficient to allow for collisional coupling to the next ionisation stage
(\citealt{2011A&A...528A..87M}; cf.~the mean electron kinetic energy
$E=\frac{3}{2}k_{\mathrm{B}}T=0.75\,\mathrm{eV}$ for solar photospheric
conditions).  

\subsubsection{Radiative transitions} 
\label{sect:method-atoms-radiation}

\begin{figure*}
\includegraphics[width=5in]{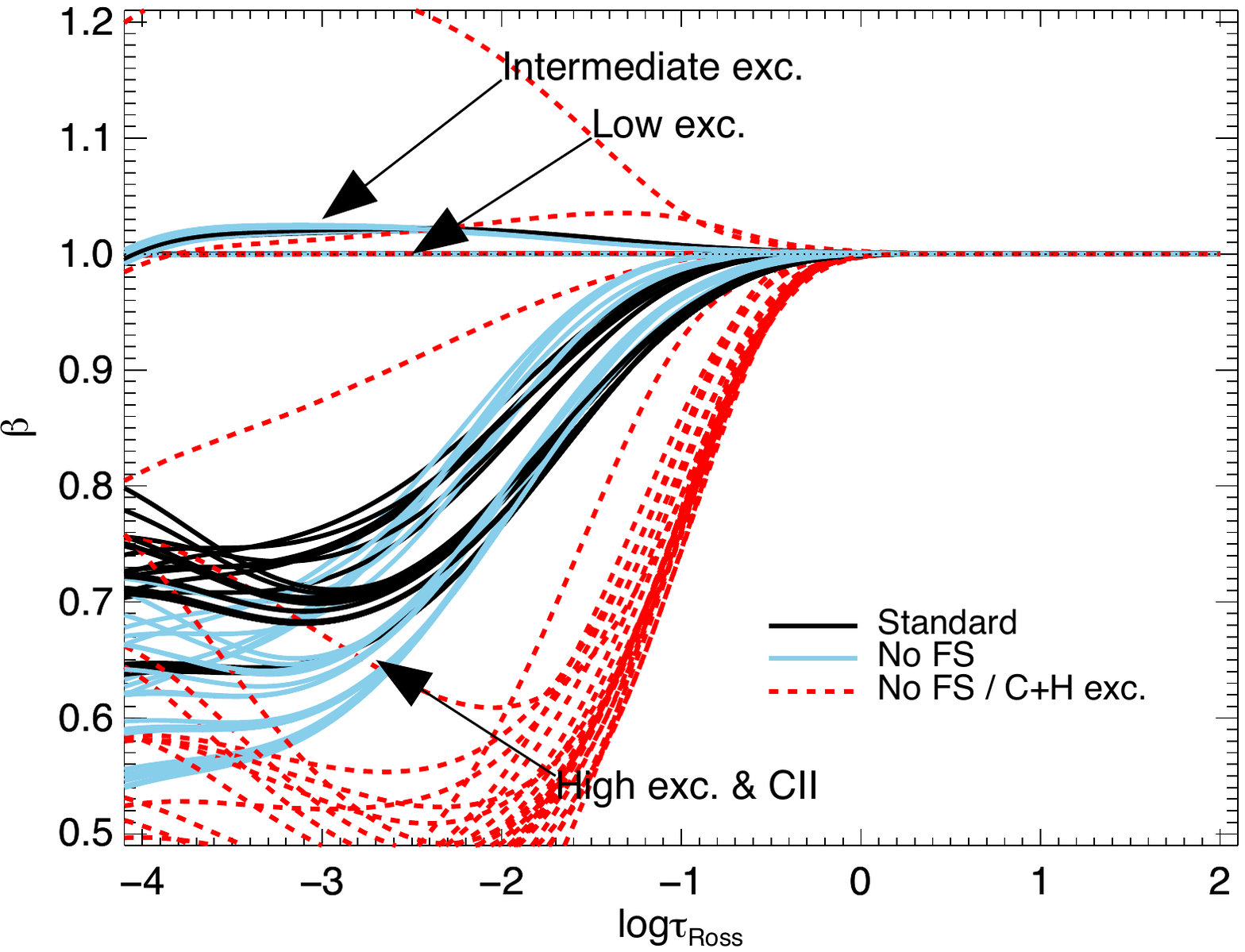}
\caption{Departure coefficients for levels of neutral C
 as well as the ground level of singly-ionised C
 in a $\mtd{}$ model of the solar atmosphere,
    with levels labelled by excitation energy.
 Results are shown for a model atom where most fine structure is resolved 
    (``Standard''), where fine structure is merged
    (``No FS''), and where, in addition,
    excitation by hydrogen collisions is omitted
    (``No FS / C+H exc.''). From \citet{2019A&A...624A.111A}.}
\label{fig:fs}
\end{figure*}

The radiative rates $R_{ij}$ appearing in the equations of statistical
equilibrium (\sect{sect:method-spectra-nlte}) depend on the Einstein
coefficients or oscillator strengths for the bound-bound radiative transitions,
and the cross-sections for the bound-free radiative transitions, together with
the radiation field.  
NIST ASD contains a critical
compilation of experimental and theoretical data, including accuracy rating of transition probabilities, and is again the usual starting
point in the current context. 
The VALD repository can also be very useful, containing a wealth of
experimental and theoretical data compiled from the literature.
It can be worth scanning literature directly for the
latest results from the various groups measuring precise data
\citep[e.g.][]{2021JQSRT.27107703B,2022ApJS..259...44G,
2023A&A...672A.197B,2023MNRAS.519.4040C,2023ApJS..265...42D,
2023JQSRT.29808484Z}, and groups calculating extended data
of sufficient overall accuracy for stellar spectroscopy 
\citep[e.g.][]{2023ApJS..265...26L,2023A&A...674A..54L} based on for example
Multi-Configuration Dirac-Hartree-Fock (MCDHF) methods
\citep[e.g.][]{2019CoPhC.237..184F}.
To complete the model atom, such data
must often be complemented with the results of semi-empirical
or fully theoretical calculations, for example
from the Kurucz database, or from TOPbase, TIPbase,
and NORAD-Atomic-Data that employ the $R$-matrix method
\citep{2011rmta.book.....B}, as well as data compiled within CHIANTI. \fig{fig:atom} shows the subset of radiative transitions in \ion{Fe}{I} that have experimentally measured oscillator strengths, which in this case is 
only $0.5\%$ of the more than half a million fine-structure transitions with 
semi-empirical values from the Kurucz database.

Theoretical line data, for example from TOPbase, are often
calculated without fine structure.  The Einstein coefficients
or oscillator strengths can be redistributed over fine
structure levels under the assumption of LS-coupling (see for example 
Sect.~10.16.6 of \citealt{2006sham.book..175M}). 
Even if the fine structure levels are 
merged in order to reduce the complexity of the model 
atom as we describe in \sect{sect:method-atoms-reduction},
it may be important to consider this splitting in the calculation of the 
bound-bound opacity profiles. This is
especially the case when the multiplet components are unsaturated, but
resolvable and separated. In this situation, neglecting fine structure then
leads to a line that is significantly too strong, corresponding to an
angle-averaged intensity $J_{\nu}$ that is too small 
(see Appendix B of \citealt{2015A&A...583A..57S}).
This could, for example, lead to larger departures 
from LTE for some species due to enhanced photon losses (\fig{fig:fs}).

For photoionisation cross-sections, it is often necessary to use the results
of theoretical calculations.
For lighter elements these are usually drawn from TOPbase or 
NORAD-Atomic-Data.  Other data
also employing the $R$-matrix method can be found
in the literature \citep[e.g.][]{2017A&A...606A.127B,2019ApJ...874..144S}.
There are now also results for a number of important species based on
the $B$-spline $R$-matrix method \citep[BSR; e.g.][]{2013JPhB...46k2001Z,
2019PhRvA..99b3430Z}. 
Typically, for excited levels of heavier elements, it is necessary
to resort to hydrogenic approximations \citep[e.g.][]{2012A&A...540A..98M};
however, the error in this approximation
can be large enough to lead to, for example,
predict absorption instead of emission for \ion{Si}{II} lines
in hot stars \citep[][]{2022AstL...48..303M}.
As mentioned above, the energies in these theoretical
calculations are typically not of sufficient accuracy
for stellar spectroscopy, and for large deviations
the cross-sections should be shifted so that the predicted thresholds
match the experimental ones. Sharp
autoionisation resonances in the data, while potentially important to the non-LTE
solution, are similarly offset from their true frequencies, and also do not include for example
thermal broadening.  Thus it is justified to apply some
Gaussian smoothing to the data
\citep[e.g.][]{1998ApJS..118..259B,2003ApJS..147..363A,2003ApJS..146..417L}.  This also aids in
putting all of the photoionisation cross-sections onto a common frequency grid,
which helps reduce the computational cost of the post-processing
calculations \citep[e.g.][]{2008ApJ...682.1376B}.

\subsubsection{Collisional transitions}
\label{sect:method-atoms-collisions}

The collisional rates $C_{ij}$ appearing in the equations of 
statistical equilibrium
(\sect{sect:method-spectra-nlte}) depend on both the number density
of perturbers, and the collisional rate coefficients (which is the Maxwellian-averaged
collisional cross-section: $\langle\sigma v\rangle$). Electrons are important
perturbers in late-type stars because of their low masses and thus large mean
thermal velocities, and also because, as per the Massey criterion
\citep{1949RPPh...12..248M}, the collisions for
optical transitions should be close to resonance
\citep{2016A&ARv..24....9B}.  
As discussed in \citet{1993PhST...47..186L} and \citet{2011A&A...530A..94B},
hydrogen collisions were initially estimated to be of lesser importance,
noting that hydrogen atoms move more slowly
by a factor of $\sqrt{m_{\mathrm{e}^{-}}/m_{\mathrm{H}}}\sim 10^{-2}$, 
and that the transitions corresponding to
optical wavelengths are in the adiabatic regime and so should have much smaller
cross-sections per the Massey criterion.
However, this is offset by the
large abundance of hydrogen atoms: $n_{\mathrm{H}}/n_{\mathrm{e^{-}}}\sim10^{4}$
in the upper layers of the solar photosphere, with even larger ratios in cooler
and more metal-poor stars \citep{1984A&A...130..319S}.
Moreover, hydrogen collisions move closer to
resonance towards smaller wavelengths \citep{1993PhST...47..186L} and for this
reason one expects levels with very small energy differences, for example
Rydberg levels and fine structure levels, to be close to relative LTE in
late-type stellar atmospheres (collision data are now available to test this in
some cases; e.g.~\citealt{2023MNRAS.522.1265Y}).
In all, in light of quantum mechanical data for the collision
cross-sections (discussed below), the current consensus 
is that (de-)excitation and ionisation or recombination by
electron collisions, and (de-)excitation and charge transfer
or mutual neutralisation by hydrogen collisions,
are of critical importance for non-LTE studies of late-type stars.

For electron collisions, data for (de-)excitation 
($\mathrm{X}_{i}+\mathrm{e}^{-}\leftrightarrow \mathrm{X}_{j}+\mathrm{e}^{-}$) 
and ionisation or recombination
($\mathrm{X}_{i}+\mathrm{e}^{-}\leftrightarrow \mathrm{X}_{j}^{+}
+2\mathrm{e}^{-}$) based on $R$-matrix
calculations can be found via TIPbase, NORAD-Atomic-Data, CHIANTI, and Open-ADAS
for many species of interest.  Data based on more advanced
methods are also available for some species,
for example employing the BSR method mentioned above \citep{2022ApJS..259...52T}
or the
Convergent Close Coupling method \citep[CCC; e.g.][]{1992PhRvA..46.6995B,2020JPhB...53a5204V}.
It is reassuring that BSR and CCC collision
data have been demonstrated to agree to a remarkably high level
\citep{2017A&A...606A..11B,2019ADNDT.127....1D,2019A&A...627A.177R}.
In general, however, data from $R$-matrix and
more advanced calculations do not extend to Rydberg levels, and neutral
species of heavier elements are less well represented overall.  As such, it is
necessary to also employ more approximate methods.  
For excitation, the most common recipes are the semi-empirical
\citet{1962ApJ...136..906V} recipe and the semi-classical impact parameter
method as formulated by \citet{1962PPS....79.1105S}.
Their accuracies depend on the species and transitions in question;
for \ion{K}{I}, the former has errors of around a factor of ten,
with larger errors for transitions corresponding to longer wavelengths
(Fig.~2 of \citealt{2019A&A...627A.177R}).  The latter approach is 
expected to be more reliable overall, particularly for transitions between
Rydberg levels \citep{2016A&ARv..24....9B}, 
as also suggested by recent modelling
of the \ion{Mg}{I} infrared emission lines that are highly sensitive to the
collisional data \citep{2015A&A...579A..53O,2018ApJ...866..153A}.  Both
approaches employ the Bethe approximation, namely they use the oscillator
strength $f_{ij}$ and are thus only defined for permitted transitions. Some
discussion of different approximations for forbidden and intercombination
transitions can be found in \citet{2016A&ARv..24....9B}.  Similarly, the recipes
of \citet{1970ARA&A...8..329B} or \citet{1962amp..conf..375S} are often used for
ionisation; the former typically has errors of around a factor of two.

For hydrogen collisions, the availability of reliable quantum mechanical data are much
more limited. The challenge is that collisions in late-type stellar atmospheres
fall in the low energy regime and calculations must take into account the
structure of the quasi-molecule that forms as the hydrogen perturber and the
atom or ion approach each other \citep{2011A&A...530A..94B}.  
So-called ``full quantum'' (FQ) methods do this in two steps: first, using
quantum chemisty calculations to describe the potential energies and the
couplings for fixed nuclei \citep[e.g][]{1999JPhB...32...81C,
1999JPhB...32.5451D,2010CPL...488..145G}; and second, using these potentials and
couplings in quantum mechanical scattering calculations
\citep[e.g.][]{2003PhRvA..68f2703B,2010PhRvA..81c2706B, 2012PhRvA..85c2704B}.
Relatively extended FQ rate coefficients are available for low-lying levels of
neutral Li, Na, and Mg
\citep{2003A&A...409L...1B,2010A&A...519A..20B,2012A&A...541A..80B}.  In light
of these data it is now understood that charge transfer and mutual
neutralisation ($\mathrm{X}_{i}+\mathrm{H}\leftrightarrow
\mathrm{X}_{j}^{+}+\mathrm{H}^{-}$), dominate the overall statistical
equilibrium for Li and Na \citep[e.g.][]{2021ApJ...908..245B}.  
For Mg, however,
(de-)excitation ($\mathrm{X}_{i}+\mathrm{H}\leftrightarrow
\mathrm{X}_{j}+\mathrm{H}$) cannot be neglected  \citep{2015A&A...579A..53O}.
In fact, by perturbing the rate coefficients of different transitions,
it has been demonstrated for several species
that (de-)excitation by hydrogen collisions
can be the most important ingredient in the model atom
\citep[e.g.][]{2017MNRAS.464..264A,2018A&A...616A..89A,2019A&A...624A.111A,
2020A&A...636A.120A}. For 
Na and Li, the FQ data for mutual neutralisation compare well against
recent measurements from the Double Electrostatic Ion Ring Experiment
\citep[DESIREE;][]{2013RScI...84e5115S,2020PhRvA.102a2823E,
2021PhRvA.103c2814E} as well as from UCLouvain
\citep{2019ApJ...883...85L}, at least to the level where the discrepancies do
not significantly affect non-LTE abundance determinations
\citep{2021ApJ...908..245B}. The analogous comparisons
for Mg, however, have highlighted the 
practical challenges of carrying out complete and 
detailed FQ calculations for complex
species \citep{2022PhRvL.128c3401G,2023PhRvL.130b9901G}.

Recent years have seen the development of so-called ``asymptotic models'' for
inelastic hydrogen collisions
\citep[][]{2013PhRvA..88e2704B,2016PhRvA..93d2705B}.  These models should not be
confused with FQ models described above: rather than quantum chemistry
calculations and quantum mechanical scattering calculations, asymptotic models
employ a more simplified description of the potentials and couplings (for
example, linear combinations of atomic orbitals) and of the collision dynamics
(usually based on the semi-classical Landau–Zener model).  Nevertheless,
comparisons with the FQ data for Li, Na, and Mg show good
agreement for (de-)excitation and charge transfer or mutual neutralisation
processes, at least for the low-lying levels included in the FQ dataset and for
the processes with the largest rate coefficients, suggesting
order of magnitude accuracies  \citep{2016A&ARv..24....9B}. 
Asymptotic model data are now available for many species
(e.g.~for Fe, \citealt{2019MNRAS.483.5105Y}
and \citealt{2018A&A...612A..90B}; see also other papers by these authors
and collaborators).

The asymptotic models predict zero rate coefficients for highly
excited levels with energies less than $0.754\,\mathrm{eV}$ to the ionisation
threshold (corresponding to the electron affinity of hydrogen).  This is because
these models describe a particular interaction mechanism, namely electron
transfer between the atom and the hydrogen perturber via radial couplings at
avoided ionic crossings \citep{2011A&A...530A..94B}.  For levels above this
energy, the valence electron cannot bind to the hydrogen perturber.  The free
electron model of \citet{1991JPhB...24L.127K}, describing momentum transfer
between the perturber and the outer electron is appropriate to use here
\citep[e.g.][]{2015A&A...579A..53O}, and codes are available to calculate these
rates \citep{2017ascl.soft01005B}. 
Even below this threshold, asymptotic models predict low 
rate coefficients for collisional interactions happening at long
and short internuclear distances, where other interaction mechanisms dominate.
Thus, adding the coefficients from the asymptotic model and the free electron
model together may lead to a better overall description of hydrogen collisions;
as discussed in \citet{2018A&A...616A..89A}, 
the two models describe different interaction mechanisms and 
thus do not double count rates.
 Such an ``asymptotic+free'' approach has been shown to better reproduce the
CLVs of lines of \ion{O}{I}
\citep{2018A&A...616A..89A,2021MNRAS.508.2236B} and \ion{C}{I}
\citep{2019A&A...624A.111A};
this was also ultimately adopted by \citet{2019A&A...631A..80B}
in their study of \ion{Mn}{I} lines.
The caveat is that the free electron model is valid in the limit 
of high effective principal quantum numbers; extending to lower $n^{*}$ is an
extrapolation, but arguably necessary in the absence of a more complete theory.

In the absence of asymptotic model data for inelastic hydrogen collisions,
it is still common to use the Drawin
recipe (see Appendix A of \citealt{1993PhST...47..186L}).
As discussed in \citet{2011A&A...530A..94B} the recipe does not
describe the
physics in the low-energy regime relevant here.
It is thus usual to scale
the Drawin rate coefficients with a scale factor $S_{\mathrm{H}}$,
that can be calibrated astrophysically 
\citep[e.g.][]{2023MNRAS.524.3526M}, or possibly by comparison
with the rate coefficients of some similar species.
Another approach might be to use 
the general empirical fitting method presented in \citet{2018A&A...618A.141E}, 
the results of which could be combined with the free electron model as described above.

\subsubsection{Model atom reduction}
\label{sect:method-atoms-reduction}

Comprehensive model atoms as described above are usually too complex to be
used for 3D non-LTE abundance analyses, given the high
cost of the post-processing calculations.  Reducing the complexity of the level
structure sets the stage for further reductions, and a common first step is to
merge fine structure levels: levels differing in energy and
total angular momentum $J$ can be collapsed into single
terms labelled by their configurations as well as $S$, $L$, and $P$
(here assuming LS-coupling). The simplest approach is as follows:
\phantomsection\begin{IEEEeqnarray}{rCl}
    \label{eq:collapse-g}
    g_{\mathrm{I}}&=&
    \sum_{i\in \mathrm{I}} g_{i} \\
    \label{eq:collapse-e}
    E_{\mathrm{I}}&=& \sum_{i\in \mathrm{I}} E_{i}\,\frac{g_{i}}{g_{I}}
\end{IEEEeqnarray}
This exactly conserves the total LTE populations in the limit of zero
energy differences between the fine structure sublevels.  
The corresponding assumption is that the fine structure sublevels are in relative
LTE, meaning that their departure coefficients $b_{i}\equiv
n_{i}/n_{i,\mathrm{LTE}}$ are identical. This holds if, for example, the
levels are efficiently coupled via inelastic hydrogen collisions
(\sect{sect:method-atoms-collisions}).

Analogously, levels with different configurations and quantum numbers can be
merged into super levels \citep{1989ApJ...339..558A}.  As before, the assumption
is that the merged levels are in relative LTE.  It therefore makes sense to
merge similar levels together, for example those that are close in energy, those
that have the same core configurations, those in the same spin system $S$ and
with the same parity $P$, or those that are close in orbital angular momentum
$L$.  In general, the merging should be at least validated a posteriori via
non-LTE radiative transfer calculations on 3D RHD 
model atmospheres, or $\mtd{}$ or 1D
columns, that are representative of the stars of interest. 
In some cases, very simple level
structures can be constructed in this way
depending on the lines and stars of interest (e.g.~the
five level atom of \citealt{2013ApJ...772...89L}, used to study the \ion{Mg}{II}
H and K lines in the solar atmosphere).

Reducing the number of radiative transitions in the model atom is an effective
way of reducing the overall computational cost of the
post-processing calculations.  For levels merged
into terms and super levels, one can merge
transitions into super transitions.
There are multiple ways to approach this, that give slightly different results.
For example, for lines $i\rightarrow j$ 
with oscillator strengths $f_{ij}$ connecting components of the super levels $I$ and $J$,
the oscillator strength of the super line $f_{IJ}$ could be defined as follows:
\phantomsection\begin{IEEEeqnarray}{rCl} 
    \label{eq:collapse-f}
    f_{IJ}&=&
    \left(\sum_{i\in I}g_{i}\,
    \exp{\left(-E_{i}/\left(k_{\mathrm{B}}T_{\mathrm{a}}\right)\right)}
    \right)^{-1}\, \sum_{j\in J} \sum_{i\in I} 
    f_{ij}\, g_{i}\,
    \exp{\left(-E_{i}/\left(k_{\mathrm{B}}T_{\mathrm{a}}\right)\right)} 
\end{IEEEeqnarray} 
Here $T_{\mathrm{a}}$ is a typical temperature for the
line-forming region of the atmosphere (e.g.~$5500\,\mathrm{K}$).
All other quantities can then be determined consistently based on
these merged values to ensure the correct asymptotic behaviour. 
Bound-free transitions can be merged in a similar fashion, for example
replacing $f_{ij}$ with the photoionisation cross-sections $\sigma_{ij}$; 
this is identical to Eq.~18.160 of \citet{2015tsaa.book.....H} 
when the occupation probabilities are set to unity.

Even when lines in the model atom 
are merged as per \eqn{eq:collapse-f}, 
for important lines (\sect{sect:method-atoms-radiation}) it
is desirable to explicitly consider the splitting of 
bound-bound opacity for the calculation of $J_{\nu}$.
For example, a particular radiative transition can be described as
a composite of Voigt profiles of different strengths and centred on different
wavelengths. \citet{2015A&A...583A..57S} describe another efficient
approach for the \ion{O}{I} $777\,\mathrm{nm}$ triplet,
where the oscillator strength of the collapsed line profile
(\eqn{eq:collapse-f}) is artificially reduced by a factor of three,
only for the calculation of $J_{\nu}$.

A complementary approach to the reduction of the model atom is based on
consideration of the radiative bracket,
$n_{i}\,R_{ij}-n_{j}\,R_{ji}$.
One can carry out non-LTE radiative transfer
calculations on representative 3D RHD 
model atmospheres, or $\mtd{}$ or 1D columns to
calculate these quantities. Radiative transitions with
small absolute radiative brackets can be identified and excluded from
the model atom, while retaining the collisional coupling.  This approach was adopted by 
\citet{2017MNRAS.468.4311L} and \citet{2022A&A...668A.103M}
for the study of Fe and Ti line formation, respectively.  Investigations into more complex
ways of merging and removing transitions, perhaps 
in a semi-automated way to minimise changes in departure coefficients
\citep[e.g.][]{2008ApJ...682.1376B} or with the aid of machine learning
algorithms, could be beneficial for more accurate 3D non-LTE modelling of
complex species.

Unlike for radiative transitions, including more collisional transitions does
not significantly impact the cost of 3D non-LTE spectrum synthesis.
Nevertheless, it may be necessary to merge collisional transitions associated
with super-levels.  This can be done in an analogous way to radiative
transitions: for example, by using \eqn{eq:collapse-f} or
and replacing $f_{ij}$ with the upwards
rate coefficient.  For a given collisional transition, it is advisable to
specify only either the upward or downward (merged) rate coefficient in the
model atom, and then to calculate the rate of the reverse process consistently
in the spectrum synthesis code via detailed balancing. This ensures that the
solution converges to LTE in the deep layers, and in the limit of very large
collisions.

\subsubsection{Cost}
\label{sect:method-atoms-cost}

Compared to the 3D RHD models (\sect{sect:method-atmospheres-cost}) and the 3D non-LTE spectrum synthesis (\sect{sect:method-spectra-cost}), it is less relevant to talk about computational costs for the model atoms, because most non-LTE modellers interested in stellar abundances are primarily compiling data already available in the literature, rather than carrying out atomic and molecular structure calculations themselves.  Nevertheless, the human time for even this task can be considerable; a whole PhD thesis can easily be based around constructing and using a model atom for just one or two elements.  The most time consuming tasks include linking together datasets in a consistent way when these datasets have different formats and often with sometimes inconsistent labelling or different physical assumptions; and reducing the complexity of the atom and validating this reduction.  Semi-automated approaches are possible for light elements by utilising the databases listed in \sect{sect:method-atoms}. However, these do not always contain the best available data; especially, as noted above, they do not contain data for inelastic hydrogen collisions.

\subsection{Spectrum synthesis}
\label{sect:method-spectra}

\begin{figure*}
\includegraphics[width=5in]{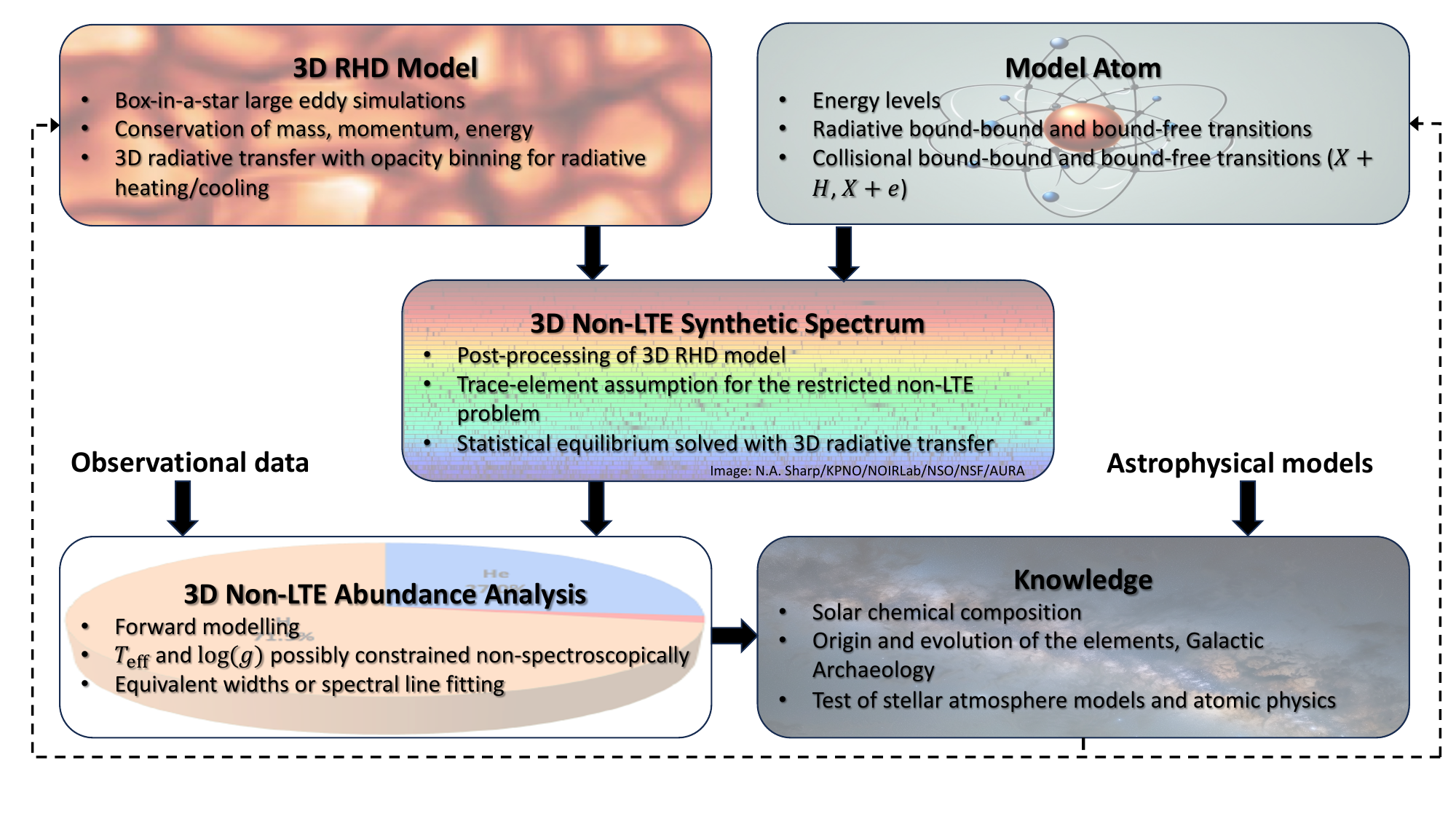}
\caption{Highly simplified schematic of the workflow for 3D non-LTE
    abundance analyses.}
\label{fig:schematic}
\end{figure*}

Abundance analyses proceed via forward modelling
(\fig{fig:schematic}), that is to say one starts with constructing
realistic synthetic spectra with different parameters, and then fits them to
observational data to deduce stellar properties.
These synthetic spectra are usually constructed via
post-processing of model atmospheres: the model atmospheres are
constructed first with a simplified treatment of radiative transfer (with
opacity bins, and with a Planck source function, possibly with
some treatment for isotropic, coherent scattering in the continuum),
that is assumed to be sufficient for describing the
radiative heating and cooling rates; and the synthetic spectra are calculated in
a separate step, often with an independent code, by employing detailed radiative
transfer calculations with typically much higher frequency resolution and with
more accurate atomic, ionic, or molecular data relevant to the spectral lines of
interest.

Furthermore, it is common to invoke the trace-element approximation,
and to only consider the restricted non-LTE problem.  The first approximation
means that differences between the elemental abundances of a suitably
chosen model atmosphere (for example with $\feh$ chosen to match observations,
and the other elements appropriately scaled relative to the solar composition)
and those of the star do not correspond to significant errors in the model
atmosphere structure. The analysis can then proceed by varying the
abundances only in the post-processing stage, without recomputing the model
atmospheres themselves.  The trace-element approximation is also commonly used in
1D LTE analyses.  The
restricted non-LTE problem is closely related; changes to the absorber
populations and hence opacities due to non-LTE modelling are assumed to be
sufficiently small to not significantly affect the model atmosphere structure.

\subsubsection{Codes}
\label{sect:method-spectra-codes}
There are a handful of codes that have been used for 3D non-LTE abundance analyses of late-type stars.  Most results have been based on either the frequency- and domain-decomposed MPI-parallelised codes \multitd{} \citep{1999ASSL..240..379B,2009ASPC..415...87L} and its offshoot \balder{} \citep{2018A&A...615A.139A}; or \nltetd{} \citep{2007A&A...473L..37C}, the latest version of which solves the full 3D non-LTE radiative transfer problem \citep{2015A&A...583A..57S,2017MmSAI..88...22S}.  We also refer the reader to the MPI-parallelised code \porta{} \citep{2013A&A...557A.143S}, that has to our knowledge not yet been used for 3D non-LTE abundance analyses and the 3D LTE codes \linfortd{}, which can read departure coefficients \citep{2017MmSAI..88...82G}, \scate{} \citep{2011A&A...529A.158H}, \optimtd{} \citep{2009A&A...506.1351C}, or \asset{} \citep{2008ApJ...680..764K}. Several other relevant codes use the so-called 1.5D approximation, whereby each column of the 3D RHD model atmosphere is treated independently for the solution of the statistical equilibrium.  The quality of this approximation depends on the line, species, and star as well as the desired accuracy (e.g.~Sect.~3.3 of \citealt{2016MNRAS.455.3735A} for a discussion about the \ion{O}{I} $777\,\mathrm{nm}$ lines; and Sect.~3.2.3 of \citealt{2017A&A...597A...6N} for a discussion of various lines in an ultra metal-poor star).  Several important studies of line formation in the Sun have used the 1.5D code \nataja{} \citep{2001ApJ...550..970S}.  Readers interested in developing or using codes themselves might also look at the open-source, MPI-parallelised 1.5D non-LTE codes \citep{2015A&A...574A...3P}, and \nicole{} \citep{2015A&A...577A...7S}. As we discussed in \sect{sect:method-atmospheres-codes}, 3D LTE synthetic spectra tend to be in good agreement when using different radiative transfer codes and model atmospheres. Unfortunately, systematic comparisons of 1.5D or 3D non-LTE calculations have not yet been performed, and one expects larger differences owing in particular to different treatments of background opacity (\sect{sect:method-spectra-opacity})

\subsubsection{Overview of post-processing in 3D LTE}
\label{sect:method-spectra-overview}

Post-processing in 3D non-LTE, like that in 3D LTE, 1D non-LTE, and 1D LTE,
usually begins by calculating the EOS and extinction and emissivity
coefficients in LTE.  Two or more
state variables, for example temperature and gas density, are extracted from the
model atmospheres as a function of 1D or 3D geometric space.  For the 3D RHD
models, several temporal snapshots are considered individually.  Next, given these quantities as well as 
the elemental abundance ratios (which, under
the trace-element approximation, do not necessarily correspond exactly to that
with which the model atmosphere was constructed), the EOS and 
the monochromatic LTE opacities are re-determined (\sect{sect:method-atmospheres-micro}).  
The careful treatment of line profiles is crucial for spectral line fitting;
metal-line profiles are usually treated with Voigt profiles with collisional
broadening by hydrogen atoms best accounted for via ABO theory
\citep{2000A&AS..142..467B}, while hydrogen line profiles can include self-broadening
and Stark broadening for example via the \hlinop{} suite of software
(\citealt{2016A&ARv..24....9B}, and references therin).

The radiation field and hence the emergent intensity can be calculated by
solving the radiative transfer equation, given the model atmosphere, EOS, and opacities.  We refer the reader to \citet{2003rtsa.book.....R} and
\citet{2015tsaa.book.....H} for an introduction into radiative transfer in
stellar atmospheres, and, for example, \citet{2003ASPC..288....3A} and
\citet{2021ApJ...912...63D} for an overview of the associated challenges.  The
latter authors suggest that short characteristics methods, which are more
efficient for multidimensional problems and parallelise well, are sufficiently
accurate provided one is interested in spatially-integrated spectra.
\citet{2013A&A...549A.126I}, \citet{2013A&A...557A.143S}, and
\citet{2017MmSAI..88...22S} describe the implementation of such methods on 3D
cartesian grids.  For box-in-a-star 3D RHD models as well as for semi-infinite
1D models, one solves the radiative transfer equation along several outgoing
rays of different inclinations $\mu=\cos\theta$, where the vertical ray at
$\mu=1$ corresponds to light emergent from the centre of the stellar disc; in 3D
one must also solve for different azimuthal angles $\phi$.  The astrophysical
flux can then be obtained via disc-integration, adopting a suitable angle
quadrature \citep{2020A&A...636A..24S,2021A&A...645A.101J}.  Abundance analyses
based on spectral line fitting (rather than equivalent widths) must take
macroscopic broadening effects into account.  In particular, rotational
broadening can be added following \citet{1990A&A...228..203D} or
\citet{2007A&A...471..925L}, and the synthetic profiles must also be convolved
with the instrumental profile.

Post-processing of 1D and $\mtd{}$ models invoke at least two additional parameters,
microturbulence and macroturbulence
\citep[e.g.][]{2017A&A...597A..16P}.  They are necessary to
account for spectral line broadening primarily due to the velocity gradients,
on scales much smaller than or much larger than an optical depth,
respectively. Despite their names, they are
explained by stellar granulation rather than by turbulence
\citep{1997A&A...328..229N}, and
are thus predicted from first principles by the 3D RHD simulations
\citep{2000A&A...359..729A}.  Not having to deal with these imperfect fudge
parameters \citep{2013MSAIS..24...37S} is a major advantage of 3D abundance
analyses.

\subsubsection{Solving for the statistical equilibrium}
\label{sect:method-spectra-nlte}

Moving to non-LTE brings about additional complications. We once again recommend
\citet{2003rtsa.book.....R} and \citet{2015tsaa.book.....H} for an introduction
into this topic.  In brief, rather than assuming that absorber populations are
given by the Saha ionisation equation and the Boltzmann distribution, one solves
for the time-independent equations of statistical equilibrium:
\phantomsection\begin{IEEEeqnarray}{rCl}
\label{eq:statisticalequilibrium}
    n_{i} \sum_{j\neq i} \left(R_{ij}+C_{ij}\right) -
    \sum_{j\neq i} n_{j} \left(R_{ji}+C_{ji}\right) &=& 0
\end{IEEEeqnarray}
Here, $n_{i}$ are the number densities of atoms or ions in a particular level
$i$, $R_{ij}$ and $C_{ij}$ are the total radiative and collisional rates
(with dimensions $T^{-1}$) from level $i$ to level $j$,
which are calculated given the radiative transition probabilities or 
cross-sections (together with the angle-averaged intensity $J_{\nu}$)
and collisional rate coefficients described in the model atom.
Sect.~9.2
of \citet{2015tsaa.book.....H} explains how 
\eqn{eq:statisticalequilibrium} follows from the
Boltzmann transport equation.  A more general form of this equation has a
source term on the right hand side, $\mathrm{d}n_{i}/\mathrm{d}t = \partial
n_{i}/\partial t + \nabla\cdot\left(n_{i}\bm{v}\right)$; this is set to zero
under the assumption that the radiative and collisional transitions happen on
much shorter timescales than the dynamical timescales in the stellar atmosphere.
To a good approximation, the particles follow Maxwellian velocity distributions
(see Chapter 4 of \citealt{2015tsaa.book.....H}); as such 
\eqn{eq:statisticalequilibrium} reflects the competition between
inelastic collisions that bring the overall system closer to LTE,
and radiative processes that in general drive the system
away from LTE. The equations are closed by number conservation:
\phantomsection\begin{IEEEeqnarray}{rCl}
    \label{eq:numberconservation}
    \sum_{i} n_{i} &=& N
\end{IEEEeqnarray}
Here, $N$ is the total number density of particles in all of the levels
of the atoms, ions, and possibly molecules described in the model atom.  The
quantity $N$ is extracted from the LTE EOS (and
excludes any nuclei locked away in species not 
explicitly described by the model atom; this could be, for example,
molecules and more than doubly-ionised species).
\eqn{eq:statisticalequilibrium} and \eqn{eq:numberconservation} are solved at
every gridpoint in the model atmosphere.  The radiative rates require 
$J_{\nu}$, which is obtained by solving the
radiative transfer equation. The radiative transfer equation, however, depends
on the extinction and source function which depend on the absorber populations
$n$.  As such, the equations are non-linear; they are also non-local, as the
radiation couples different parts of the atmosphere together.

\multitd{}, \balder{}, and \nltetd{} solve the non-linear
\eqn{eq:statisticalequilibrium} as well as the radiative transfer equation
iteratively, preconditioning them 
(analogous to linearising them; \citealt{1997ApJ...490..383S})
as per Multi-level Accelerated Lambda
Iterations (MALI; \citealt{1992A&A...262..209R}, see also \citealt{1994A&A...292..599A})
which is a generalisation of the method presented in \citealt{1986JQSRT..35..431O}.
It is common to use a local (diagonal) approximate operator,
amounting to Jacobi iterations; this is simple to implement 
and parallelise well in 3D, compared to 
more complex approaches, for example Gauss-Seidel
iterations \citep{1995ApJ...455..646T} that have better convergence
properties.  The MALI approach is usually combined
with acceleration methods \citep{1991ASIC..341....9A} that can help increase the
convergence rate and prevent the iterations from stabilising away from the true
solution. In the near future,
significant improvements to convergence rates could be had via the use
of multigrid methods \citep{1997A&A...324..161F,2017A&A...599A.118B},
non-stationary methods \citep{2009A&A...507.1815P,2016EJPh...37a5603L}, or
asynchronous iterations implemented within the \dispatch{} framework
\citep{2018MNRAS.477..624N,2019ASPC..519...93N}.

The overall number of iterations depends on how close the initial guess is to
the converged solution.  For a given snapshot of a 3D RHD model atmosphere, the
usual approach is to start with LTE populations (as well as $J_{\nu}=B_{\nu}$ if
background scattering is included).  It should be possible to improve the initial
guess based on the converged solutions from other snapshots of the same model, and possibly
from snapshots of other models, perhaps using machine learning algorithms
\citep{2022A&A...658A.182C}.

\subsubsection{Sampling the 3D RHD models}
\label{sect:method-spectra-sampling}

There are a number of compromises between accuracy and speed
that must be made when carrying out 3D non-LTE spectrum synthesis calculations.
Of the order of two to five snapshots are sufficient for analyses based on
equivalent widths or abundance corrections to achieve a precision of 0.01\,dex \citep{2017A&A...597A...6N,Diaz_stagger}. It
is common to use up to 20 snapshots for analyses of profiles of weak lines
\citep{2022A&A...668A..48D}, but half that number has been shown to suffice \citep{Diaz_stagger}. Models in the \stagger{}-grid and \cifist{}-grid
tend to be downsampled in the horizontal by factors ranging from two to four
\citep{2015A&A...583A..57S,2017MNRAS.464..264A}; again introducing an abundance error smaller than 0.01\,dex \citep{2017MNRAS.464..264A,Diaz_stagger}. The tests of
\citet{2017A&A...597A...6N} suggest that higher horizontal resolution is needed
in 3D non-LTE compared to in 3D LTE. Conversely, it is desirable to increase
the vertical resolution of the 3D RHD models, by trimming away the
optically-thick layers and interpolating onto a new depth scale with finer
resolution of the steep continuum-forming regions \citep{2018A&A...615A.139A}.

Typically the angle-averaged intensity $J_{\mathrm{\nu}}$ is calculated using
around $15$ rays or $30$ directions on the unit sphere
(e.g. \citealt{2016A&A...586A.156K} use $17$ rays).
\citet{2017MNRAS.464..264A} suggest to use the Lobatto quadrature for the
integration over the cosine of the polar angle $\mu=\cos\theta$ (i.e.~the
Gaussian quadrature but which is pinned on both of the integration limits,
namely $\mu=1$ and $\mu=-1$) and, for the inclined rays, trapezoidal integration
over the azimuthal angle $\phi$.  One advantage of this simple approach is that it results in the Lobatto
quadrature in the limit of a horizontally homogenous, semi-infinite 1D
atmosphere, and should give results close to the optimal Gauss quadrature.
Nevertheless, it may be that
the angle quadratures tailored for 3D applications proposed by
\citet{2020A&A...636A..24S} and \citet{2021A&A...645A.101J} give higher
accuracy for the same number of rays.




\subsubsection{Background opacities and scattering}
\label{sect:method-spectra-opacity}

The opacity contributed by background species can have a significant impact on
the statistical equilibrium of the element under consideration.  In particular,
recent results \citep[e.g.][]{2015MNRAS.454L..11A,2021MNRAS.500.2159W} have
demonstrated that it can be important to consider the opacity from background
lines that overlap (blend with) the bound-bound transitions in the model
atom.  However, the computation of background opacity due to millions of atomic/ionic lines and billions of molecular lines, for $10^{6}$ grid points of a 3D
RHD model atmosphere snapshot and $10^{4}$ frequency points of a model atom is
computationally expensive.  As such, to save time these can be precomputed and
interpolated on the fly. \balder{} does this on a regular grid of logarithmic
temperature and density, and a frequency grid that corresponds exactly to that
of the model atom, for the chemical composition corresponding to that of the
particular 3D RHD model.

It is common to consider some background continuum scattering processes, in
particular Rayleigh scattering from \ion{H}{I} on the red wing of the
Lyman-$\alpha$ line.  This is treated assuming the scattering is coherent and
isotropic, such that the background continuum source function becomes
$S_{\nu}=\epsilon_{\nu}\,B_{\nu}+\left(1-\epsilon_{\nu}\right)J_{\nu}$, with
background continuum photon destruction probability $\epsilon_{\nu}$
\citep[e.g.][]{2011A&A...529A.158H}. Following \citet{2010A&A...517A..49H} and
\citet{2020A&A...642A..62A}, one can make a similar approximation for background
lines via Eq.~22 of \citet{1962ApJ...136..906V} that can be modified to estimate
$\epsilon_{\nu}$ for a given line due to electron collisions and under the
assumption of a two-level atom.  This approach probably underestimates
$\epsilon_{\nu}$ by neglecting hydrogen collisions; this could easily be
estimated via the Drawin recipe (see Appendix A of
\citealt{1993PhST...47..186L}).  However, it may be better simply to treat
background lines in true absorption ($\epsilon_{\nu}=1$) given the limitations
of the two-level atom, and that coherent scattering is a poor approximation for
weak lines in stellar photospheres (we note here that complete frequency
redistribution is a good approximation at least for the solar photosphere;
\citealt{1992A&A...265..268U}).  Perhaps it is instead worth moving
away from the single-element approach altogether, and to develop more efficient
algorithms for multi-element non-LTE calculations, which are now
feasible at least in 1D (see below).  Including scattering in  
the source function requires updating $J_{\nu}$ (or the emissivity $\eta_{\nu}$)
after every MALI iteration.  We note that the algorithm of
\citet{1992A&A...262..209R} can be modified to also use preconditioning for this
background component, and this can slightly improve the overall convergence rate
when these processes strongly affect the statistical equilibrium.

\subsubsection{Cost}
\label{sect:method-spectra-cost}
The cost of 3D non-LTE spectrum synthesis with respect to 1D LTE can be estimated under the assumption that the radiative transfer dominates the overall cost (which is often the case for simple model atoms) and that there are no overheads when scaling up the calculations and when doing 3D radiative transfer. For an entire spectrum of lines of a given species or element, the non-LTE to LTE cost is then given by the number of iterations to reach convergence, which can be around $10$ to $50$.  The 3D to 1D cost can be estimated via the number of snapshots, horizontal gridpoints, and azimuthal angles, $N_{t}\times N_{x}\times N_{y}\times N_{\phi}$, which is at least $10^5$. In all, the 3D non-LTE to 1D LTE cost comes to around $10^{6}$ to $10^{7}$ for an entire spectrum of a given species or element. Thus a 1D LTE analysis taking around 1 hour on a single CPU would amount to several million CPU hours in 3D non-LTE, necessitating the use of a supercomputer. If one is only interested in performing the abundance analysis with just a single line, these relative cost increase by an amount proportional to the number of lines in the model atom, which can be hundreds or sometimes thousands; the 3D non-LTE to 1D LTE cost then can reach around $10^{9}$.

\subsection{Astrophysical validation}
\label{sect:method-validation}

The physics underpinning 3D non-LTE models are far more sophisticated and
realistic than the various simpler alternatives (1D LTE, 1D non-LTE, 
$\mtd{}$ LTE, $\mtd{}$ non-LTE, 3D LTE).  
Nevertheless, it is worth asking if this translates into more accurate
abundance results overall.
Various astrophysical tests of the 3D RHD models have been performed
using the Sun as a testbench,
for which $\mu$-resolved observations \citep[e.g.][]{2023A&A...673A..19E} with high signal-to-noise
ratios and resolving powers can be used; it is also the only star for which the
uncertainties in $\teff$ and $\lgg$ can be neglected (in this context).
The continuum limb darkening is better reproduced by 3D RHD models than
by theoretical and semi-empirical 1D models as well as $\mtd{}$ models
\citep[e.g.][]{2008ApJ...680..764K,2013A&A...554A.118P}, giving confidence to
the mean temperature stratification predicted by the simulations. Intensity fluctuations \citep{2013A&A...554A.118P} as well as line shifts
and asymmetries, none of which can be predicted by 1D models, are
well-reproduced by 3D RHD models
\citep[e.g.][]{2000A&A...359..729A,2009LRSP....6....2N,2009A&A...507..417P},
giving confidence to the gas dynamics.  Similarly, the profiles of strong
\ion{H}{I} Balmer and Paschen lines are better reproduced by 3D RHD models as discussed in Sect.\,\ref{sect:results-parameters-balmer}.

Solar CLVs of spectral lines are a probe of coupled 3D non-LTE effects. 
Spectral lines observed towards the limb form higher up in the photosphere (in a
cooler, less dense environment) than those observed at disc-centre.  Lines of
different excitation potential, and in different ionisation stages, and
molecular lines, exhibit different CLVs, to first order due to their temperature
and pressure dependences described by Saha-Boltzmann statistics.  As such, CLVs
are sensitive to the mean structure of the atmosphere.  They are also sensitive
to the fluctuations, due in part to the non-linearities, but in this case more
because towards the limb one begins to see the granules edge-on
\citep{2009LRSP....6....2N,2011ApJ...736...69U}; as a consequence, in 1D one
requires higher values of microturbulence to reproduce observations towards the
limb \citep{2022SoPh..297....4T} even though 1D inversions of the solar
disc-centre intensity have microturbulence values that decrease with height
\citep{1974SoPh...39...19H}.  Solar CLVs offer an opportunity to validate the
model atom and the non-LTE modelling: departures from LTE are larger for lines
observed towards the limb, where gas densities are lower and thus collisions are
too infrequent to maintain Saha-Boltzmann statistics.  In particular, the
ionisation fraction decreases towards the cooler upper photosphere such that the
ratio $n_{\mathrm{H}}/n_{\mathrm{e^{-}}}$ increases steeply with height, and
thus the CLV can be used to test the description of hydrogen collisions
\citep{2004A&A...423.1109A,2009A&A...508.1403P,2015A&A...583A..57S} insofar as
these are the largest uncertainty in the model atom. In practice, the method strives to reproduce the disc centre and limb spectra with a uniform abundance. 
By virtue of such tests for
\ion{O}{I} and \ion{C}{I} lines in 3D non-LTE,
\citet{2018A&A...616A..89A,2019A&A...624A.111A} suggest to use the
asymptotic+free approach for hydrogen collisions
(\sect{sect:method-atoms-collisions}), although this should be explored for more
species.  Despite these uncertainties, it is clear that 3D non-LTE outperforms
1D or $\mtd{}$ LTE or non-LTE and 3D LTE \citep[e.g.][]{2017MNRAS.468.4311L} as also seen for \ion{Na}{I} (\citealt{2023arXiv231205078C}, and \sect{sect:results-planets-transmission}).

For stars other than the Sun, analogous tests as those described above may soon
be possible with the benefit of high fidelity observations of exoplanet transits
\citep[e.g][]{2017A&A...605A..90D,2018A&A...616A..39M}. In
general, however, astrophysical validation for such stars comes from various
consistency checks on inferred abundances.  By similar arguments to the solar
CLVs, the coupled 3D non-LTE effects influence the predicted strengths of
spectral lines due to different levels of different species and forming at
different heights in the atmosphere; the inferred abundances should all agree
with each other to within the observational uncertainties.  One can explore a wider parameter space by also considering
benchmark stars for which stellar parameters are well-constrained on an absolute
scale for example by means of interferometry and parallaxes
\citep[e.g.][]{2018MNRAS.475L..81K,2020A&A...640A..25K}, as discussed in Sect.\,\ref{sect:results-parameters}.

Lastly, we note that the scatter in abundance-abundance planes can be dominated
by systematics caused by the 1D or LTE assumptions. Analyses of large samples of
stars thus offer a further way to validate the 3D non-LTE modelling, and reduced scatter can potentially reveal new and exciting astrophysics, as we elaborate on in \sect{sect:results}. We stress that results from 1D models have various parameters that are calibrated
to fit the observations, and as such,
even when 1D or $\mtd{}$ (non-)LTE abundances appear to give consistent results in
terms of excitation, ionisation, or molecular equilibrium, the average abundance
may still lie further from the absolute truth than 3D non-LTE.


\subsection{3D non-LTE effects on abundances}
\label{sect:method-effects}

We present an overview of the physics behind some common 3D non-LTE effects.
Specific results for different species and elements, and their astrophysical
implications, are given in \sect{sect:results}; moreover, general advice for
astronomers working with 1D LTE abundances are provided in
\sect{sect:discussion}.

\subsubsection{3D effects}
\label{sect:methods-effects-threed}

\begin{figure*}
\includegraphics[width=5in]{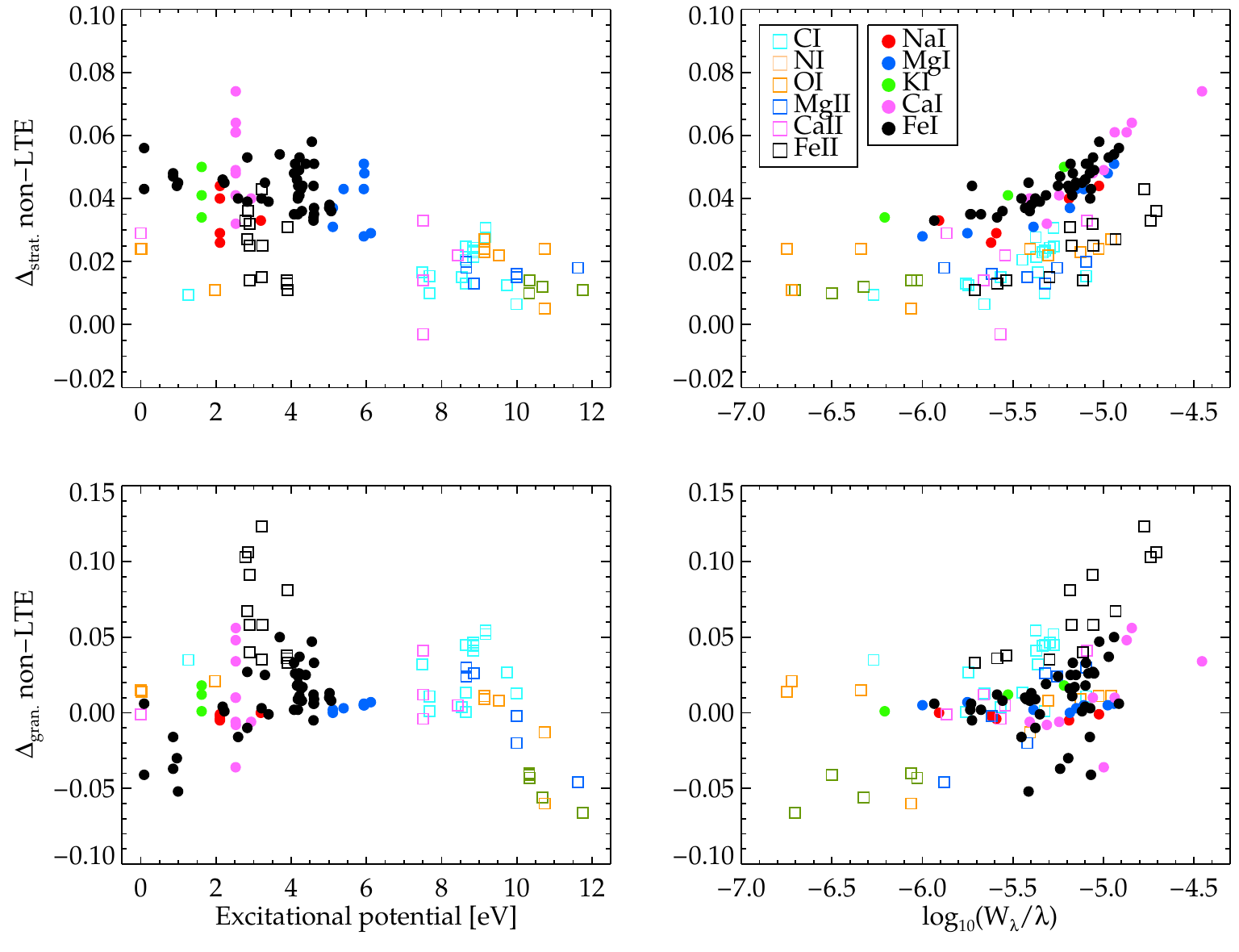}
\caption{The difference in solar abundances determined under different assumptions; the top panels represent the influence of the atmospheric stratification by showing the difference between abundances determined in $\mtd{}$ and 1D ($\Delta_{\rm strat.}$), and the lower panels represent the impact of granulation by showing the difference between 3D and $\mtd{}$ abundances ($\Delta_{\rm gran.}$). All abundances have been computed in non-LTE, and are based on analyses of the disc-centre intensity. The data have been compiled from \citet{2019A&A...624A.111A,2020A&A...636A.120A} and \citet{2021A&A...653A.141A}. Majority species are marked with open squares, minority species with filled bullets.}
\label{fig:adiffsun}
\end{figure*}

Before discussing the 3D effects on profile-integrated
equivalent widths, we first note that
there is a large literature on spectral line shapes
\citep[e.g.][]{1981A&A....96..345D,1982ARA&A..20...61D}.  The flow of gas due to stellar
granulation, overshooting into the convectively-stable regions, impart net
blueshifts, line asymmetries, and line broadening 
when considering spatially-integrated or
disc-integrated spectra.  This field has seen renewed interest due to their
degeneracies with radial velocity signatures of exoplanets
\citep[e.g.][]{2023MNRAS.525.3344D}.  We refer the reader to other literature
for a discussion of these effects
\citep[e.g.][]{2019A&A...624A..57L,2021A&A...649A..16D}. 
The severity and possibly the direction of the abundance corrections when
inferred via spectral line profile fitting
will depend on the adopted line masks and the treatment of rotational velocity. The line shape is particularly important for measurements of isotopic ratios, as discussed in \sect{sect:results-origin-li} and \sect{sect:results-origin-barium}.

For a first order understanding of the 3D effects on equivalent widths, it
is helpful to separately consider the ``stratification effects'' $\Delta_{\mathrm{strat.}}$ due to differences
in the mean atmospheric stratification, and
the ``granulation effects''
$\Delta_{\mathrm{gran.}}$ due to temperature and velocity inhomogeneities.
The former may be tested by comparing 
$\mtd{}$ abundances to 1D abundances,
and the latter by comparing 3D to $\mtd{}$ 
(all in non-LTE).
\fig{fig:adiffsun} shows an overview of $\Delta_{\rm strat.}$ and $\Delta_{\rm gran.}$
for eight of the elements that have been consistently analysed in 3D non-LTE in the solar disc-centre intensity. We note that the trends identified with line strength in these plots strongly depend on the adopted microturbulence in the 1D and $\mtd{}$ analyses, here $1$\,km/s.
We also emphasize that $\mtd{}$ models are not unambiguously defined from their parent 3D models, but depend strongly on the averaging method, in particular the 
reference depth scale (geometric, column mass, optical depth) as demonstrated by \citet{2013A&A...560A...8M}, as well as the
choice of which quantities to average and which (if any) to recalculate from the EOS. The $\mtd{}$ models represented in \fig{fig:adiffsun} have been computed by averaging 
the logarithmic gas density, and the gas temperature raised to the fourth power,
on surfaces of equal Rosseland optical depth \citep{2018A&A...616A..89A}.

The stratification effects $\Delta_{\rm strat.}$ show a clear trend, with increasingly positive values with increasing line strength, essentially independently of excitation potential, with minority species more affected than majority species. This is likely a consequence of the slightly shallower temperature gradient of the $\mtd{}$ model (line weakening) combined with a slight reduction in electron density (line strengthening of majority species).
In metal-poor stars, the effect of the mean stratification is more prominent. Although for example the Sun is approximately in radiative equilibrium \citep{2009LRSP....6....2N},
towards lower metallicities radiation plays a smaller role on the energy
balance, and the upper layers are able to cool efficiently as the upflows overshoot
into the convectively-stable regions and expand 
(the 1D models are too hot). 
For example, \citet{2012MNRAS.427...27B} show that \ion{Fe}{I} are typically strengthened and \ion{Fe}{II} lines typically weakened in $\mtd{}$-models of metal-poor stars.

The granulation effects show more complex behaviour.  For weak lines in the Sun,
\fig{fig:adiffsun} shows that $\Delta_{\rm gran.}$ is close to zero.  
The exception is for minority species of
low excitation potential ($<1$\,eV), and majority species
of high excitation potential ($>10$\,eV).
These are also the most temperature-sensitive
lines: For minority species, the ratio between line opacity and continuous opacity (when the latter is dominated by $H^-$) scales approximately with $\exp{(I-(E_{i}+E_{\mathrm{H^{-}}}))/k_BT}$, where $I$ is the ionisation potential, $E_i$ the excitation potential, and $E_{\mathrm{H^{-}}}=0.754\,\mathrm{eV}$ is the electron affinity of hydrogen. Low excitation lines are therefore more temperature sensitive. For majority species, both neutral and ionized, the ratio scales as $n_{\mathrm{e^{-}}}T^{3/2}\exp{(-(E_i+E_{\mathrm{H^{-}}}))/k_BT}$, explaining why high-excitation lines of majority species react more strongly to temperature changes. Typically, the larger the temperature-sensitivity of the line under investigation, the larger the impact of granulation. \citet{2013A&A...559A.102D} and \citet{2013A&A...549A..14K} draw similar conclusions based on fictitious weak-line analysis in LTE in red giants, but they also show that different elements can react very differently at low metallicity. Partially or fully saturated lines are expected to be more strongly sensitive to granulation in general, because the atmospheric inhomogeneities become more pronounced at higher layers, where desaturation and strengthening in 3D by fluctuations should lead to negative
$\Delta_{\mathrm{gran.}}$ compared to when neglecting velocity fields; however, the calibration
of the microturbulence parameter makes this ambiguous and can produce positive $\Delta_{\mathrm{gran.}}$ (see for example Fig.~3 of
\citealt{2022A&A...668A..68A}). 

Although not illustrated in \fig{fig:adiffsun}, extreme granulation effects
can be found for molecular lines.  Inhomogeneities in the 3D models
leads to enhanced molecule formation in the cool
pockets of gas, leading to extreme negative abundance corrections for CO
\citep{2011ApJ...736...69U,2021A&A...656A.113A}. In the most metal-poor stars, the stratification effect can also be significant, but the granulation effect dominates \citep{2013A&A...559A.102D}.

\subsubsection{Non-LTE effects}
\label{sect:method-effects-nlte}

One can begin a physical interpretation of 
non-LTE effects on multi-level atoms by considering two-level
and three-level atoms, as presented by \citet{2003rtsa.book.....R} and in Chapter 14
of \citet{2015tsaa.book.....H}.
The two fundamental effects are photon losses and photon pumping. 
Via bound-bound transitions, these lead to over-deexcitation and over-excitation of populations relative to LTE;
while via bound-free transitions they lead to over-recombination and over-ionisation. The effects are driven by splits between $J_\nu$ and $B_\nu$ that occur naturally across the spectrum even in LTE, because of how the Planck function's sensitivity to temperature varies with wavelength. 

Photon losses occur because photons manage to scatter large distances when they are emitted via a bound-bound or bound-free transition, instead of being immediately reabsorbed as per thermodynamic equilibrium. As described by \citet{2003rtsa.book.....R}, the loss of photons by back-scattering to deeper layers is not compensated for by the reverse effect in the lack of incoming light, and therefore the emergent intensity is reduced by such processes. The imbalance between
photoexcitations and photoemissions shifts the equilibrium relative to
LTE with fewer atoms (or ions or molecules) sitting in the upper level of the
bound-bound or bound-free transition, and more sitting in the lower level. 
Photon pumping is the opposite: here an excess of photons are absorbed compared
to that expected from the local Planck function, these photons typically
scattering out of the deeper, hotter layers of the atmosphere,
and is typically very effective for UV transitions
because the Planck function has a steep temperature gradient
at such wavelengths.

In a multi-level atom, the non-LTE effects of individual transitions may enhance or cancel each other; for example can a 'ladder' of bound-bound transitions with high scattering probability lead to efficient photon losses, referred to as photon suction \citep[e.g.][]{1994A&A...288..860C}. To understand which radiative transitions are most influential for the statistical equilibrium of a given model atom and atmosphere, we recommend to start by inspecting the size of the radiative bracket \citep[e.g.][]{2012MNRAS.427...50L}. The strengthening or weakening of a given line can be understood by a combination of the change in opacity (proportional to the departure coefficient of the lower level) and the source function (proportional to the ratio between the upper and lower departure coefficient in the Wien regime).

The same effects that are frequently studied in atoms are also of relevance to molecules.
Quantitative studies of non-LTE
effects on molecules have been limited by their complexity and the lack of
required data. \citet{1973ApJ...181.1039T} and
\citet{1975MNRAS.170..447H} have shown that ro-vibrational
transitions within the ground electronic state of diatomic molecules are
collisionally controlled and thus should not show departures from LTE, and this
appears to be borne out in calculations for the Sun
\citep{1989ApJ...338.1033A,2000ApJ...541.1004S,2000ApJ...536..481U}.  On the
other hand, \citet{1975MNRAS.170..447H} show that excited electronic states may be
radiatively coupled to the ground state, rather than collisionally; and using a
simple two-level atom, \citet{2001A&A...372..601A} demonstrate abundance effects
of the order $0.2\,\mathrm{dex}$ for electronic UV lines of the OH molecule.
Species with small dissociation energies may be susceptible to
over-photodissociation, and recent calculations by \citet{2023A&A...670A..25P}
suggest this could be significant for CH ($E_{\mathrm{dis.}}=3.47\,\mathrm{eV}$).
Accurately quantifying these effects requires reliable cross-sections for
excitation and ionisation by collisions with electrons and with hydrogen atoms,
which are still lacking. 

Another kind of departure from LTE was recently investigated by \citet{2023A&A...675A.146D}. The authors challenged the hypothesis that the longer chemical timescales of metal-poor stars may cause deviations from molecular equilibrium, and thereby bias abundance determination under this assumption, by allowing time-dependent chemistry of CO, OH, CH, CN, $\rm C_2$ in 3D hydrodynamic atmospheres of metal-poor dwarfs. They conclude that the molecules are generally in equilibrium throughout the photospheres, but that deviations can be present in shallow layers, in particular for stars with C/O ratios much greater than unity. However, the deviations are unlikely to contribute significantly to spectroscopic measurements because the key diagnostic species CH and OH are formed in deeper layers.

\subsubsection{Coupling of 3D and non-LTE effects}

Although we have discussed
3D and non-LTE effects separately, in reality the two
are coupled. This coupling can be quantified in
terms of abundances for example by comparing the 3D non-LTE result
$A_{\mathrm{3N}}$ with the approximation that is obtained by separately applying
3D LTE and 1D non-LTE abundance corrections to the 1D LTE result,
namely $A_{\mathrm{approx.}}=\left(A_{\mathrm{3L}}-A_{\mathrm{1L}}\right)+
\left(A_{\mathrm{1N}}-A_{\mathrm{1L}}\right)+ A_{\mathrm{1L}}$.  For \ion{Fe}{I}
lines in the ultra metal-poor G-dwarf SDSS J102915.14+172927.9
\citep{2011Natur.477...67C}, one finds, for Fe, $A_{\mathrm{3N}}=3.28$, and
$A_{\mathrm{approx.}}=\left(-0.10\right)+ \left(0.25\right)+2.80=2.95$
\citep{2023A&A...672A..90L}.  
Similarly, for the K-giant SMSS J031300.36-670839.3 \citep{2014Natur.506..463K}  
one finds upper limits on the Fe abundance of $A_{\mathrm{3N}}=0.97$, and
$A_{\mathrm{approx.}}=\left(-0.34\right)+ \left(0.61\right)+0.16=0.43$
\citep{2017A&A...597A...6N}. For Li in a very metal-poor dwarf \citep[][Sect.\,\ref{sect:results-origin-li}]{2021MNRAS.500.2159W}, $A_{\mathrm{3N}}=2.02$ and 
$A_{\mathrm{approx.}}=\left(-0.21\right)+ \left(-0.07\right)+2.00=1.72$.
Thus, $A_{\mathrm{approx.}}$ is $0.3-0.5\,\mathrm{dex}$ too low, reflecting how the steeper gradients in the 3D RHD models enhance the non-LTE overionisation of the
minority neutral species \citep{2016MNRAS.463.1518A}. 
For the Sun, both 3D and non-LTE effects are smaller and so is the inferred error if lines are selected with care. $A_{\mathrm{approx.}}$ is accurate to within $\pm0.03\rm\,dex$ for several species according to \citet[][Table A1]{2021A&A...653A.141A}. For Ba, \citet{2020A&A...634A..55G} show that $A_{\mathrm{approx.}}$ would underestimate the abundance by 0.05\,dex.
 
In general, for species with significant departures
from LTE, 1D (non-)LTE abundances are often closer to 
3D non-LTE than are the 3D LTE abundances. 
Nevertheless, we stress that 3D LTE should be preferred when departures from LTE
are negligible, such as for lines of \ion{Fe}{II} \citep{2022A&A...668A..68A} or
the forbidden [\ion{O}{I}] lines \citep{2016MNRAS.455.3735A}.

\subsubsection{Magnetic fields}
\label{sect:method-effects-mhd}

Aside from the simplified treatment of radiative transfer
(\sect{sect:method-atmospheres-micro}),
it could be that the most important missing physics in the context of 3D non-LTE
stellar abundances comes from magnetic fields.  The use of 3D RMHD model
atmospheres for abundance analyses of late-type stars analyses have so far been
largely explorative, and limited to the Sun.  Pilot studies were carried out by
\citet{2010ApJ...724.1536F,2012A&A...548A..35F} and \citet{2015ApJ...802...96F}, who
performed 3D RMHD simulations with a net vertical magnetic field using a
modified version of \stagger{} \citep{1996JGR...10113445G}.  They found for
example large effects on \ion{Fe}{I} spectral lines corresponding to positive
abundance corrections of the order $0.05$ to
$0.09\,\mathrm{dex}$ for simulations with mean unsigned vertical flux density of
around $100\,\mathrm{G}$ (see Fig.~2 of \citealt{2015A&A...579A.112S}).

However, local dynamo simulations with zero net
vertical magnetic flux density might better reflect the actual case of the solar
photosphere \citep{2014ApJ...789..132R}.
\citet{2015A&A...579A.112S,2016A&A...586A.145S} as well as
\citet{2015ApJ...799..150M} have analysed the 3D RMHD \muram{} local dynamo
simulations of \citet{2014ApJ...789..132R}, which
have mean unsigned vertical flux
density of around $80\,\mathrm{G}$ near the optical surface.
They quantified the MHD
effects on the equivalent widths of spectral lines of different species
(\ion{C}{I}, \ion{N}{I}, \ion{O}{I}, \ion{Fe}{I}).  The net 
effects are small; \citet{2015ApJ...799..150M} report abundance changes of the order
$+0.004\,\mathrm{dex}$ overall for \ion{Fe}{I} lines, while
\citet{2015A&A...579A.112S} report slightly more severe effects of the order
$+0.014\,\mathrm{dex}$.  It should be noted that the effects grow for
lines that form higher up in the atmosphere.
At least in these studies, the MHD effects on spectral lines are mainly the
indirect ones, namely due to changes in the atmosphere itself, which tend to
weaken the lines.
The direct effect due to Zeeman broadening tends to act in
the opposite direction, strengthening lines and reducing inferred abundances.

\section{RESULTS}
\label{sect:results}

In this section, we quantify 3D and non-LTE effects for late-type stars based on recent literature, and discuss the impact in four research fields: 
the Sun (\sect{sect:results-sun}), 
stellar parameters (\sect{sect:results-parameters}),
the origins of the elements (\sect{sect:results-origin}) 
and planet-host stars (\sect{sect:results-planets}).

\subsection{Solar modelling problem}
\label{sect:results-sun}

High-quality spectroscopic observations of the quiet Sun today allow for the determination of the photospheric composition of 62 elements \citep{2021A&A...653A.141A}. The abundance of a handful of other elements can be detected in the solar wind (the noble gases) or be derived from molecular lines in sunspots (F, Cl, In, Tl). The solar chemical composition not only provides a fundamental reference frame for many fields of stellar and galactic astrophysics, but is crucial to model its interior structure and evolution.  
In the 20th century, spectroscopic analysis of the Sun pointed to a metal-to-hydrogen mass fraction $Z/X=0.0231$ \citep{1998SSRv...85..161G} or higher. The decade to follow saw a downward revision of this value by approximately $10-20\%$ \citep{2009ARA&A..47..481A,2011SoPh..268..255C}, thanks to many improvements on the 3D and non-LTE modelling and atomic data, but a strong consensus between competing groups is still lacking. Recently, a claim was made that the photospheric metallicity is in fact very close to the 25 year old canonical value, $Z/X=0.0225$ \citep{2022A&A...661A.140M}, based on $\mtd{}$-modelling. 

\begin{figure*}
\includegraphics[width=4in]{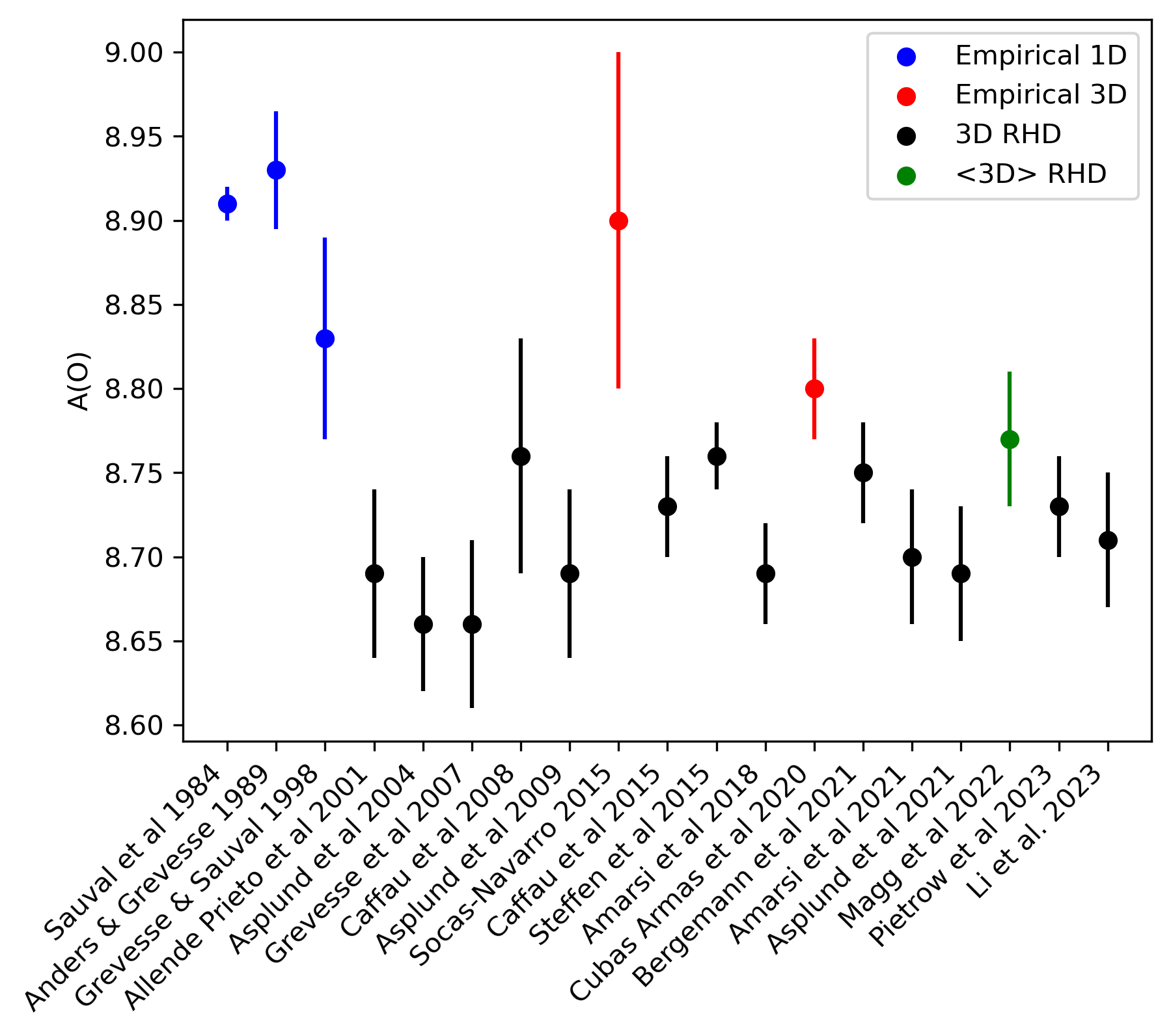}
\caption{The evolution of the Solar photospheric O abundance over the last 40 years, 
colour-coded by the type of model atmosphere employed in the analysis.
}
\label{fig:OSun}
\end{figure*}

The reduction of the abundances of a handful of key elements caused a reduction of opacity in the solar interior, with the consequence that standard solar models could no longer well reproduce the sound speed profile as determined from helioseismic inversions of the p-mode oscillations \citep[e.g.][]{2008PhR...457..217B}.  The offset is particularly large, of order $1\%$, in the radiative zone just below the convective envelope. The largest contributions, $>80\%$ in total, to the Rosseland opacity in this region are provided by O, Ne, and Fe and the abundances of these elements thus have the largest impact on the sound speed \citep{2014ApJ...787...13V}. Fig.\,\ref{fig:OSun} depicts how the use of different model atmospheres have affected the inferred solar O abundance over time. There are no detectable Ne lines forming in the solar photosphere, and so the solar Ne abundance must be inferred through indirect means, for instance via spectroscopy of the transition region \citep{2018ApJ...855...15Y} and of solar flares \citep{2015ApJ...800..110L}, or 
via measurements of the solar wind \citep{2009GeCoA..73.7414H,2019M&PS...54.1092B}.  However, as discussed in \citet{2021A&A...653A.141A}, care must be taken to account for fractionation relative to the solar photosphere due to the so-called FIP effect. 
Important metal opacity is also provided by C, N, Mg, Si, S, and Ni, while remaining elements contribute individually by less than $\sim1\%$ in this region \citep{2016ApJ...821...45K}. To shed light on the solar modelling problem, it is of particular interest to understand how 3D RHD models and non-LTE line formation have affected the solar abundances of the largest opacity contributors, and what the remaining uncertainties are and we comment on these elements individually in Sect. \ref{sect:appendix-sun-cn} to \ref{sect:appendix-sun-fe}.  

\subsubsection{Carbon and nitrogen}
\label{sect:appendix-sun-cn}
The 3D non-LTE line formation of atomic lines of \ion{C}{I} and \ion{N}{I} in the Sun was recently investigated for the first time by \citet{2019A&A...624A.111A,2020A&A...636A.120A}, with subsequent updates \citep{2021MNRAS.502.3780L,2023ApJS..265...26L,2021A&A...653A.141A}. Here, we summarize the main conclusions. Both species have highly excited ($7-9$\,eV for C, $10-12$\,eV for N) permitted atomic lines that can be used for abundance determination. Photon losses in these lines create sub-thermal line source function and small negative non-LTE effects that do not exceed $-0.07$\,dex for C and $-0.02$\,dex for N (using disc centre intensity). Larger negative effects were found in 1D for infrared lines of C by  \citet{2010A&A...514A..92C} and \citet{2015MNRAS.453.1619A}; however, these studies predated the arrival of asymptotic models for inelastic hydrogen collisions \citep{2019A&A...625A..78A}. The low-excitation forbidden [\ion{C}{I}] line at 872\,nm is formed in LTE. 

As evident from \fig{fig:adiffsun}, the granulation effect, as measured by $\Delta_{\rm gran}=$ 3D abundances$-\mtd{}$ abundances, has the opposite sign for C \citep[$+0.02$\,dex,][]{2010A&A...514A..92C,2019A&A...624A.111A} and N \citep[$-0.05$\,dex][]{2009A&A...498..877C,2020A&A...636A.120A}, as a consequence of the difference in lower-level excitation potential. \citet{2022A&A...661A.140M} analysed a subset of three atomic \ion{C}{I} lines and two \ion{N}{I} lines in $\mtd{}$ LTE and found abundances that 
are $\sim0.10$\,dex higher than \citet{2019A&A...624A.111A} and \citet{2021MNRAS.502.3780L} for C and $\sim0.15$\,dex higher than \citet{2020A&A...636A.120A}, when differences in oscillator strengths have been accounted for. On first look, these differences are too large
to be explained by granulation effects and the agreement is significantly better when 1D LTE \marcs{} models are used. The high result of \citet{2022A&A...661A.140M} may be related to the treatment of the blending CN lines or to the limitations and possible ambiguity connected to $\mtd{}$ models (\sect{sect:methods-effects-threed}). 

The 3D LTE line formation calculations of $\rm C_2$, CH, CO, NH and CN by \citet{2021A&A...656A.113A} highlight their larger sensitivity to granulation, in particular for CO where $\Delta_{\rm gran}=-0.2$\,dex. The C abundance inferred from molecular lines agrees well with the 3D non-LTE abundance from atomic lines, while the N abundance inferred from molecular lines is $\sim0.1$\,dex higher than the most recent 3D non-LTE N abundance from atomic lines \citep{2020A&A...636A.120A}. The difference cannot be explained by uncertainties in atomic oscillator strengths \citep{2023ApJS..265...26L}.  The CLV of molecular lines may be indicative of their sensitivity to non-LTE, and so should be confronted with 3D LTE and eventually non-LTE modelling. 

\nocite{1984ApJ...282..330S}
\nocite{1989GeCoA..53..197A}
\nocite{1998SSRv...85..161G}
\nocite{2001ApJ...556L..63A}
\nocite{2004A&A...417..751A}
\nocite{2007SSRv..130..105G}
\nocite{2008A&A...488.1031C}
\nocite{2009ARA&A..47..481A}
\nocite{2015A&A...577A..25S}
\nocite{2015A&A...579A..88C}
\nocite{2015A&A...583A..57S}
\nocite{2018A&A...616A..89A}
\nocite{2020A&A...643A.142C}
\nocite{2021MNRAS.508.2236B}
\nocite{2021A&A...656A.113A}
\nocite{2023A&A...672L...6P}
\nocite{2023A&A...674A..54L}

\subsubsection{Oxygen and nickel}
\label{sect:appendix-sun-o}
Despite being the third most abundant element, O has few atomic lines forming in the photosphere that are strong enough and clean enough to be used as abundance diagnostics \citep{1978MNRAS.182..249L,2001ApJ...556L..63A}. The most useful lines are the low-excitation forbidden [\ion{O}{I}] 630\,nm line, and the high-excitation permitted \ion{O}{I} 777\,nm triplet. infrared transitions of the OH molecule is another alternative that has been explored. Fig.\,\ref{fig:OSun} shows an overview of the evolution of the O abundance over the previous 40 years. 

The [\ion{O}{I}] 630\,nm line couples to the ground state of neutral O, which has a dominant level population in the solar atmosphere and is not significantly perturbed by non-LTE effects. However, the line is strongly blended with a \ion{Ni}{I} line, separated by only 0.003\,nm, that first had its transition probability determined experimentally, with an uncertainty of $0.06$\,dex, by \citet{2003ApJ...584L.107J}. Combined with the uncertainty of the solar Ni abundance, the strength of the blend is still fairly ill-constrained and its treatment impacts significantly on the O abundance. Studies that perform 3D LTE analyses and constrain the strength of the blend using the CLV have reported $\rm A(O)=8.69\pm0.05$ based on \stagger{} \citep{2001ApJ...556L..63A} and $\rm A(O)=8.73\pm0.02\pm0.05$ based on \cobold{} \citep{2015A&A...579A..88C}. Fixing the blend strength using the Ni abundance determined from unblended lines ($\rm A(Ni)=6.20\pm0.04$), \citet{2021A&A...653A.141A} finds $\rm A(O)=8.70$, with $0.02\,\mathrm{dex}$ variation between
different solar intensity atlases. \citet{2021MNRAS.508.2236B} adopt a slighter higher Ni abundance based on meteorites ($\rm A(Ni)=6.23\pm0.04$), but show that 3D non-LTE calculations significantly weaken the Ni blend, when employing the scaled Drawin formalism for hydrogen collisions, and finally arrive at $\rm A(O)=8.77\pm0.05$. However, using an improved version of the same model atom, \citet{2022A&A...661A.140M} later showed that Ni lines form close to LTE in the Sun. An even higher value, $\rm A(O)=8.80\pm0.03$, was recently reported by \citet{2020A&A...643A.142C}, based on an LTE analysis with a semi-empirical model atmosphere and spatially resolved observations. To make further progress, the Ni abundance and 3D non-LTE Ni line formation of the Sun must be better constrained. Hydrogen collisions were recently published by \citet{2022ApJ...926..173V}, but tailored calculations of photo-ionisation cross-sections are still lacking. The \ion{Ni}{I}-\ion{Ni}{II} ionisation balance is not a reliable diagnostic in the Sun \citep{2015A&A...573A..26S}, but the observed CLV-dependence of unblended \ion{Ni}{I} lines can and should be challenged by new models to verify the magnitude of non-LTE effects.  

The \ion{O}{I} 777\,nm triplet lines originate from the meta-stable level $\rm 2p^33s ^5S_2^o$ at $9.15\,\mathrm{eV}$. That the assumption of LTE line formation was inadequate to model these lines was realized already more than 50 years ago based on their observed CLV \citep{1968SoPh....5..260A}, pointing to an over-population of the lower level and sub-thermal line source functions. Further improvements in the model-fits to spatially resolved data were found with a combined 3D and non-LTE treatment by \citet{1995A&A...302..578K} and several authors since. A milestone was reached following the work by \citet{2018A&A...610A..57B}, after which several studies could demonstrate that the CLV of the O triplet can be well reproduced by 3D non-LTE calculations without using the efficiency of hydrogen collisions as a free parameter. It is reassuring that they advocate abundances that agree to within $\rm 0.03\,dex$; $8.69\pm0.03$ by \citet{2018A&A...616A..89A} was renormalised to $8.71\pm0.03$ with the new atomic data presented by \citet{2023A&A...674A..54L}, compared to $8.74\pm0.03$ by \citet{2021MNRAS.508.2236B} and $8.73\pm0.03$ by \citet{2023A&A...672L...6P}. The remaining difference may partly be attributed to the choice of observational data \citep{2021MNRAS.508.2236B}. It should also be noted that the \ion{O}{I} 777\,nm triplet lines
are strongly sensitive to the rate of (de-)excitation by hydrogen collisions;
these recent analyses are based on the asymptotic+free approach described in 
\citet{2018A&A...616A..89A}, but a more fundamental understanding of the relevant
physics would be desirable (\sect{sect:method-atoms-collisions}).

\citet{2021A&A...656A.113A} recently re-determined the 3D LTE solar O abundance from  
OH rovibrational transitions of OH ($|\Delta\nu|=1$) at around $\rm 3-4\mu m$.  The mean abundance $\rm A(O)=8.70\pm0.04$ is in agreement with the atomic indicators.  They are also in agreement with the mean results from rotational ($|\Delta\nu|=0$) transitions in the far infrared ($\rm 9-12\mu m$) as well as the weak first overtone lines ($|\Delta\nu|\geq 1$) at $\rm 1.5-1.7\mu m$, although for these latter diagnostics there are noticeable trends in line-by-line abundances with excitation potential and line strength probably related to missing physics in the 3D RHD model at the shallow surface layers as well as uncertainties in the equivalent widths, respectively. Both the $\mtd{}$ and the semi-empirical 1D \citet{1974SoPh...39...19H} model (hereafter HM model) return higher abundances, due to their neglect of temperature inhomogeneities.

In summary, the recommended O abundance for the Sun has been revised downward by at least 0.1\,dex in the 21st century (\fig{fig:OSun}), thanks to the combination of several effects that coincidentally act in the same direction;  in particular the recognition of the Ni blend in the [\ion{O}{I}] 630\,nm line, the use of non-LTE rather than LTE models for the \ion{O}{I} 777\,nm triplet and the use of 3D, rather than 1D, models for OH lines.

\subsubsection{Magnesium, silicon, and sulphur}
\label{sect:appendix-sun-mgsis}
Full 3D non-LTE solar calculations have also recently been performed using high-quality model atoms for Mg and Si \citep{2017MNRAS.464..264A,2021A&A...653A.141A}.  Differences between 3D LTE and $\mtd{}$ LTE, as well as their differences with LTE results from the HM model, are small for both elements, of order $0.01$\,dex \citep{2015A&A...573A..25S,2021A&A...653A.141A}. The non-LTE effects for the selected lines in the solar disc-centre intensity are only $\pm0.01\rm\,dex$, which is (as expected) slightly smaller than that in the solar disc-integrated flux
reported in 1D 
for Mg \citep{2022A&A...665A..33L} and Si \citep{2020MNRAS.493.6095M}. We caution that more severe effects exist for other lines.  Indeed, these latter sources report 1D non-LTE corrections that reach $-0.1\rm\,dex$ for both elements; and the 3D non-LTE versus 1D LTE correction
for the $571\,\mathrm{nm}$ \ion{Mg}{I} line is estimated to be $0.06\,\mathrm{dex}$ in the Sun
\citep{2023arXiv231207768N}.

Despite this relatively weak modelling sensitivity, the solar Si abundance is debated: \citet{2022A&A...668A..48D} report a value of $7.57\pm0.04$ based on a 3D LTE analysis with a $-0.01$\,dex non-LTE correction applied from \citet{2017MNRAS.464..264A}, which is significantly higher than the $7.51\pm0.03$ reported by \citet{2021A&A...653A.141A}, mainly as a consequence of the choice of oscillator strengths and line fitting technique. An even higher value but with a large scatter was found by \citet{2022A&A...661A.140M}, $7.59\pm0.07$, based on a $\mtd{}$ non-LTE analysis of the solar disc-integrated flux. 
For Mg, there is a greater consensus between groups, with several recent 3D \citep{2021A&A...653A.141A}, $\mtd{}$ \citep{2017ApJ...847...15B}, and 1D \citep{2018ApJ...866..153A} non-LTE values in the range $7.54-7.57$.  However, this agreement is partly coincidental as different model atmospheres and \ion{Mg}{I} lines were used. For example, limiting the comparison to 1D \marcs{} calculations for the same four \ion{Mg}{II} lines, with almost identical gf-values, the average difference between \citet{2018ApJ...866..153A} and \citet{2021A&A...653A.141A} is 0.17\,dex in LTE and 0.09\,dex in non-LTE. Part of the reason may be the use of disc-integrated flux in the former study and the disc centre intensity in the latter study, but in general, caution is advised for abundance determination using \ion{Mg}{II} lines because of the unconstrained hydrogen collisional cross-section and the large sensitivity of this majority species on the electron number density \citep{2021A&A...653A.141A}. Since the Mg and Si abundances are very close in absolute numbers, the Mg/Si ratio in the solar photosphere cannot yet be definitely established to be above or below unity, which is important in the context of exoplanet mineralogy (\sect{sect:results-planets}).

The solar abundance of S has not yet been calculated using consistent 3D non-LTE calculations. Indeed, no study has even determined the 1D non-LTE abundance of this important refractory element using recently published hydrogen collisional cross-sections \citep{2020ApJ...893...59B}. This is of high priority, in particular considering the known non-LTE sensitivity of the near-infrared \ion{S}{I} triplet \citep[e.g.][]{2005PASJ...57..751T}.

\subsubsection{Iron}
\label{sect:appendix-sun-fe}
\begin{figure}
\includegraphics[width=5in]{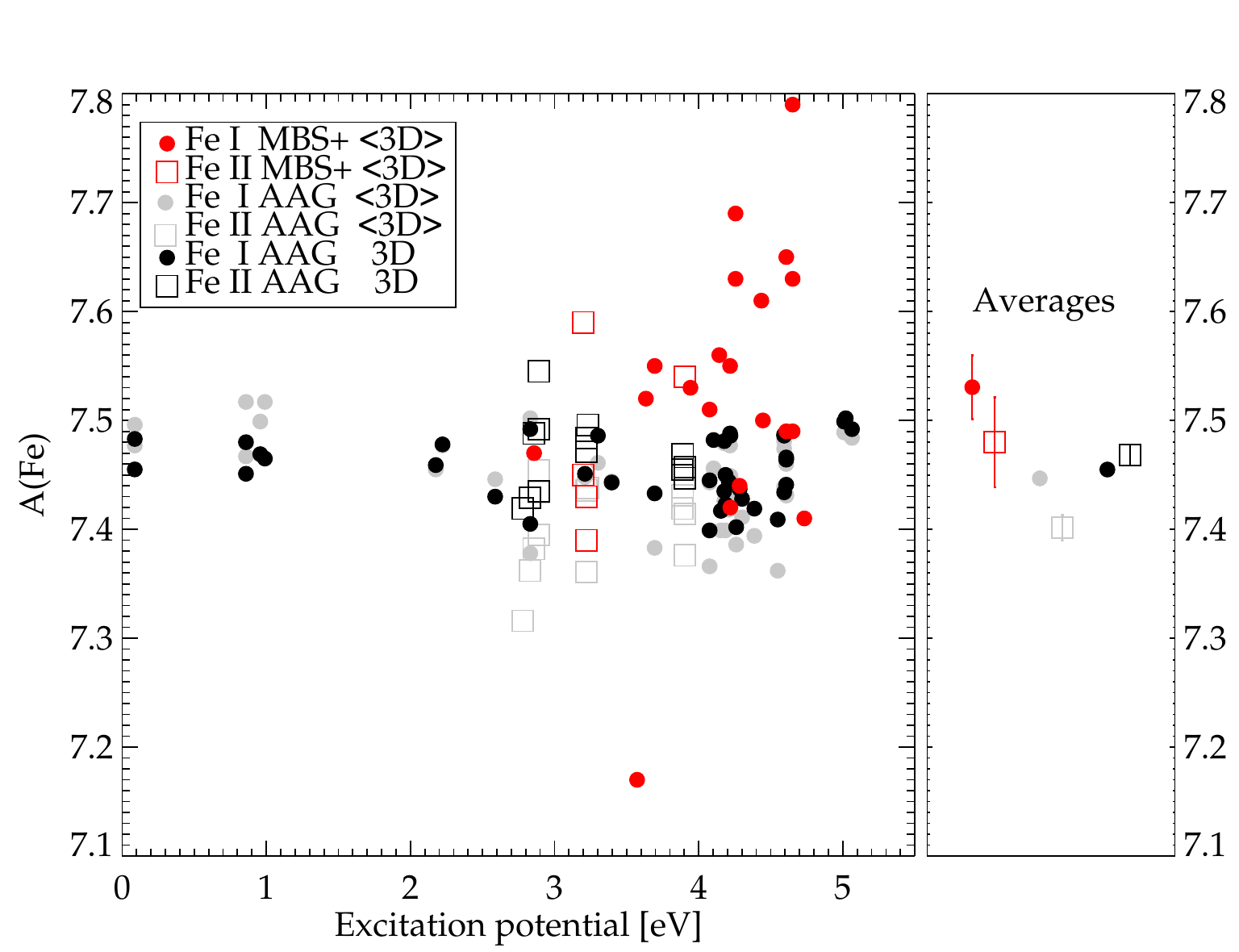}
\caption{The left-hand panel compares the Fe line abundance results of \citet[][MBS+]{2022A&A...661A.140M} and \citet[][AAG]{2021A&A...653A.141A}. All abundances have been computed in non-LTE. Simple averages and standard errors are shown in the right-hand panel. The recommended Fe abundances are $\rm A(Fe)=7.51\pm0.06$ (MBS+) and $\rm A(Fe)=7.46\pm0.04$ (AAG).
}
\label{fig:FeSun}
\end{figure}

A high-resolution optical and near-infrared solar atlas can resolve thousands of \ion{Fe}{I} lines and at least a hundred \ion{Fe}{II} lines, allowing a critical selection of lines based on blending properties and atomic data quality. Fe lines thereby allow powerful diagnostic tests to support claims of preferred Fe abundance for our star; consistent abundances should be determined from different lines regardless of their excitation potential, ionisation stage, line strength, central wavelength, viewing angle, and so on.

It is well known that the neutral Fe atom is over-ionized compared to LTE in the solar atmosphere, an effect that is boosted by over-excitation of the lowest levels \citep[e.g.][]{2012MNRAS.427...50L}. \citet{2011A&A...528A..87M} emphasized the need for non-LTE studies to include highly excited theoretically predicted levels of neutral Fe in the atom (see Fig.\,\ref{fig:atom}), to achieve realistic coupling to the next ionisation stage and remove the need for artificial thermalization of the upper levels \citep[e.g.][]{2001A&A...366..981G}. Nevertheless, prior to the arrival of asymptotic model calculations of collisional cross-sections for excitation and charge transfer with neutral hydrogen \citep{2018A&A...612A..90B}, model atoms had to be astrophysically calibrated.
Using either the Drawin formula with $S_{\mathrm{H}}$ between $0.1$ and $1.0$ or the rates from \citet{2018A&A...612A..90B}, perhaps combined with those computed with the recipe
of \citet{1991JPhB...24L.127K}, recent work in 1D and 3D agree that the average non-LTE impact on \ion{Fe}{I} lines in the Sun is positive and of order $0.01-0.03\rm\,dex$ \citep[e.g.][]{2011A&A...528A..87M,2012MNRAS.427...27B,2015A&A...573A..26S,2015ApJ...808..148S,2017MNRAS.468.4311L,2021A&A...653A.141A}, with effects for individual lines ranging from slightly negative to at most $+0.1\rm\,dex$, depending on the atmosphere and the model atom. In the Sun, the non-LTE effect on \ion{Fe}{II} lines is consistently reported to be less than $0.01\rm\,dex$ in both 1D and 3D \citep{2012MNRAS.427...50L,2017MNRAS.468.4311L,2022A&A...668A..68A}.

The first quantitative determination of the solar Fe abundance using a high-resolution 3D simulation ($200\times200\times82$ meshsize) was published by \citet{2000A&A...359..743A}. The average value determined from \ion{Fe}{II} lines using 3D RHD models and 3D LTE or non-LTE radiative transfer have since then varied in the range $7.45-7.53$ \citep{2000A&A...359..743A,2011SoPh..268..255C,2015A&A...573A..26S,2021A&A...653A.141A}, which is $0.04-0.1\rm\,dex$ higher than theoretical 1D models and $0.04-0.07\rm\,dex$ higher than corresponding $\mtd{}$ models \citep[e.g.][]{2012MNRAS.427...27B,2021A&A...653A.141A}. As seen in \fig{fig:adiffsun}, the weakening of synthetic \ion{Fe}{II} lines in 3D reflects both on the shallower mean temperature stratification of 3D RHD models compared to 1D models in line-forming regions and on the effect of granulation and temperature inhomogeneities. 

Consistent 3D non-LTE calculations for Fe using realistic model atoms have only recently become possible (see Sect.\,\ref{sect:method-atoms-reduction}). \citet{2017MNRAS.468.4311L} determined the solar abundance using both \ion{Fe}{I} and \ion{Fe}{II} lines with the result $7.48\pm0.04\rm\,dex$, which was later improved further by \citet{2021A&A...653A.141A} to $7.46\pm0.04\rm\,dex$. In these studies, it is shown that 3D RHD modelling outperforms $\mtd{}$ and 1D modelling in reproducing the CLV of Fe lines \citep[also demonstrated in LTE by][]{2009A&A...508.1403P} and the excitation and ionisation balance of Fe (\fig{fig:FeSun}). However, earlier studies \citep[e.g.][]{2011A&A...528A..87M,2012MNRAS.427...27B}, have demonstrated that excellent excitation and ionisation balance is also achievable with 1D and $\mtd{}$ models, which reflects the large influence that line selection and fitting, treatment of blends, as well as the choice of atomic data, have on such conclusions.
\fig{fig:FeSun} compares the \ion{Fe}{I} and \ion{Fe}{II} non-LTE line-by-line abundances inferred using 3D and $\mtd{}$ models by \citet{2021A&A...653A.141A}, and using an $\mtd{}$ model by \citet{2022A&A...661A.140M}. The line selection is very different, only four lines overlap, but they have a systematic offset ($0.07\rm\,dex$) that is representative of the difference between the recommended Fe abundances. The upward skew in the latter results may reflect the influence of blends or treatment of microturbulence. We note that excitation balance cannot be verified by the \citet{2022A&A...661A.140M} selection and the substantial scatter in the data ($0.12\rm\,dex$ standard deviation) prevents conclusions on a satisfactory ionisation balance to better than $0.04\rm\,dex$ precision.   


\subsubsection{Recommendations}
\label{sect:sun-recommendations}
In summary, solar 3D non-LTE abundance determinations have been performed for Li, C, N, O, Na, Mg, Al, Si, K, Ca, Fe and Ba \citep[e.g.][and references therein]{2021A&A...653A.141A} as well as Mn \citep{2019A&A...631A..80B}. We argue that 3D LTE modelling for molecules and 3D non-LTE modelling from atoms can result in consistent abundances for available abundance diagnostics without empirical tuning, for the elements for which this has been tested, with the exception of N. Nevertheless, discrepancies between recommended abundances from different studies can be up to $0.1\rm\,dex$: importantly, these are likely not caused by the choice of 3D atmosphere or model atom. Instead, they can often, but not always, be attributed to differences in line selection and fitting, consideration of blends, source of oscillator strengths for diagnostic lines, and choice of observational data. As solar interior models improve in sophistication 
(e.g.~\citealt{2021LRSP...18....2C}, and references therein), abundance and opacity changes of only a few percent can significantly alter our understanding of the solar modelling problem, hence a stronger consensus is desirable. We list general and element-specific advice in bullet-point form below.

Simply raising the solar metallicity does not resolve all facets of the solar modelling problem: as well as the sound speed profile, the observed neutrino fluxes and the Li depletion relative to meteorites need to be consistently explained  \citep{2023A&A...669L...9B}. Neutrino fluxes are sensitive to the assumed model metallicity, but measurements of observed fluxes from the CNO cycle are not yet precise enough to provide strong constraints \citep{2018Natur.562..505B}.  In connection to these issues, it is interesting to consider the radiative opacities below the base of the convection zone \citep{2015Natur.517...56B,2023Atoms..11...52P}, 
pebble-accretion in the pre-main sequence phase \citep{2022A&A...667L...2K},
as well as macroscopic transport \citep{2022NatAs...6..788E}.
As demonstrated by \citet{2023arXiv230813368B}, helioseismic inversions of the adiabatic exponent in the convective envelope, independent of the opacities in the radiative interior, in fact favour a low metal mass fraction that is in line with \citet{2021A&A...653A.141A}. 
This underlines the need for further theoretical and experimental work before strong conclusions about the origins of the solar modelling problem can be drawn; it seems improbable, however, that the discrepancies can be resolved simply by adjusting the abundances alone. Our recommendations to make further progress on the solar debate from a stellar-atmosphere theory perspective are as follows:

\begin{itemize}
\item 3D non-LTE modelling for S, Ni, and CNO-bearing molecules should be performed.
\item Model validation using the observed CLV is desirable for more spectral lines, for example using the data presented in \citet{2023A&A...673A..19E}.
\item The impact on the 3D model atmosphere from abundance changes \citep{2023A&A...677A..98Z} or RMHD simulations should be further investigated \citep[e.g.][]{2015A&A...579A.112S,2016A&A...586A.145S}.
\item There is still a need for higher quality atomic data, for example concerning inelastic hydrogen collisions for \ion{Mg}{II}, photoionisation cross-sections for \ion{Ni}{I}, and oscillator strengths for diagnostic lines of \ion{Fe}{II}.
\item Further theoretical and experimental studies into the physics of inelastic hydrogen collisions, for example with the DESIREE experiment \citep{2021ApJ...908..245B,2022PhRvL.128c3401G} are necessary.
\end{itemize}

\subsection{Stellar parameters}
\label{sect:results-parameters}

\begin{figure*}
\includegraphics[width=5in]{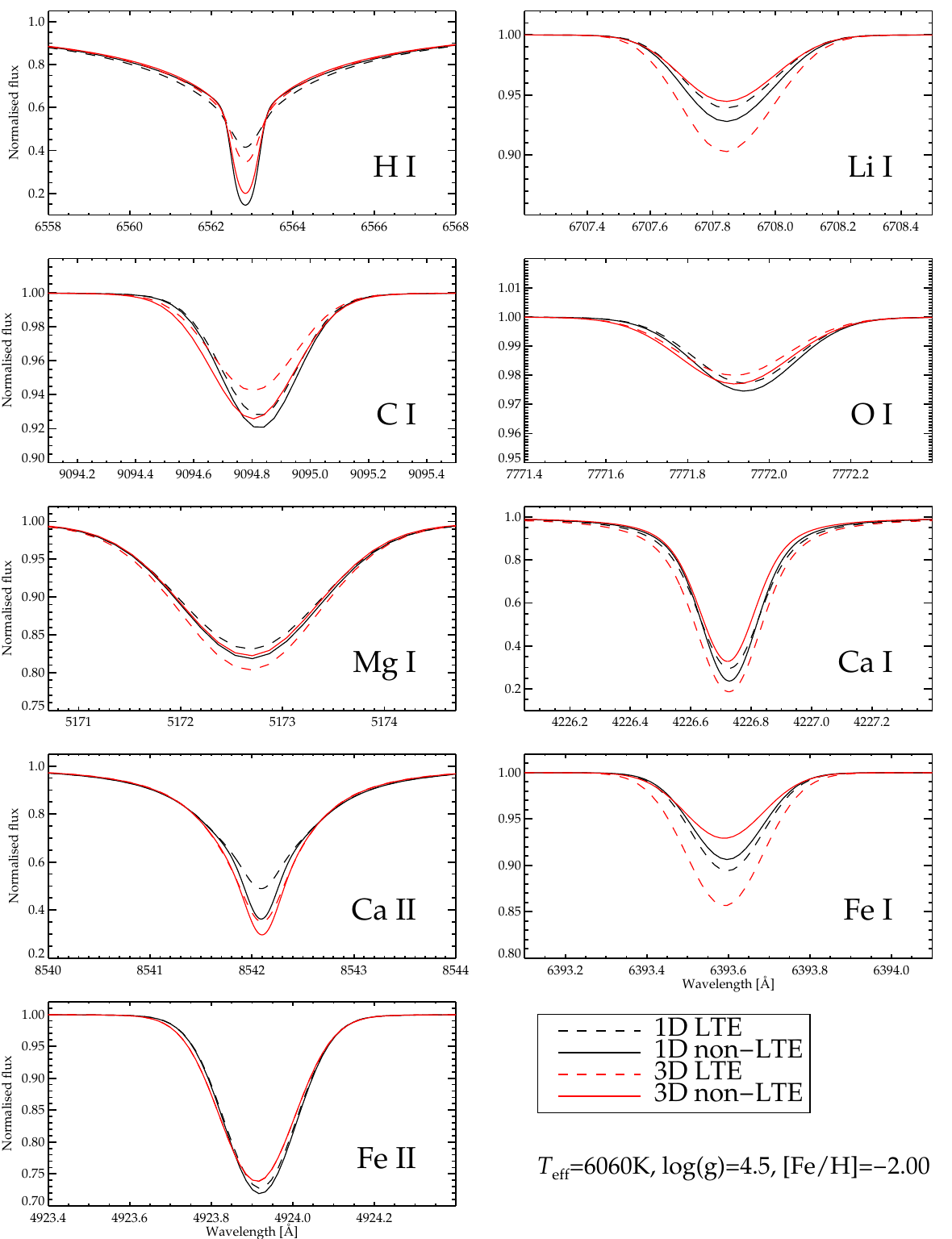}
\caption{Synthetic line profiles computed under different assumptions for a metal-poor dwarf star, with stellar parameters as indicated to the bottom right. The figure has been produced from the calculations by \citet[][\ion{H}{I}]{2018A&A...615A.139A}, \citet[][\ion{Li}{I}]{2021MNRAS.500.2159W}, \citet[][\ion{C}{I} and \ion{O}{I}]{2019A&A...630A.104A}, 
\citet[][\ion{Mg}{I}]{Matsuno_mg}, \citet[][Ca]{2013A&A...554A..96L}, and \citet[][Fe]{2022A&A...668A..68A}.
}
\label{fig:t60g45m20}
\end{figure*}

Before turning to abundance analyses for stars other than the Sun in \sect{sect:results-origin} and \ref{sect:results-planets}, we shall discuss stellar-parameter determination using 3D RHD models and non-LTE calculations. As elaborated on below, spectroscopic stellar parameters inferred from Balmer lines and excitation/ionisation-balance of Fe are indeed model dependent. However, the improved models show great promise to overcome systematic uncertainties and allow the community to fully leverage the reddening-independence and the high precision that is technically achievable with spectroscopy. Throughout this section and next, we refer to \fig{fig:t60g45m20} that shows examples of how 1D LTE, 1D non-LTE, 3D LTE, and 3D non-LTE lines profiles may differ for a very metal-poor dwarf star. 

\subsubsection{Balmer lines}
\label{sect:results-parameters-balmer}

The Balmer lines are prominent in optical spectra of solar-type stars; H$\alpha$ in particular, situated as it is in a relatively line-free region, is often used to determine stellar $\teff$. In the era of large surveys, the large data volumes necessitate parameter optimization via data-driven methods,
in which the observed flux in every (unmasked) pixel contributes to maximize precision. This ought to give the very broad Balmer lines a large influence on the $\teff$ determination, in particular at low metallicity, where there are fewer metal lines with competing influence. However, these lines are prone to uncertainties in the continuum definition \citep[e.g.][]{2002A&A...385..951B}.

Despite the inherent freedom to tune the parameters of the mixing length theory, most importantly the mixing length $\alpha$, 1D models struggle to reproduce the Balmer line profiles predicted by 3D RHD models. The outer wings form deep in the atmosphere and are often stronger in 3D than in 1D, due to the steeper temperature gradients of the hot upflows in this region, while the inner wings of strong Balmer lines can experience the reversed situation at shallower depths \citep{2018A&A...615A.139A,2022A&A...661A..76B}. The final influence on $\teff$ thus depends critically on the choice of line, fitting method, and mixing length, and 1D models errors can reach up to $300\,\mathrm{K}$ \citep{2009A&A...502L...1L}.
To disentangle the influence of atmospheric inhomogeneities with the difference in average atmospheric structures in 1D and 3D LTE, \citet{2009A&A...502L...1L} also computed $\mtd{}$ Balmer line profiles. Interestingly, they find that the full 3D profiles, including the effects of granulation, are substantially different from $\mtd{}$ in many cases. Even the sign of the temperature correction found for 3D$-$1D can differ from $\mtd{}-$1D, e.g., $-65$\,K and $+106$\,K, respectively, for H$\beta$ in a very metal-poor turn-off star ($\alpha=1.0$). \citet{2019A&A...624A..10G} proposed an empirical $\feh$-dependent positive correction to H$\alpha$-$\teff$ inferred from 1D models of dwarfs and subgiants, equal to 28K at $\feh=0$ and $139$\,K at $\feh=-0.7$. They find good qualitative agreement of this trend with expected 3D$-$1D effects, and conclude that 3D models are superior in this respect.    

\citet{2018A&A...615A.139A} investigated 3D non-LTE line formation of Balmer lines over a large grid of atmospheres. They find photon pumping by Ly$\alpha$ to be an important non-LTE effect, over-populating the first excited state of neutral hydrogen, which is the lower level of the Balmer series. In combination with several other influential bound-bound and bound-free transitions among highly excited levels, the non-LTE effect can reach $\pm100$\,K for the H$\alpha$ line wings; the non-LTE effects are more muted for the wings of H$\beta$ and H$\gamma$ that form deeper in the atmosphere. For metal-poor turn-off stars, 3D and non-LTE effects act in the same direction and 1D LTE models of H$\alpha$ underestimates $\teff$ by 150\,K for typical choices of mixing length parameters (see \fig{fig:t60g45m20}). For the Sun, 3D non-LTE modelling of H$\alpha$, H$\beta$ and H$\gamma$ returns $\teff$-values that are hotter than 1D models, but still slightly too low at 5709-5722\,K \citep{2018A&A...615A.139A}, which \citet{2021A&A...650A.194G} suggest is caused by a small normalisation distortion in the adopted solar atlas \citep{2019A&A...624A..10G}.
$\mtd{}$ models cannot well reproduce the 3D line profiles in the Sun, either in LTE \citep{2009A&A...502L...1L} or non-LTE \citep{2013A&A...554A.118P}. \citet{2023A&A...679A.110G} find excellent agreement within $\pm46$\,K between 3D non-LTE H$\alpha$-temperatures, using the synthetic profiles by \citet{2018A&A...615A.139A}, and the infrared flux method for an extended sample of metal-poor and metal-rich stars, including red giants.

The central $\pm0.1$\,nm has the largest $\teff$ and non-LTE sensitivity, but the Balmer line cores are poorly reproduced by both 1D and 3D photospheric models, because of the chromospheric influence \citep[e.g.][]{2012ApJ...749..136L}, necessitating RMHD models including non-equilibrium hydrogen ionisation. \citet{2016A&A...594A.120B} found empirical correlations between the width of the H$\alpha$ line core in red giants and other fundamental stellar parameters; $\feh$, $\lgg$ and, interestingly, stellar mass. However, in the lack of RMHD chromospheric simulations for such stars, the authors forego speculation about the underlying cause of the observed correlations. The first RHD model of a late-type giant (Aldebaran, $\teff/\lgg/\feh=4000/1.5/0.0$) that extends to the chromosphere was computed by \citet{2017A&A...606A..26W}, revealing a dynamic and shock-heated layer above the photosphere. However, they show that the 3D non-LTE line profile of H$\alpha$ is still far too narrow to reproduce the observed core width, which could be due to missing physics in the model atmosphere itself, not least magnetic fields.

\subsubsection{Fe lines}
\label{sect:results-parameters-fe}

A common method to determine stellar parameters spectroscopically is to exploit the excitation equilibrium of the numerous \ion{Fe}{I} lines present in late-type spectra and the ionisation equilibrium of \ion{Fe}{I} and \ion{Fe}{II} lines. However, 1D LTE spectroscopic parameters are known to underestimate both $\teff$ and $\lgg$ for metal-poor stars \citep[e.g.][and references therein]{2013ApJ...769...57F}. The determination of $\lgg$ has seen significant improvement with 1D non-LTE calculations \citep[e.g.][]{2003A&A...407..691K,2023MNRAS.524.3526M}, which capture the over-ionisation of \ion{Fe}{I} and raise the Fe abundances determined from these lines, while non-LTE effects on \ion{Fe}{II} lines are negligible. \citet{2012MNRAS.427...50L} illustrate how the effect on $\lgg$ varies over the FGK star parameter space, and can exceed $1\rm\,dex$ for hot, metal-poor giants. However, $\teff$ inferred from the excitation balance of \ion{Fe}{I} lines tend to be even lower than in LTE in metal-poor stars, as a consequence of the erroneous temperature stratification and neglect of inhomogeneities. 1D non-LTE studies have attempted to mitigate this problem by excluding the lines with lowest excitation potential \citep{2023AJ....165..145L} or by combining photometric $\teff$ with spectroscopic $\lgg$ \citep{2013MNRAS.429..126R}. As shown by \citet{2013MNRAS.429.3645S}, the model-dependence of $\teff$ and $\lgg$ furthermore leads to severely underestimated ages and overestimated masses and distances from 1D LTE analyses. 

Since the first 3D LTE analysis studies of \ion{Fe}{II} lines, it has been known that the impact on Fe abundances is significant and can be both positive and negative, depending on line properties and stellar parameters \citep[e.g.][]{2002ApJ...567..544A,2004A&A...415..993N,2007ApJ...671..402K}. Weak \ion{Fe}{II} lines are typically affected by $\pm0.1$\,dex, as shown for red giants using fictitious lines \citep{2007A&A...469..687C,2011A&A...529A.158H,2013A&A...559A.102D}. This behaviour was later confirmed for both dwarfs and giants using real lines \citep{2019A&A...630A.104A}, while saturated lines have larger corrections that also can be both positive and negative. Significant impact is not only seen for low-metallicity stars, but a turn-off star like Procyon is affected by of order $+0.1$\,dex \citep{2002ApJ...567..544A,2022A&A...668A..68A}. The LTE assumption for \ion{Fe}{II} line formation in 3D models has been shown to be accurate within $\sim0.02$\,dex for $\feh>-2.5$, but non-LTE effects reach $\sim+0.1$\,dex for extremely metal-poor turn-off stars \citep{2016MNRAS.463.1518A,2022A&A...668A..68A} and thereby exacerbate the difference with respect to 1D LTE in such stars.

\ion{Fe}{I} lines are affected by both 3D and non-LTE effects that can act in the same or opposite directions, emphasizing the need for consistent modelling, in particular when non-LTE effects are large
\citep{2005ApJ...618..939S,2016MNRAS.463.1518A,2017A&A...597A...6N,2017MNRAS.468.4311L,2022A&A...668A..68A,2023A&A...672A..90L}. The combined impact on average metallicities is of order $+0.3$\,dex for very/extremely metal-poor turn-off stars \citep{2016MNRAS.463.1518A,2022A&A...668A..68A}, increasing to $+0.5$\,dex at $\feh<-4$ \citep{2023A&A...672A..90L}. A slightly smaller $+0.2$\,dex effect was reported for the red giant HD122563 \citep{2016MNRAS.463.1518A}, but caution is advised given the significant ionisation imbalance remaining in 3D non-LTE, which is expected to improve with an updated 3D model that use the $-0.2$\,dex revision of the parallax-constrained $\lgg{}$ \citep{2018MNRAS.475L..81K}.
The upper limit to $\feh$ for the most Fe-poor star known, also a red giant, was raised by $+0.8$\,dex \citep{2017A&A...597A...6N}. In the most recent 3D non-LTE study by \citet{2022A&A...668A..68A}, it is seen that \ion{Fe}{I} lines of low and high excitation are differently impacted, such that the abundance slope with excitation potential increases and flattens the negative gradients seen for the benchmark metal-poor stars HD84937 and HD140283 in 1D LTE. The effect corresponds to a $\teff$-change of $250-300$K and reflects on the complex and coupled 3D non-LTE mechanisms. The low excitation lines of \ion{Fe}{I}, when the species is in minority, are the most sensitive to the steeper temperature stratification and inhomogeneities that strengthen the lines \citep{2007A&A...469..687C,2013A&A...559A.102D,2015A&A...579A..94G}, but they are also most affected by non-LTE effects in 3D models that weaken them \citep{2016MNRAS.463.1518A}. The high-excitation \ion{Fe}{I} lines have a behaviour that is more similar to \ion{Fe}{II} lines.

The full impact of 3D non-LTE modelling on stellar parameters can be exemplified with G64-12, an extremely metal-poor dwarf/turn-off star used as benchmark star in many studies. Fe line formation in 
3D non-LTE \citep{2016MNRAS.463.1518A} gives approximately $\teff=6450$\,K, $\lgg=4.3$ and $\feh=-3.0$, in line with the infrared flux method $T_{\rm eff}=6463$\,K \citep{2010A&A...515L...3M}, while 1D LTE spectroscopic analysis returns $\teff=6100$\,K, $\lgg=3.9$ and $\feh=-3.6$ \citep{2023ApJ...953...31S}; i.e., a $0.6$ dex error in $\feh$.

\subsubsection{Photometry}
\label{sect:results-parameters-photometry}
The impact of 3D RHD models on photometric colours have been investigated by \citet{2018A&A...611A..11C} using the \stagger{}-grid
(with grids of spectra available via the POLLUX database:
\citealt{2010A&A...516A..13P}),
and \citet{2018A&A...613A..24K} and \citet{2018A&A...611A..68B} using the \cifist{}-grid. Some degree of sensitivity of continuum fluxes to granulation is in fact expected simply from the non-linear temperature-dependence of the Planck function, with cooler models and shorter wavelengths being most affected \citep[see Appendix A in][]{2018A&A...613A..24K}. The \cifist{} results show that 1D narrow-band colours, such as $g-z$, can underestimate $\teff$ by up to 200\,K for the hot end of the grid, with a smaller effect in the opposite direction seen for the cooler end. Excluding UV-centered bands, the maximum effect on the investigated colours is small, but non-negligible at $\pm0.03$\,mag. However, as cautioned by the authors, the treatment of scattering as true absorption in the $\cifist{}$-grid may overestimate the effect.

\subsection{Origin of the elements}
\label{sect:results-origin}

Stars act as cosmic element factories, forging elements through nuclear reactions, and redistributing them into the universe through powerful winds and explosions. The ongoing cycle of stellar birth, evolution, and death enriches the cosmos with heavy elements and provides the raw materials for new stars and planets. The observed element abundances in stars in our Galaxy serve to constrain the ingredients of this process, e.g., the star formation history, the hierarchical assembly of the Galaxy, and the mass-dependent yields of stars, and allow us to validate and calibrate models of Galactic chemical evolution. 
In the newborn and metal-poor universe, massive stars were the main producers of new elements, forged either during hydrostatic burning or in core-collapse  supernovae (SNe) and hypernovae (HNe) explosions that return the elements to the interstellar medium. We focus mainly on the metallicity regime $-4<\feh<-1$, in which the chemical pattern of Galactic stars reflect on the initial mass function of star formation and the yields from an accumulated large number of events, rather than the stochastic effects of individual explosions. The limit in metallicity where chemical enrichment from delayed sources, such as AGB stars and thermonuclear supernovae (SNIa), becomes noticeable is element and model-dependent, e.g., $\feh\gtrsim-2.5$ for N, but $\feh\gtrsim-1$ for $\alpha$-elements \citep[e.g.][]{2020ApJ...900..179K}. 

We limit the discussion to the origin of Li, C, N, O, $\alpha$-elements, Fe-peak elements, and the neutron-capture element Ba, in the light of 3D non-LTE calculations. The main focus on metal-poor stars is a reflection on the larger modelling errors that are often associated with their 1D LTE abundances: reduced UV line blocking leads to  amplification of non-LTE effects, and there are also larger differences between the 1D and 3D temperature stratifications at lower metallicities (\sect{sect:method-effects}). The use of the Sun as reference star to cancel systematic errors also works less well compared to the solar-metallicity regime, thus making the need for realistic stellar atmosphere and line formation modelling all the greater.

\subsubsection{Lithium}
\label{sect:results-origin-li}
More than two decades ago, Li was the first element targeted with 3D non-LTE modelling in a metal-poor dwarf and subgiant \citep{2003A&A...399L..31A}. It is also the first element for which a grid of 17 models was produced for such stars \citep{2010A&A...522A..26S}, a number that has today grown to cover 70 \cobold{} models \citep{2020A&A...638A..58M} and 195 \stagger{} models for metal-poor and metal-rich dwarfs and giants \citep{2021MNRAS.500.2159W}. This astonishing progress is partly a consequence of the simple term structure of this alkali element, making grid computations numerically feasible, and the reliability of the atomic data in particular with full quantum mechanical data for the inelastic hydrogen collisions \citep{2003A&A...409L...1B}, which help justify carrying out expensive 3D non-LTE calculations.  It also of course reflects on the great astrophysical interest in Li. The Big Bang origin of the dominant isotope $\rm^7Li$, combined with both production and depletion by stars and stellar environments give rise to a complex abundance pattern that is correlated with stellar, and possibly planetary, parameters, but also involves Li-rich and Li-poor outliers \citep[see, e.g., review by][]{2021FrASS...8....6R}. 

The non-LTE line formation of the resonant \ion{Li}{I} 670.7\,nm line follows a simple pattern: when the line is weak, the corrections are small and depend mainly on the photoionisation/photorecombination balance, while at larger line strength, significant photon losses cause increasingly negative corrections until the point of maximum saturation. \citet{2021MNRAS.500.2159W} uncovered an important systematic error affecting all previous work that has neglected background line opacity for bound-bound transitions \citep[e.g.][]{2009A&A...503..541L,2020A&A...638A..58M}. If blending metal lines are not included in the calculation of radiative bound-bound rates, UV \ion{Li}{I} lines cause significant photon pumping, which leads to non-LTE corrections that are too positive or not sufficiently negative by up to $+0.15$\,dex for a test case. For giants and cool dwarfs around $\feh=-2$, the assumption of 1D non-LTE is often accurate to within $\pm0.03$\,dex, while cancellation of 3D and non-LTE effects coincidentally lead to similarly accurate 1D LTE results for warm dwarfs and TO stars (see Fig.\,\ref{fig:t60g45m20}). However, we caution that these comparisons depend on both metallicity and Li abundance, as well as the microturbulence adopted
for the 1D analysis.

\citet{2021MNRAS.500.2159W} analyse how their 3D non-LTE results impact particularly important Li abundance measurements. The Spite plateau of unevolved metal-poor stars experiences a small decrease ($-0.03\,\mathrm{dex}$) compared to 1D LTE at $\feh=-2$. At $\feh=-3$, corrections are close to zero, while they revert sign and reach $+0.1$\,dex at $\feh=-4$. This has a flattening effect on the plateau but likely not sufficiently large to completely erase the positive slope with increasing metallicity \citep[see e.g.][]{2023MNRAS.522.1358N}, which instead should be understood in the context of atomic diffusion and internal mixing. Similar conclusions on the robustness of the Spite plateau were drawn by \citet{2020A&A...638A..58M}. Recently, \citet{2020MNRAS.497L..30G} uncovered a systematic $0.4$\,dex difference in 1D non-LTE Li abundance between stars on the hot and cool sides of the so called Li dip that is centered on $\teff=6500\,\mathrm{K}$ on the main sequence, suggesting that the cool stars have indeed undergone depletion by approximately this amount. Due to the similarity in stellar parameters, the relative difference is not expected to change significantly in 3D non-LTE, but the absolute plateau levels may experience a slight increase. As discussed by \citet{2021MNRAS.500.2159W}, the effect on the RGB stars is more severe, reducing Li abundances on the lower RGB compared to 1D LTE by approximately $-0.15$\,dex at $\feh=-2$. When corrected for the large dilution caused by the first dredge-up, the initial Li abundances are broadly in agreement with the canonical Spite plateau. In Li-rich giants, the 670.7\,nm line can be very strong, and abundance corrections have been shown to reach $-0.6$\,dex; clearly these are important to implement to understand the underlying cause of the enrichment \citep[e.g.][]{2022arXiv220902184K}.

Several studies have shown that the $\rm^6Li/^7Li$ isotopic ratio can be constrained from the shape of the 670.7\,nm line in high-resolution spectra and several significant detections at a level of $5\%$ were reported for stars on the Spite plateau at $-3<\feh<-1$ \citep[e.g.][and refereces therein]{2006ApJ...644..229A}. However, any significant detection of the lighter isotope in unevolved very metal-poor stars is difficult to reconcile with a standard Big Bang nucleosynthesis scenario, in particular when simultaneously considering stellar depletion of this element, prior to and during the main sequence. \citet{2007A&A...473L..37C} demonstrated how the impact of convective motions on the line profile shape is degenerate with the amount of $\rm^6Li$, in LTE as well as non-LTE. A reappraisal of the Li isotopic ratios in halo stars with 3D non-LTE calculations were later performed by \citet{2013A&A...554A..96L} and \citet{2022MNRAS.509.1521W}, transforming previous detections to upper limits at $<2\%$, thereby removing the tension with the cosmological predictions. As shown, for example, by \citet{2017A&A...604A..44M}, non-LTE modelling in both 1D and 3D can still lead to significant detection of the lighter isotope in a solar-metallicity subgiant, suggesting a connection to planet engulfment or stellar activity.   

\subsubsection{Carbon and $\alpha$-element abundances from atomic lines}
\label{sect:results-origin-calpha}

\begin{figure*}
     \centering
     \begin{subfigure}[b]{0.48\textwidth}
         \centering
         \includegraphics[width=\textwidth]{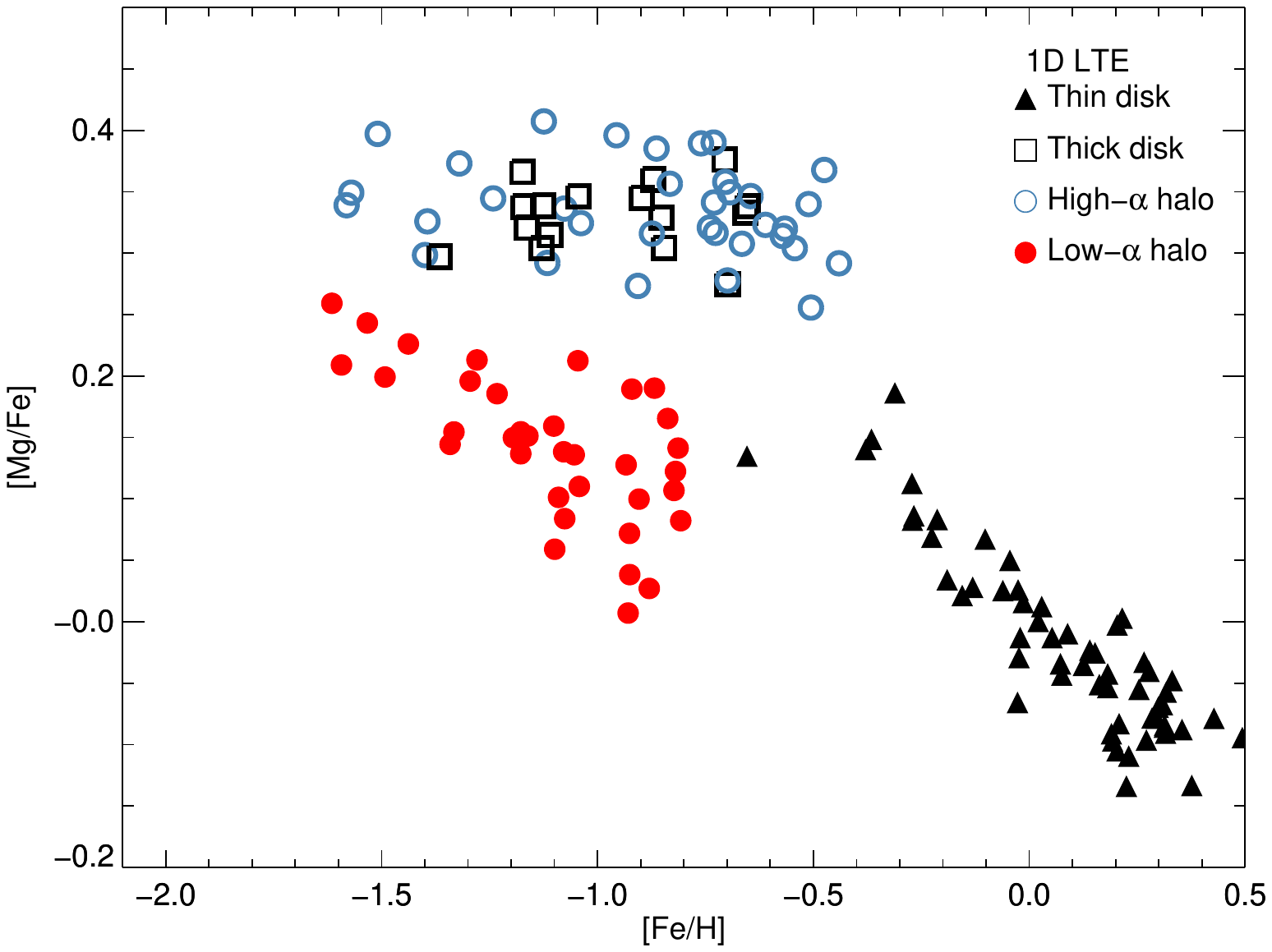}
     \end{subfigure}
     \hfill
     \begin{subfigure}[b]{0.48\textwidth}
         \centering
         \includegraphics[width=\textwidth]{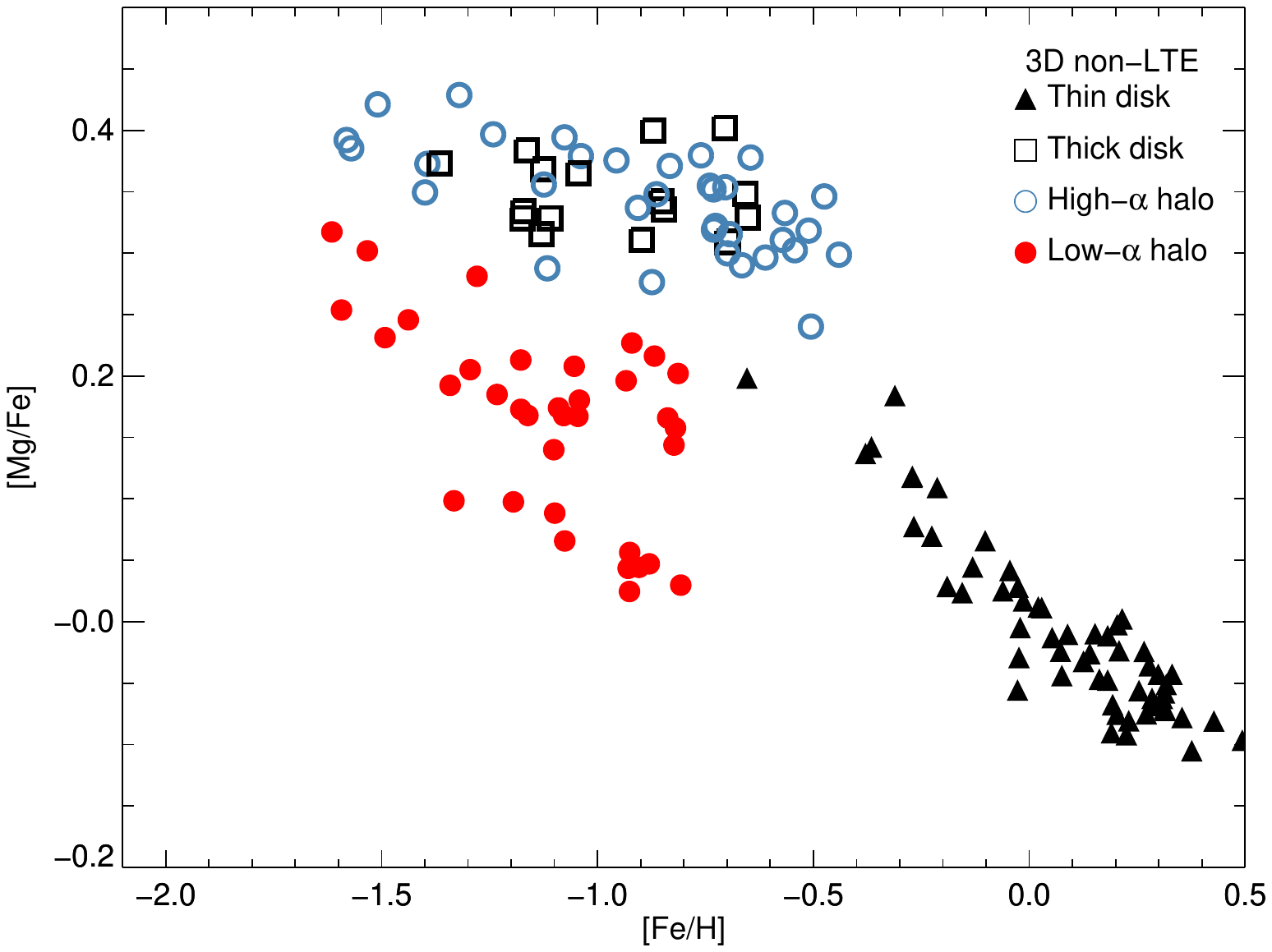}
     \end{subfigure}
     \hfill
        \caption{The [Mg/Fe] ratios of different Galactic populations obtained from 1D LTE (left) and 3D non-LTE (right) analysis. Two separate tracks are uncovered in the low-$\alpha$ halo from the more accurate abundances displayed in the right-hand panel, implying that two separate accretion events have taken place \citep{2023arXiv231207768N,Matsuno_mg}.}
        \label{fig:MgFe}
\end{figure*}

The so called $\alpha$-elements, O, Mg, Si, S, and Ca, are over-produced with respect to Fe by core-collapse SNe and HNe, while their building block C is less so. However, the abundance patterns predicted by Galactic chemical evolution models have many uncertainties. First, the $\rm[\alpha/Fe]$-ratios in metal-poor stars are affected ($0.1-0.5\rm\,dex$ for [O/Fe]) by assumptions made for the mass range of failed SNe, for which the entire CO-core fall back onto the black hole, and the upper mass limit of HNe \citep{2010A&A...522A..32R,2020ApJ...900..179K}. Secondly, the assumed fraction of HNe affects the tilt of the $\rm[\alpha/Fe]$ trends with metallicity before the onset of SNIa, which can be either steep ($\sim0.6$\,dex decline over $\rm-4<\feh<-1$) in the absence of HNe or shallow, almost flat, if all stars $M>20 M_\odot$ explode as HNe \citep{2010A&A...522A..32R}. Thirdly, mixing and fallback of HNe also influences yields \citep{2002ApJ...565..385U}, but is treated as a free parameter. Ratios between alpha elements are much less sensitive to the mass ranges, but can, e.g., be affected by changes in the $\rm^{12}C(\alpha,\gamma)^{16}O$ nuclear reaction rate.

The metallicity range $\rm \feh>-3.5$ can be probed by atomic C and O lines, for which both 3D and non-LTE effects have been investigated. 1D non-LTE effects for the highly excited allowed transitions are slightly negative in this metallicity regime for unevolved stars, of order $-0.1\rm\,dex$ \citep{2016ApJ...833..225Z,2019A&A...622L...4A}, placing the [O/Fe] values at $\sim0.6-0.7$ and the [C/Fe] values at $\sim0.1-0.2$. Consistent 3D non-LTE calculations have not dramatically altered the picture, but cause a further decrease of order $0.05-0.1\rm\,dex$ for both elements \citep{2019A&A...622L...4A,2020A&A...643A..49H} and display a slight increasing trend towards lower metallicity, in reasonable agreement with the predictions by \citet{2020ApJ...900..179K} that are based on an assumed HNe fraction of 0.5 at low metallicity. \citet{2004A&A...416.1117C} reported a similar mean [O/Fe] value for extremely metal-poor giants based on 1D LTE analysis of the 630\,nm line. The authors estimate a $-0.23\rm\,dex$ correction for 3D effects, following \citet{2002A&A...390..235N}, but more recent work have shown that 3D corrections to [\ion{O}{I}] line abundances are in fact very small in extremely metal-poor giants \citep{2015A&A...576A.128D,2018MNRAS.475.3369C}. Importantly, the trend of the [C/O] ratio with metallicity is constant at $-0.6$ in 3D non-LTE \citep{2019A&A...630A.104A}, in agreement with, e.g., \citet{2020ApJ...900..179K}.  In contrast, in 1D LTE [C/O] shows a striking upturn towards low metallicity. This was previously interpreted as a possible signature of Pop III stars \citep{2004A&A...414..931A}, or of rapid rotation in low-metallicity massive stars \citep{2006A&A...449L..27C}; although more recent GCE models including rotation by \citet{2018MNRAS.476.3432P} suggest that while there is a small positive effect on the absolute C and O levels, rotation does not strongly impact the tilt of the [C/O] ratio.

Several recent studies place the [Mg/Fe] plateau in very metal-poor Milky Way stars around $\sim0.3$ in 1D non-LTE, which is considerably lower than [O/Fe] \citep{2016ApJ...833..225Z,2017ApJ...847...16B,2017A&A...608A..89M,2020A&A...642A..62A,2020A&A...643A..49H}. Non-LTE effects are typically minor for Mg, but become increasingly positive with decreasing metallicity for unsaturated lines, and increasingly negative with line strength until saturation \citep[e.g.][]{2016A&A...586A.120O}. The large positive ($0.3-0.4\rm\,dex$) non-LTE effects predicted for extremely metal-poor turn-off stars by \citet{2010A&A...509A..88A}, which would raise [Mg/Fe] to 0.6, are not confirmed by later works. However, the same studies that predict smaller non-LTE corrections have also revealed an ionisation imbalance of 0.2\,dex in the sense that \ion{Mg}{II} lines give higher abundances than \ion{Mg}{I} for metal-poor unevolved stars \citep{2018ApJ...866..153A,2022A&A...665A..33L}.   

Sample 3D non-LTE calculations were performed for a mildly metal-poor dwarf and very metal-poor giant model by \citet{2017ApJ...847...15B} with the result that Mg abundances based on \ion{Mg}{I} lines can be both higher and lower in 3D non-LTE compared to 1D non-LTE, largely dependent on the microturbulence. Recently, \citet{2023arXiv231207768N} used a grid of 3D$-$1D non-LTE corrections for the \ion{Mg}{I} 571.1\,nm in dwarfs \citep{Matsuno_mg}, with differential abundance corrections
of $-0.03$\,dex for the coolest stars to $+0.06$\,dex in the warmest, most metal-poor stars around $\feh=-1.5$, on top of a solar correction of $+0.06\,\mathrm{dex}$.
 Thus, \citet{2023arXiv231207768N} demonstrate the first 3D non-LTE Mg abundances for a large sample of stars, in this case halo dwarfs. Combined with 3D LTE \ion{Fe}{II}-based Fe abundances \citep{2019A&A...630A.104A}, the [Mg/Fe]-ratio is reported to be stable around $\sim0.3$, but split between slightly higher and lower values in two halo populations with different chemical enrichment histories (\fig{fig:MgFe}). The 3D non-LTE [O/Mg]-values are not affected by this bimodality, but are uniformly centered on $\sim0.2$, in good agreement with the predictions by \citet{2020ApJ...900..179K}. At even lower metallicities, tailored studies of two ultra metal-poor stars have shown that 3D non-LTE Mg abundances are $0.3-0.5$\,dex higher than 1D LTE, and $0.1-0.2$\,dex higher than 1D and $\mtd{}$ non-LTE for a dwarf and a giant \citep{2017A&A...597A...6N,2023A&A...672A..90L}. However, the corresponding effect on Fe is even larger (\sect{sect:results-parameters-fe}), and [Mg/Fe] is therefore effectively reduced. 


The [Ca/Fe] plateau in very metal-poor stars is at a very similar level to Mg in 1D non-LTE in the Milky Way, i.e., at approximately $\sim0.3-0.4$, including similar corrections to neutral lines in 1D \citep[e.g.][]{2016ApJ...833..225Z,2017A&A...608A..89M,2019MNRAS.485.3527S,2020A&A...642A..62A,2020A&A...643A..49H}. A 3D non-LTE investigation was made by \citet{2013A&A...554A..96L} for four unevolved stars in the range $\rm-3.5<\feh<-2$, indicating that 3D non-LTE abundances are higher than 1D non-LTE by 0.2\,dex for the neutral resonance 422.67\,nm line (see Fig.\,\ref{fig:t60g45m20}) and 0.1\,dex higher for neutral subordinate lines. The \ion{Ca}{II} triplet, on the other hand, usually has negative corrections in 1D \citep[e.g.][]{2019MNRAS.485.3527S,2022ApJ...928..173O}, which may cancel or decrease even further in 3D \citep{2013A&A...554A..96L,2017A&A...597A...6N,2023A&A...672A..90L}. Reassuringly, a 0.5\,dex Ca ionisation imbalance seen in 1D LTE for an ultra metal-poor dwarf is strongly reduced in 1D non-LTE and removed completely in 3D non-LTE \citep{2023A&A...672A..90L}. A similar degree of improvement was seen in 1D non-LTE for a larger sample of ultra metal-poor stars by \citet{2019MNRAS.485.3527S}.

\subsubsection{Carbon, nitrogen, oxygen, and magnesium abundances from molecular lines}
\label{sect:results-molecules}

\begin{figure*}
     \centering
     \begin{subfigure}[b]{0.48\textwidth}
         \centering
         \includegraphics[width=\textwidth]{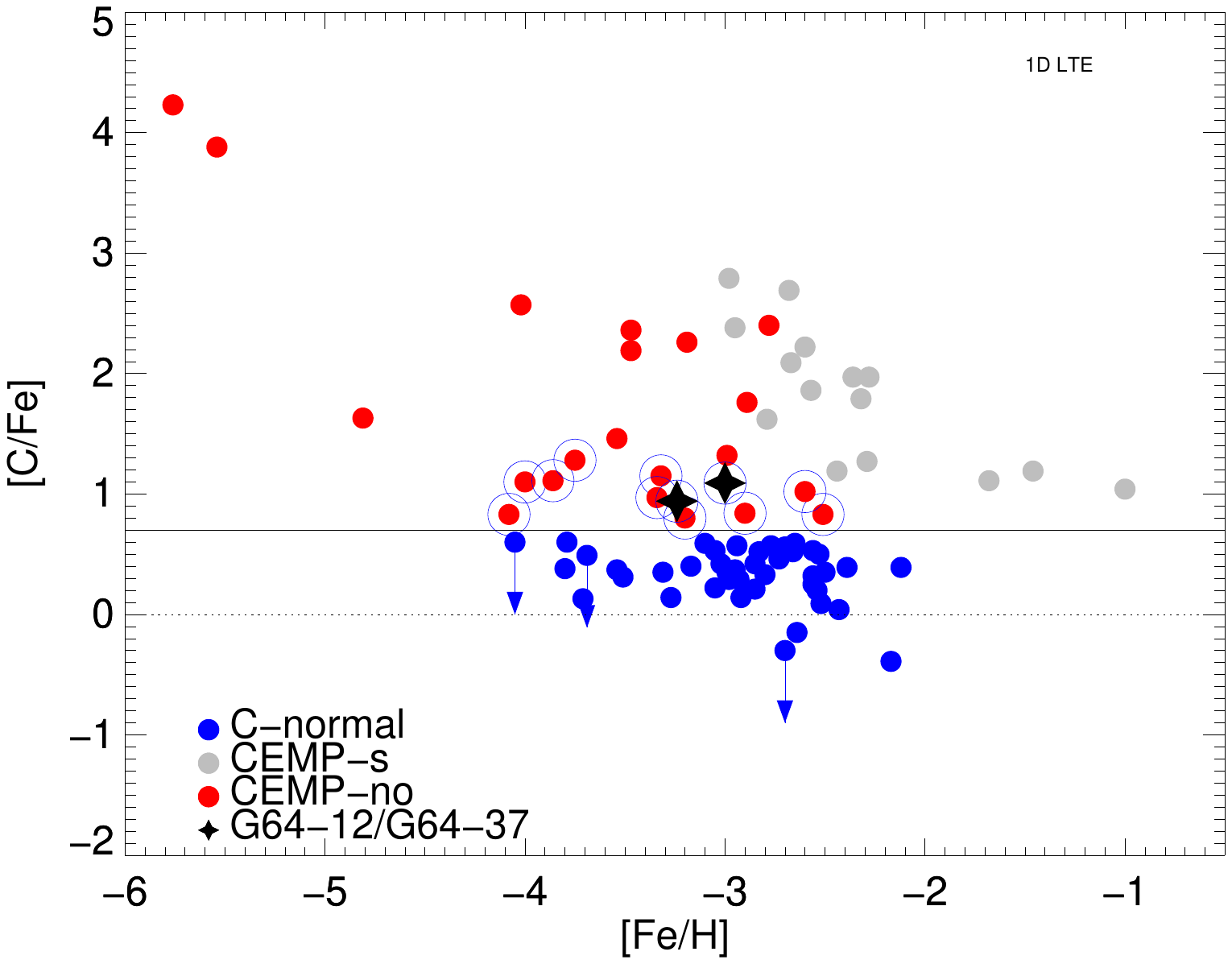}
     \end{subfigure}
     \hfill
     \begin{subfigure}[b]{0.48\textwidth}
         \centering
         \includegraphics[width=\textwidth]{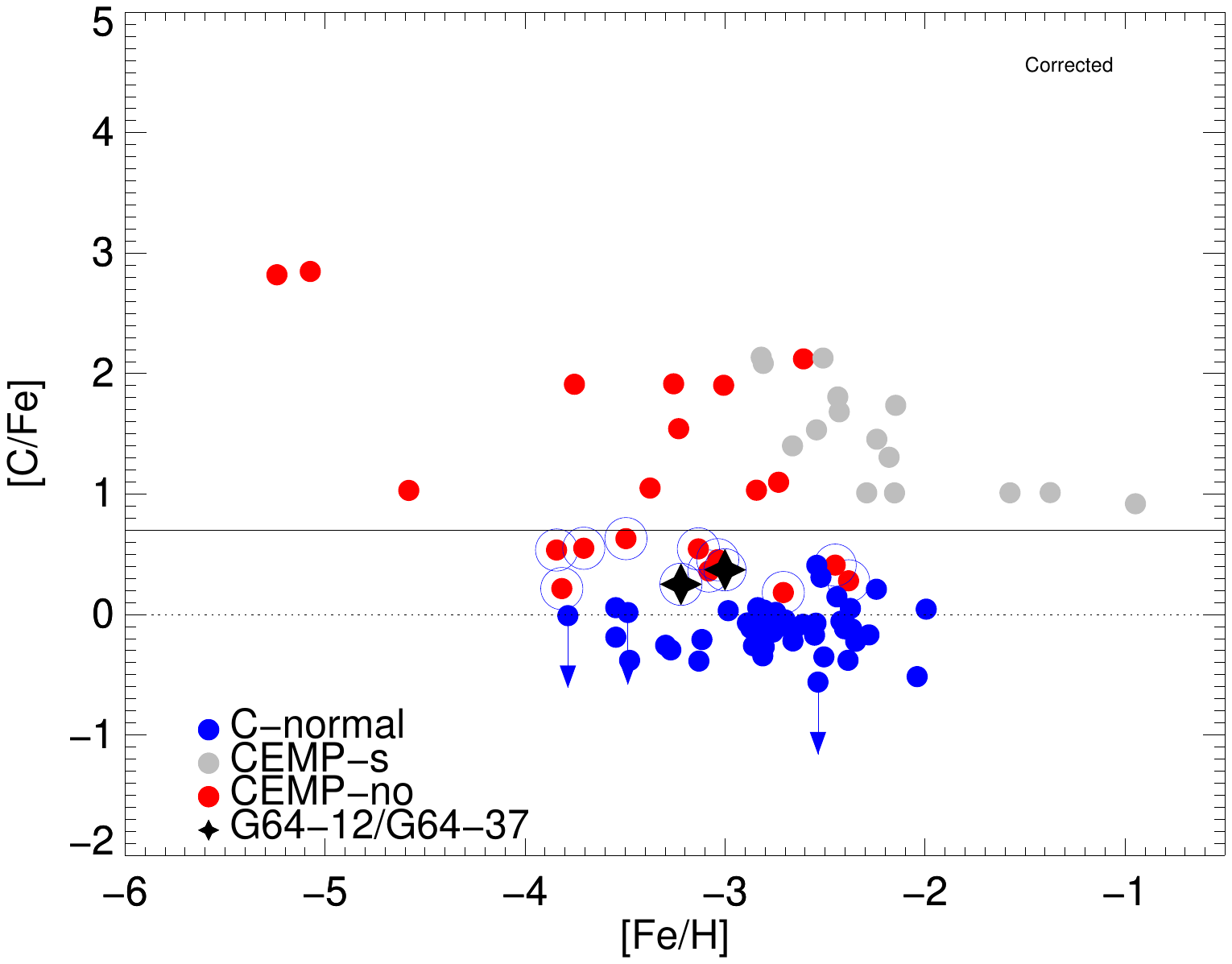}
     \end{subfigure}
     \hfill
        \caption{The left-hand panel shows 1D LTE abundances for C and Fe for the metal-poor star sample ($\teff\rm<6600$\,K and $\lgg>2$) of \citet{2013ApJ...762...27Y} and two stars (G64-12 and G64-37) from \citet{2016ApJ...829L..24P} and \citet{2019ApJ...879...37N}. The right-hand panel shows C abundances corrected using 3D LTE line formation of the CH-band, and Fe abundance corrected for 1D non-LTE effects. The stars marked with blue open circles would be categorized as C-enhanced in the left panel, but C-normal in the right panel.}
        \label{fig:CEMP}
\end{figure*}

Since the first discovery of metal-poor stars with anomalously strong CH features, C-enhanced metal-poor stars (CEMP stars) has continued to further our understanding of the very first stellar generations. Of particular importance in this context is the subclass CEMP-no, that are characterized by high abundances of light elements (C, N, O, Na, Mg, Al), but lack enhancement of neutron-capture elements \citep[][and references therein]{2019ApJ...879...37N}. The fraction of CEMP stars grow steadily toward lower metallicity and they dominate the ultra metal-poor regime.  

The only available CNO abundance indicators in $\feh<-4$ stars are molecular bands, hence it is of great importance to understand the systematic uncertainties associated with them in model spectra. The first 3D RHD models that were created for metal-poor stars \citep{2001A&A...372..601A,2006ApJ...644L.121C} showed that they have a profound impact on molecular line formation. As clearly illustrated by \citet{2013A&A...559A.102D}, the effect of the added granulation is larger than the change in temperature stratification. The contribution functions of molecules like CH, NH and OH extend to much shallower layers, strengthening synthetic line profiles and reducing abundances in 3D. CNO abundance corrections down to $-1\rm\,dex$ have been predicted for extremely metal-poor dwarfs \citep{2008A&A...480..233G,2010A&A...519A..46G,2016A&A...593A..48G} and giants \citep{2007A&A...469..687C,2011A&A...529A.158H,2013A&A...559A.102D}. Some notable differences in the size of the predictions for different stars do exist, but there is agreement on increasingly large negative corrections with decreasing metallicity. CH is often less affected by 3D effects than NH and OH, which in turn are less affected than CN and CO \citep[e.g.][]{2013A&A...559A.102D}.

It has also been recognised that the C/O ratio plays an important role in metal-poor 3D atmospheres, because of the atoms' preference to form CO \citep{2007A&A...469..687C,2016A&A...593A..48G}. When C and O abundances are optimized simultaneously or constrained by atomic lines, corrections between $-0.2$ to $-0.7$\,dex for the UV OH bands and $+0.05$\,dex to $-0.7$\,dex for the CH bands have been reported for several stars \citep{2010A&A...513A..72B,2015ApJ...806L..16B,2017A&A...600A..26S,2018MNRAS.475.3369C}. \citet{2016A&A...593A..48G} quantified the impact of the C/O ratio in dwarfs; for example, when increasing the ratio from 0.2 to 4, the predicted OH lines become weaker while the CH lines become stronger and the 3D correction to the C abundance becomes more negative by $-0.5$\,dex at $\feh=-3$. \citet{2017A&A...598L..10G} and \citet{2018IAUS..334..364S} further pointed out the importance of consistency between the abundances used to construct the 3D RHD model atmosphere and the abundances used to synthesize the spectrum. However, the effects are at the $0.1-0.3$\,dex level and are reported at high C/O-ratios that are not necessarily representative of ultra metal-poor stars.

The agreement between \ion{C}{I} and CH lines was investigated by \citet{2019ApJ...879...37N}, using 1D non-LTE corrections for \ion{C}{I} by \citet{2009A&A...500.1143F} and 3D LTE corrections to CH based on several literature studies. They find good consistency down to $\feh=-2.5$, but  that the atomic lines give $\sim0.5$\,dex lower abundances than the molecules for the extremely metal-poor TO stars G64-12 and G64-36. In Fig.\,\ref{fig:CEMP}, we show the 1D LTE CH-based abundances of \citet{2013ApJ...762...27Y} and compare them to 3D LTE abundances using new corrections based on \stagger{}-models and \scate{}. It can be seen that modelling assumptions indeed impact the classification of stars as C-enhanced, according to the traditional $\rm[C/Fe]>0.7$ limit, in agreement with the conclusion of \citet{2019ApJ...879...37N}. However, our corrected [C/Fe] values are generally higher. For G64-12 and G64-36, we find an average $\rm[C/Fe]=0.19$, which is only $0.05-0.1$\,dex higher than the values reported for atomic C lines by \citet{2019A&A...630A.104A}, after correcting for differences in stellar parameters. 

For O, the agreement between atomic and molecular abundance indicators is less satisfactory. For the metal-poor giant HD122563, studies report almost solar [O/Fe] values based on the UV OH lines, $0.3-0.5\rm\,dex$ below the abundances derived from the [\ion{O}{I}] 630\,nm line \citep{2017A&A...599A.128P,2018MNRAS.475.3369C}. 
For the metal-poor TO star HD84937, \citet{2017A&A...600A..26S} report reasonable agreement between 1D non-LTE abundances from the \ion{O}{I} triplet and 1D LTE abundances from UV OH lines; applying an estimated 3D effect of $-0.7$\,dex to the latter would again bring the [O/Fe] close to solar, in stark disagreement with the atomic lines. Missing continuous opacity in the the UV region may be part of the problem, or non-LTE effects over-predicting the strength of synthetic OH lines in LTE. \citet{2017A&A...600A..26S} speculate that the presence of a chromosphere, as implied by core emission in Mg H \& K, may raise the temperature of the 3D model and improve the agreement. The OH infrared lines are better aligned with the atomic indicators \citep{2015A&A...576A.128D}.

The N abundances of extremely metal-poor stars have attracted less attention. 3D LTE analyses of the UV NH and CN bands has been performed for individual stars \citep[e.g.][]{2008ApJ...684..588F,2013A&A...552A.107S}, but not applied to large samples. This is crucial, considering the importance to understand the need for a primary source of this element and thus constrain the yields of AGB, super-AGB, and rotating massive stars at low metallicity \citep{2018MNRAS.476.3432P,2020ApJ...900..179K}. In \citet{2005A&A...430..655S,2013A&A...552A.107S} a systematic offset of 0.4\,dex between NH and CN-based N abundances were reported, as well as a possible 3D effect exceeding $-2$\,dex for the CN band in TO stars, again largely dependent on the sensitivity of C/O. 

Investigations have been made to understand the robustness of the predicted large negative 3D corrections for molecules. First, \citet{2011A&A...529A.158H} concluded that accounting for continuum scattering in red giants has a relatively small impact on molecular line based abundances, in fact making them increasingly negative by up to 0.1\,dex at $\feh=-3$. Secondly, \citet{2023A&A...670A..25P} recently performed 1D non-LTE calculations for the CH molecule and find that metal-poor stars indeed suffer from over-dissociation by radiation, which can affect abundances determined from the G band by $+0.2$\,dex in ultra metal-poor giants. We caution that this result is likely to be sensitive to the collisional cross-sections with neutral hydrogen, which were computed with the Drawin formula in the lack of alternatives. However, if the sensitivity to departures from LTE is realistic, they may have an even greater impact in 3D. Non-LTE calculations for OH lines have so far only been tentatively explored with a two-level approximation in 1D \citep{2001A&A...372..601A}. Full 3D non-LTE calculations for CNO-bearing molecules, including a systematic investigation of the impact of uncertainties in molecular data, are needed to shed further light on the abundances of the most metal-poor stars. Thirdly, deviations from molecular equilibrium was investigated by \citet{2023A&A...675A.146D} and found insignificant, as discussed in \sect{sect:method-effects-nlte} 

Several lines of MgH are present around 510\,nm in cool stars. The lines are not normally used to infer Mg abundances, but the isotopic shifts between molecular lines involving $\rm^{24}Mg, ^{25}Mg$ and $\rm^{26}Mg$ are of order 0.01\,nm and thus detectable in high-resolution spectra, allowing determination of Mg isotopic ratios. The lightest isotope is produced primarily by SNe, while the two heavier signal contribution from AGB stars to the chemical enrichment. Mg isotopic ratios determined from MgH lines have therefore been used to constrain the chemical evolution of globular clusters and the Milky Way \citep[][and references therein]{2016A&A...588A..66T}. The error introduced in the inferred isotopic ratio when modelling the shapes of MgH lines in 1D was analysed by \citet{2017ApJ...843..144T} for metal-poor dwarfs and giants, with the result that $\rm^{25}Mg/^{24}Mg$ can be underestimated by up to five percent and $\rm^{26}Mg/^{24}Mg$ by up to two percent. This is small compared to the spread reported, for example, for the globular cluster $\omega$ Centauri \citep{2013ApJ...769....8D}. [Mg/H] abundances determined from MgH lines in 1D, however, suffer effects of similar order as the CNO-bearing molecules, i.e., 3D corrections can reach $-0.7$\,dex in metal-poor dwarfs. In a recent study of $\omega$ Centauri giants, \citet{2020MNRAS.495..383R} determine 1D LTE abundances from both MgH and atomic Mg lines and find that while for some stars they are in agreement, for others they differ by up to $+0.44$\,dex. Intriguingly, the author argues that 3D and non-LTE effects, as known from literature, are not responsible for this difference, but that it reflects on a true difference in the atmospheric He abundance. Increasing the He mass fraction of the atmospheric model, relatively fewer hydrogen atoms are available to form MgH, and the lines decrease in strength to agree with the atomic ones.  
 
\subsubsection{Iron-peak elements}
\label{sect:results-origin-fepeak}

The elements with atomic number $Z=21-30$ share many important properties both in terms of nucleosynthetic origin and line formation. The first ionisation potentials fall in the range $6.6-9.4$\,eV and all the neutral atoms, except Cu and Zn, have complex term diagrams with up to four multiplets and a large number of theoretically predicted highly excited levels.  This complexity makes full 3D non-LTE calculations
extremely challenging. Several recent 1D non-LTE studies including non-hydrogenic photoionisation cross-sections (except for Ni), together with asymptotic model data for inelastic hydrogen collisions, have been performed, e.g. for Ti \citep{2022A&A...668A.103M}, Mn \citep{2020A&A...635A..38E}, Ni \citep{2023A&A...677A.151E}, Cu \citep{2022ApJ...936....4X}, and Zn \citep{2022MNRAS.515.1510S}, revealing significant over-ionisation of the neutral atoms. The typical size of the effect at $\feh=-2$ is $+0.2$\,dex, which is similar to neutral Fe, slighty lower for Ni and Zn, slightly higher for Mn and Cu. Lines of both neutral and ionized species of Fe-peak elements are available and can in principle be used to verify the ionisation equilibrium, at least for a subset, however \ion{Cu}{II} and \ion{Zn}{II} require vacuum UV observations and \ion{Sc}{I} and \ion{V}{I} are not detectable in hot metal-poor dwarfs \citep{2020ApJ...890..119C,2020ApJ...900..106O}.

The only 3D non-LTE investigation for an Fe-peak element other than Fe, for stars other than the Sun, has been performed for Mn in metal-poor dwarfs and giants by \citet{2019A&A...631A..80B} and we therefore mainly limit further discussion to this element. Substantial positive 3D non-LTE corrections are reported, such that the difference with respect to 1D (\marcs{}) LTE is approximately $+0.4$\,dex for \ion{Mn}{I} lines and $+0.15$\,dex for \ion{Mn}{II} lines, for a very metal-poor dwarf and subgiant. The numbers are similar to those found for Fe for the same stars  \citep{2016MNRAS.463.1518A,2022A&A...668A..68A}, which reflects on robustness in the [Mn/Fe] ratio. However, for the metal-poor giant HD122563, the effect reported for \ion{Mn}{I} ($+1$\,dex) and for \ion{Mn}{II} ($+0.4$\,dex), are much higher than found for \ion{Fe}{I} ($+0.17$\,dex) and \ion{Fe}{II} ($+0.08$\,dex) in \citet{2016MNRAS.463.1518A}. This difference is not expected, given that the same 3D RHD model was used in \citet{2016MNRAS.463.1518A} and \citet{2019A&A...631A..80B}. We recommend further, independent 3D non-LTE studies  to understand if [Mn/Fe] is indeed sub-solar or super-solar in this star.

Fe-peak elements are produced in the explosive nucleosynthesis of thermonuclear and core-collapse SNe, with relative yields that depend strongly on the nature of the explosion \citep[e.g.][]{2013ARA&A..51..457N}. \citet{2020A&A...635A..38E} demonstrate a close to flat [Mn/Fe] trend over several dex in metallicity, in agreement with the 3D non-LTE abundances for Mn by  \citet{2019A&A...631A..80B}, can be achieved if hypernovae are neglected and a high fraction of SNIa explode with sub-Chandrasekhar mass. However, we note that Fe was not consistently corrected for non-LTE for the lowest metallicity stars from \citet{2009A&A...501..519B} which were included in the \citet{2020A&A...635A..38E} study, and their [Mn/Fe] ratios were therefore inflated. At $\feh=-2$, [Mn/Fe] is reported at $\sim -0.1$\,dex, which is approximately 0.2\,dex higher than the 1D non-LTE [Mn/Fe] ratios found by other studies with the same Mn corrections \citep{2020A&A...642A..62A,2020A&A...643A..49H,2023arXiv231207768N}. These latter results are in turn 0.2\,dex higher than the \citet{2020ApJ...900..179K} GCE model including hypernovae and only near-Chandrasekhar mass SNIa. Additional observational evidence to support the sub-Chandrasekhar mass channel is provided by mildly super-solar non-LTE [Ni/Fe]-ratios \citep{2023A&A...677A.151E}. The suggested predominance of the sub-Chandrasekhar mass channel should be further explored with consistent 3D non-LTE calculations for Mn, Fe, and Ni.

\subsubsection{Barium}
\label{sect:results-origin-barium}
Ba is among the most well studied of the neutron-capture elements, largely thanks to the availability of strong optical absorption lines of \ion{Ba}{II}. According to our current understanding of Galactic chemical evolution, the element has multiple formation sites; at the lowest metallicities, contributions from both a rapid (r-process) channel in core-collapse SNe and/or compact star mergers as well as a slow (s-process) channel in rotating massive stars are necessary to raise the model predictions to observed levels \citep{2018MNRAS.476.3432P}. The s-process nucleosynthesis in low and intermediate mass AGB stars becomes noticeable at $\feh>-1.5$, which combined with the simultaneous Fe enrichment of SNIa produces a close to solar [Ba/Fe] ratio.

Non-LTE effects in 1D models have recently been investigated for \ion{Ba}{II} lines by several studies \citep{2019AstL...45..341M,2020A&A...643A..49H,2023MNRAS.524.3526M}, 
making use of asymptotic model data for inelastic hydrogen collisions as computed by \citet{2017A&A...608A..33B,2018MNRAS.478.3952B}. 
At extremely low metallicity, over-ionisation and over-excitation \citep[see e.g.][]{2009A&A...494.1083A} produce positive abundance corrections for \ion{Ba}{II} 455.4\,nm, that can exceed $+0.2$\,dex. At increasing metallicity and line strength, photon losses lead to more negative corrections down to $-0.2$\,dex for the resonance line and even more negative for subordinate lines \citep{2020A&A...643A..49H,2023MNRAS.524.3526M}. As shown by \citet{2020A&A...643A..49H}, the [Ba/Fe] trend in the range $-3<\feh<-1$ is shifted downwards by approximately 0.3\,dex and flattened when both elements are consistently treated in 1D non-LTE. This improves the agreement with the GCE model by \citet{2018MNRAS.476.3432P}.      

3D LTE abundance analyses of \ion{Ba}{II} lines point to corrections of up to $+0.18$\,dex for subordinate lines in metal-poor red giants \citep{2012A&A...540A.128D}, while $-0.15$\,dex was found for the resonance line in the metal-poor subgiant HD140283 \citep{2015A&A...579A..94G}. \citet{2012A&A...540A.128D} and \citet{2009BaltA..18..193M} both emphasize the strong sensitivity of the 1D abundance to the choice of the fudge parameter microturbulence.
A consistent 3D non-LTE treatment is indeed particularly important for saturated lines, but such analysis has so far only been performed for the Sun by \citet{2020A&A...634A..55G}. They demonstrate a cancellation between substantial positive 3D effects and negative non-LTE effects such that their combined impact on the Ba abundance is small for the Sun, 0.03\,dex higher than 1D LTE. This cannot, however, be extrapolated to metal-poor stars. 

In addition to the absolute abundances of Ba and other neutron capture elements, important clues to the nucleosynthesis channels involved in their creation comes from the isotopic ratios. Specifically, the ratio between odd and even Ba isotopes is different for the s-process and r-process, and this ratio can be determined from a careful profile analysis of the \ion{Ba}{II} resonance line, because the odd isotopes show significant hyper-fine splitting. The larger the contribution from the s-process, the broader the line. \citet{2019AstL...45..341M} enforced agreement in the Ba abundance from the resonance and subordinate lines in 1D non-LTE, and thereby found a clearly dominant r-process origin of the element in four metal-poor stars. In constrast, \citet{2009PASA...26..330C} used Fe lines to constrain broadening due to micro/macroturbulence (in the 1D case) and rotation and found that a 3D LTE analysis of HD140283 reduced the ratio between odd and even isotopes compared to a 1D analysis, thus favouring an s-process origin of Ba. This result was later challenged by \citet{2015A&A...579A..94G}, who found the opposite behaviour with a 3D LTE analysis, supporting a predominant r-process origin in this star. The authors speculate that the former result was caused by problems with the line selection or with the 3D RHD model used previously, e.g., erroneous temperature, insufficient resolution or opacity bins. Clearly, this type of application is the ultimate challenge for stellar atmosphere modelling and synthetic line formation; a 3D non-LTE profile analysis of the Ba isotopic ratio would certainly shed further light on the matter. 

\subsection{Planet-host stars}
\label{sect:results-planets}

We will now briefly comment on the implications that improved models of stellar spectra may have for our understanding of exoplanets and their host stars. We focus on spectral line formation in this review, but note that several studies have also been devoted to using 3D RHD models to characterize intrinsic stellar brightness fluctuations that may interfere with planetary detection \citep[e.g.][]{2022MNRAS.514.1741R}.  

\subsubsection{Chemical compositions of planet hosts}
\label{sect:results-planets-abund}
Thanks to the ever increasing number of exoplanet discoveries, numerous investigations have been made between stellar properties, such as chemical abundances, and planet occurrence and properties. First, the now well-established correlation between stellar metallicity and occurrence rate of giant planets \citep[e.g.][]{2019Geosc...9..105A}, has been extended to consider additional aspects, e.g. planetary mass, radius, orbital properties, heavy-metal content, and stellar alpha-element abundance \citep[e.g.][and references therein]{2022A&A...664A.161B}. Second, significant trends have been found in relative abundances with respect to condensation temperature between the Sun and carbonaceous chondrites on the one hand, and the Sun and solar twins on the other hand \citep{2021A&A...653A.141A,2018A&ARv..26....6N}. Several proposed explanations have been put forth, involving formation of both terrestrial and giant planets, but the observational landscape is complex, with the intertwined effects of age, chemical evolution, and radial migration clouding the picture. Third, planet properties have been linked to specific element abundances; Li, because of its unique sensitivity to stellar rotation and mixing 
\citep{2015A&A...576A..69D,2019MNRAS.485.4052C}, and C/O and Mg/Si because of their decisive influence on the planetary mineralogy \citep{2016ApJ...831...20B,2018A&A...614A..84S}.

In order to disentangle different intrinsic sources of abundance differences, the analysis for such applications as those mentioned above must in general be very precise, with systematic errors not exceeding $0.05$\,dex. However, exoplanet host stars are found all over the HR diagram, populating the main sequence from M to B dwarfs, and extending to the subgiant and red giant branches for late-type stars \citep[e.g.][]{2018ApJ...866...99B,2021A&A...646A.164J}. While systematic effects may largely cancel in differential abundance analyses of twin stars, this is obviously further from the truth the more extended the parameter space. The lowest metallicity stars known to host planets lie arguably somewhere around $\feh\approx-1.0$ \citep{2019Geosc...9..105A}. In this range, 3D non-LTE $\feh$ may differ from 1D LTE by up to $0.2$\,dex for FG dwarfs \citep{2022A&A...668A..68A} --- even more if 1D LTE spectroscopic stellar parameters are used
(\sect{sect:results-parameters-fe}) --- and must be considered, for example, when stellar abundances are compared to model predictions of the critical metallicity for planet formation \citep{2012ApJ...751...81J}.

The C/O ratio of the planet-forming environment determines whether Si forms carbide or oxide species, with a critical limit at $\rm C/O=0.8$. As discussed by \citet{2018A&ARv..26....6N}, several high-precision studies have shown that the C/O ratio of solar-type stars stays below this value. \citet{2019A&A...630A.104A} confirmed this and also showed that 3D non-LTE effects, applied differentially with respect to the Sun, decreases the C/O ratios for sub-solar metallicity stars, but increases them for super-solar metallicity. Interestingly, the 3D non-LTE abundances showed that confirmed planet-host stars have systematically higher C/O ratios than the comparison sample for $\rm[O/H]>0$, which speculatively links the C/O ratio to planet formation efficiency.      

The Mg/Si ratio governs the distribution of silicates in the planets. The solar ratio is very close to one; the value in CI-chondrites is $1.03$ \citep{2021SSRv..217...44L}, but whether it is larger or smaller than one is not definitely settled in the solar photosphere (see Sect.\,\ref{sect:appendix-sun-mgsis}). A correlation between Mg/Si ratio of planet-hosts and the planetary mass has been reported, claiming that the ratio is higher in stars with low-mass planets \citep{2015A&A...581L...2A,2022A&A...664A.161B}, a result that is not strongly influenced by 1D non-LTE calculations \citep{2017Ap.....60..325A}. However, for both elements the sensitivity to the model atmosphere can exceed the non-LTE effect for solar-type stars \citep{2015A&A...573A..25S} and we recommend that the Mg/Si of planet-hosts be reappraised with 3D non-LTE calculations. 

\subsubsection{Transmission spectroscopy}
\label{sect:results-planets-transmission}

\begin{figure*}
    \includegraphics[width=5in]{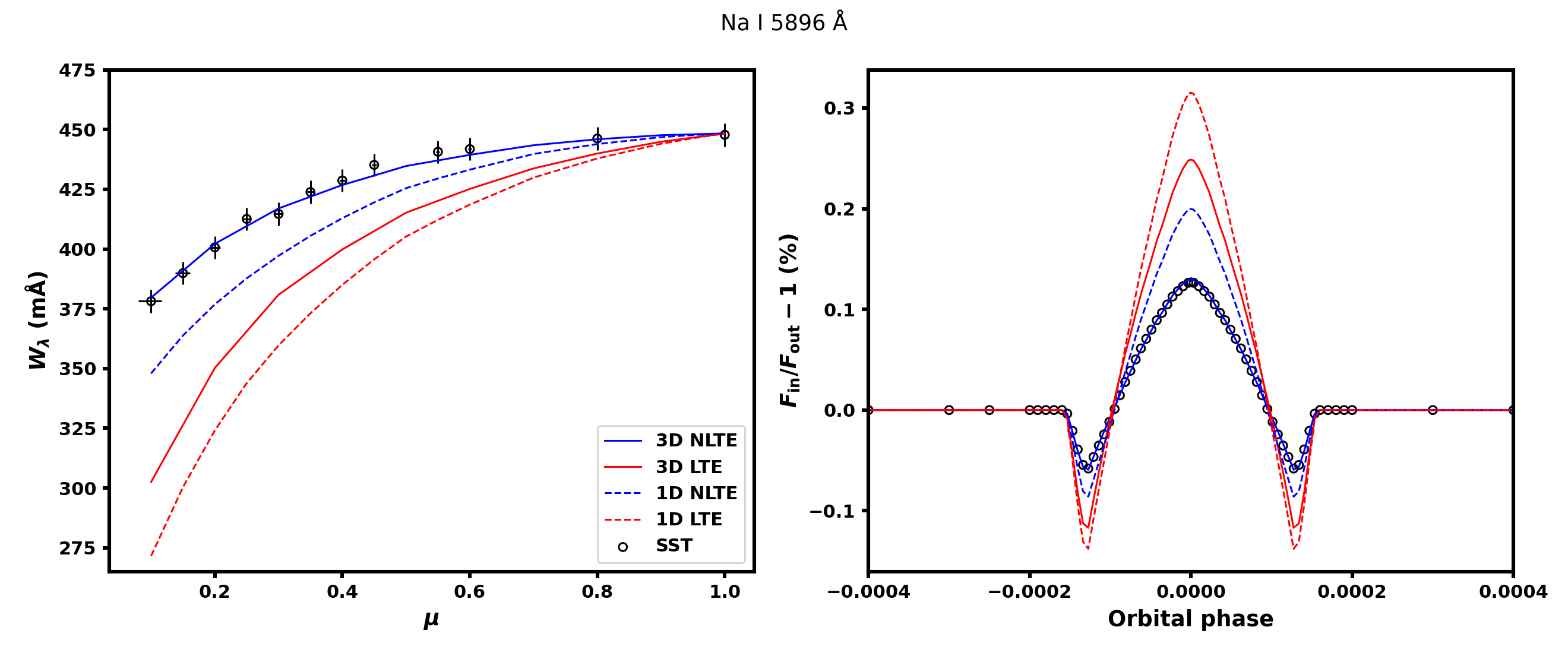}
    \caption{Observed and modelled CLV of the equivalent width of the \ion{Na}{I} 589.6\,nm line (left). All model curves have been normalised to their best-fit abundance at disc centre. The right-hand panel shows simulated transmission curves from a Jupiter-Sun edge-on system. In both panels, the 3D non-LTE model provides an excellent match to the observed data from the Swedish Solar Telescope, SST.  From \citet{2023arXiv231205078C}.}
\label{fig:NaD}
\end{figure*}

When a planet is in transit in front of its host, the light from the star is partly obscured, but also partly filtered through the planetary atmosphere. Thereby, transmission spectroscopy can help characterize the physical conditions and chemical composition at the surfaces of exoplanets. However, to accurately extract the planetary signal, the CLV of the stellar disc must be considered. Lines that are present in both the planetary and the stellar atmosphere are particularly complicated if they are blended, and erroneous planetary signal can be inferred if the stellar CLV is not accounted for \citep{2020A&A...635A.206C}. Naturally, 1D LTE model predictions of the CLV can also be expected to lead to errors, since it is well known that they cannot reproduce spatially resolved solar observation (\sect{sect:method-validation}).  
\citet{2017A&A...603A..73Y} showed that transmission curves for a solar-type star are moderately flattened by 1D non-LTE calculations for the Na D lines, a result that was later confirmed for more species and also extended to 3D LTE by \citet{2023A&A...673A..71R}. The size of the effect varies with the observational configuration of the star-planet system, e.g., the impact parameter and passband, but evidently, the difference between transmission curves modelled under different assumptions can be as large as the CLV effect itself. \citet{2023arXiv231205078C} illustrate that 3D non-LTE modelling produces the best match to the CLV of the resonance lines of Na and K in the Sun. The best-fit line profile and simulated transmission curves for a Jupiter-Sun system are shown in \fig{fig:NaD}. The calculations can readily be extended to other FGK-type stars and planetary configurations.

\section{DISCUSSION AND OUTLOOK}
\label{sect:discussion}

\begin{figure}
\includegraphics[width=5in]{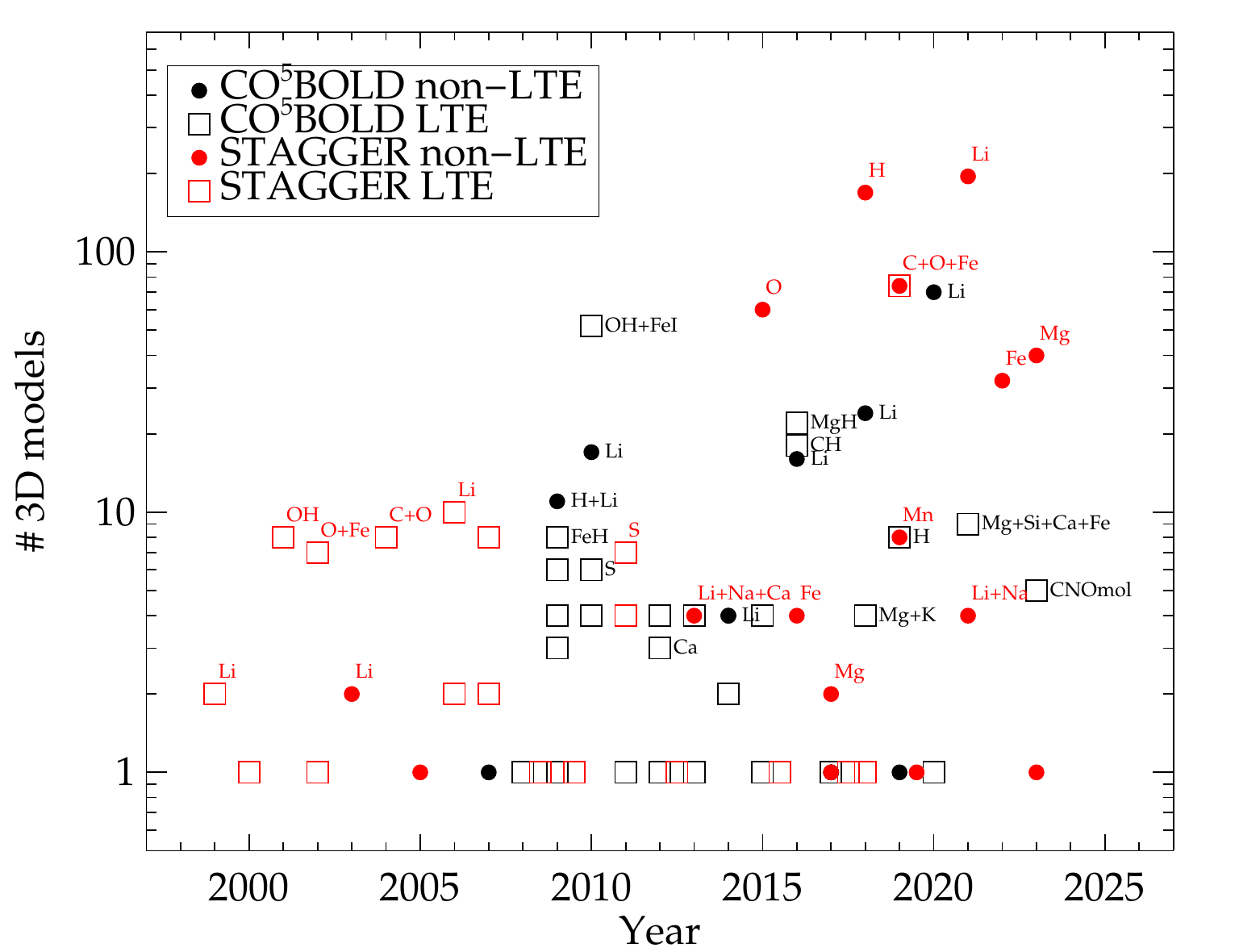}
\caption{Overview of refereed publications of 3D (non-)LTE hydrogen line formation and abundance analyses for stars other than the Sun. The y-axis represents the number of 3D RHD models used (corresponding to different values of $\teff$, $\lgg$, and $\feh$). Species names are indicated for selected studies only. The references are listed in \tab{table:Lit}.}
\label{fig:modelsyear}
\end{figure}

This century has seen tremendous improvement of model atmospheres and synthetic spectra for late-type stars. Grids of 3D RHD models have been computed and successfully validated against solar and stellar observations, for example in terms of abundance determination, CLV, spectral line shapes, and statistical properties of granulation. Applying non-LTE radiative transfer in the post-processing of snapshots of 3D RHD models is numerically expensive, but is becoming a feasible alternative for large-scale parameter and abundance analysis. \fig{fig:modelsyear} shows an overview of 3D (non-)LTE abundance analyses and hydrogen-line analyses for late-type stars other than the Sun over the past 25 years. As evident from the figure, the number of 3D RHD models used for grid calculations and quantitative spectroscopy has grown substantially, starting from a handful of models to now encompassing the full \stagger{} and \cifist{}-grids shown in Fig.\,\ref{fig:stagger-cifist-grids}. Furthermore, we see a simultaneous shift from predominantly 3D LTE calculations to 3D non-LTE. The publications used to create this figure are listed in Table \ref{table:Lit}.

\subsection{The transition from 1D LTE to 3D non-LTE}
The gradual transitioning from the use of 1D LTE models to 3D non-LTE models may look different for different types of applications. Before we proceed with some general advice in this regard, we remind the reader that quantitative stellar spectroscopy includes a variety of sources of error such as continuum placement, blends, and atomic data (oscillator strengths), that are not directly related to the 3D non-LTE models themselves (as we have discussed, for example, in the context of the solar composition; \sect{sect:results-sun}).  With the development of more realistic model atmospheres and synthetic spectra, a careful handle of these errors become all the more important.

Traditional (1D LTE) spectroscopic analysis is build on careful selection and fitting of individual lines. Line-by-line differential analyses can achieve remarkable precision, as demonstrated e.g. for solar twins and analogues with respect to the Sun 
\citep[e.g.][]{2015A&A...579A..52N,2016MNRAS.457.3934L,2018ApJ...865...68B}, but also metal-poor stars with respect to benchmark stars \citep{2016A&A...594A..43H,2017A&A...608A..46R}. The smaller the parameter space covered by a given study, the smaller the differential impact of 3D non-LTE calculations, although this depends on the science goal as well as the lines being analysed. \citet{2018A&ARv..26....6N} found that, as a rule of thumb, the forbidden O lines are reliable in a differential sense also with simplistic modelling if one limits the parameter space to 500\,K in $\Delta\teff$ and 0.5\,dex in $\Delta\lgg$ for solar-type stars. On the other hand, this is not true for an element like Li, which can have substantially different line strengths in stars with the same fundamental parameters, and for saturated lines like the \ion{O}{I} $777\,\mathrm{nm}$ triplet \citep{2017A&A...597A..34M}. 

To improve the accuracy of 1D LTE abundances, pre-computed 3D non-LTE
abundance corrections can be applied.  This is now possible for several species, on relatively extended grids of at least $40$ combinations of $\teff$, $\lgg$, and $\feh$: 
\ion{Li}{I}, \ion{C}{I}, \ion{O}{I}, \ion{Mg}{I}, \ion{Fe}{I}, and \ion{Fe}{II}. The user should be careful that the microturbulence value 
used in the 1D LTE analysis is the same as that used to calculate the 3D versus 1D abundance correction, to properly translate the 1D LTE abundance (microturbulence-dependent) to a 3D non-LTE one (microturbulence-independent).
Directly fitting pre-computed 3D non-LTE spectra to observations for individual lines is
also an option, and can include masking or empirical fitting of blends \citep[e.g.][]{2023A&A...679A.110G,Wang_li}.
For species where non-LTE effects are small (e.g. \ion{Fe}{II}), 
3D LTE abundance corrections and spectra can be used; such models
are much cheaper to calculate and do not require model atoms.

Similarly, one may also consider using 1D or $\mtd{}$ non-LTE abundance corrections
or spectra to improve the accuracy of 1D LTE abundances.
However, the 3D granulation effect, missing in both 1D and $\mtd{}$ models,
is large for temperature sensitive lines: namely, those of low excitation potential for minority species, and of high excitation potential for majority species; but cancellation effects due to the simultaneous change in mean atmospheric structure can occur. Using $\mtd{}$ models may therefore improve abundances for temperature-insensitive lines, but they capture only a small part of the full 3D effect for other lines.  The situation is worse if the lines are saturated and are thus highly sensitive to the velocity fields, crudely approximated in 1D and $\mtd{}$ models by microturbulence and macroturbulence. 
Although the use of non-LTE calculations in 1D and $\mtd{}$ models can significantly improve the ionisation balance with respect to 1D LTE, the absolute abundances can still be offset from 3D non-LTE due to the extra calibration parameters,
and such models may also perform poorly for the excitation balance and for hydrogen line profiles (Sect.\,\ref{sect:results-parameters}). 
The use of 3D LTE abundances combined with 1D or $\mtd{}$ non-LTE corrections, in the lack of full 3D non-LTE calculations, may be justified for metal-rich stars like the Sun, whose atmospheric structures are in radiative equilibrium, but can introduce large errors for metal-poor stars (\sect{sect:method-effects-nlte}). 


Unfortunately, 3D non-LTE calculations are expensive, and
the amount of available 3D non-LTE data are far from sufficient for
the demands of current large spectroscopic surveys. Significant computational gain may be accomplished by calculating non-LTE populations or departure coefficients with
(open source) 1.5D non-LTE codes (\sect{sect:method-spectra-codes}) and reading them into a 3D LTE codes such as \linfortd{}, \scate{}, \optimtd{}, or \asset{} so as to
obtain a 1.5D non-LTE disc-integrated spectrum which in some cases may satisfactorily
approximate the full 3D non-LTE result, given all the other uncertainties
\citep{2016MNRAS.455.3735A,2017A&A...597A...6N}. 
Another approach worth exploring further is to use machine learning to predict non-LTE populations from their 3D LTE counterparts \citep{2022A&A...658A.182C}. 

Spectroscopic surveys typically require high resolution
spectrum synthesis including all detectable lines over a large wavelength interval.
In the near future these surveys
will benefit from grids of hybrid 1D LTE/non-LTE spectra generated using pre-computed grids of departure coefficients. The third data release of the GALAH survey treats 13 elements in non-LTE \citep{2020A&A...642A..62A}, a number that will grow further before 4MOST \citep{2019Msngr.175....3D} sees first light. To mitigate errors in stellar parameters due to the use of 1D models for spectrum generation, photometric, astrometric and asteroseismic constraints can be folded into the analysis \citep[e.g.][]{2021MNRAS.506..150B,2022A&A...658A.147G}, in particular if colours and bolometric corrections use scaling relations based on 3D work (\sect{sect:results-parameters-photometry}). It is a safe expectation that analogous hybrid 3D LTE/non-LTE spectra will become available in the future, but to truly give superior performance, they may need to treat essentially all minority species with significant absorption in non-LTE, given that departures from LTE are generally amplified. Grids of high-resolution 3D LTE spectra with large wavelength coverage are already available \citep{2018A&A...611A..11C,2022A&A...661A..76B}.

\subsection{Future methodological developments}

To facilitate 3D non-LTE abundance analyses, it seems necessary to extend the existing grids of model atmospheres. For instance,
the current extent of the $\stagger{}$ and $\cifist{}$ grids would benefit from more complete coverage of the very metal-poor red giant branch. However, warm, low-surface gravity models have the largest run-times as discussed in Sect.\,\ref{sect:method-atmospheres-cost}, and are the most challenging to relax. The development of better techniques to initialize new models with appropriate bottom boundary conditions, perhaps involving elements of machine learning, would be welcome. This can be done by using existing boundary conditions (energy and density values) and find lines of similar entropy, taking the variation in depth and metallicity of the models into account. It is also desirable to refine the grids.
The nodes of the existing 3D \stagger{} grid are separated by up to $500$\,K in $\teff$, 0.5\,dex in $\lgg$ and $0.5-1$\,dex in $\feh$ (\fig{fig:stagger-cifist-grids}); the step-sizes in $\teff$ and $\feh$ are typically twice as large as the commonly used 1D \marcs{} grid \citep{2008A&A...486..951G}. Nevertheless, accurate interpolation of 3D non-LTE stellar spectra generated with the grid has been shown to be possible, e.g., using a method based on Gaussian process regression \citep{2021MNRAS.500.2159W}. With interpolation in four dimensions, i.e., the three stellar parameters and Li abundance, interpolation errors that correspond to 0.01\,dex in abundance can be achieved. \citet{2022A&A...661A..76B} uses an interpolation technique based on radial basis functions for the high-resolution 3D LTE full optical spectra computed for the \cifist{}-grid and reports errors of $<2\%$ in flux. These results are promising, but larger errors are to be expected for a higher-dimensional abundance space and reducing the step size of the 3D grids would certainly be beneficial for applications that require high precision.

Improvements to the physics in the 3D RHD simulations may also be important,
particularly for metal-poor stars where 3D non-LTE effects are often largest.
Improvements to the opacity binning approach (\sect{sect:method-atmospheres-micro}) could cause changes to the temperature stratifications, which are predicted to be several hundred Kelvin
cooler in 3D than in 1D for metal-poor stars.  Such changes would have
potentially large effects on temperature sensitive diagnostics 
(e.g. C abundances from CH lines, \sect{sect:results-molecules}).
Another future step is to investigate the validity of the restricted non-LTE problem by propagating the non-LTE level populations into the binned opacities.
A more immediate consideration is the input abundances of the simulations \citep{2017A&A...598L..10G,2023A&A...677A..98Z}.  The most metal-poor stars
have elemental abundance ratios that are peculiar (compared to the Sun), and it 
may be necessary to relax the trace-element assumption in order to obtain the most accurate
results.
The importance of consistency between the chemical composition used to construct the model atmosphere and the synthetic spectrum was tested for a solar-metallicity red giant by \citet[][see Appendix A]{2016ApJ...826...83T}, concluding that a significant impact, increasing toward cooler temperatures, is seen in particular for what they call primary elements (C, N, Mg, Si, Fe), but also non-negligible impact for secondary elements (O, Na, Al, Ca, Ti). A specific example of how inconsistent abundances can lead to false abundance spreads in metal-poor globular clusters was shown for the \ion{Ca}{II} triplet, whose line strengths are sensitive to the abundance of important electron donors like Mg \citep{2012MNRAS.426.2889M}. However, these studies are based on 1D atmospheres, while studies in 3D are so far limited, e.g., to the models with C/O enhanced above unity mentioned in \sect{sect:results-molecules}.

To our knowledge, all 3D non-LTE abundance results so far have considered one element at a time.  The validity of the single-element approach was investigated by \citet{2020A&A...637A..80O} in 1D non-LTE for the elements Na, Mg, K, and Ca using \tlusty{} \citep{2017arXiv170601859H}.  They found that departures from LTE for Mg affect those of Ca due to changes to the ultra-violet (UV) opacity, amounting to differences in Ca abundance of $-0.07\,\mathrm{dex}$ for the Sun, but they do not find significant effects for the other elements. However, this result is sensitive to the amount of other UV opacity included in the models, and was not confirmed by \citet{2021A&A...653A.141A}, and furthermore appears to be insignificant in metal-poor stars \citep{2022ApJ...928..173O}. Nevertheless, future work should aim to relax the single-element approximation, and ultimately move beyond the restricted non-LTE problem; this is already feasible with 1D models and realistic model atoms. 

Finally, the non-LTE solutions are only as reliable as the model atoms that they are drawn
upon (\sect{sect:method-atoms}), which in turn are reliant on accurate and complete
sets of atomic (and molecular) data --- non-LTE calculations with unrealistic model atoms may give results of lesser overall accuracy than LTE.  The fundamental theoretical and experimental 
work carried out by atomic (astro-)physicists should not be overlooked or taken for granted. In particular, non-hydrogenic photo-ionisation cross-sections are missing for several Fe-peak and neutron capture elements. 
On top of that, the physics of excitation by hydrogen collisions is still 
not satisfactorily understood, with significant impact on the 
solar O abundance for example (Sect.\,\ref{sect:appendix-sun-o}).
Ongoing theoretical and experimental investigations may help in this regard
\citep{2021ApJ...908..245B,2022PhRvL.128c3401G}.

\subsection{Concluding remarks}

In this review, we have discussed how 3D non-LTE abundances have already shed new light on a number of astrophysical problems (\sect{sect:results}), specifically:
\begin{enumerate}
\item They contributed to the lowering of the solar O abundance and helped reveal the solar modelling problem (Sect.\,\ref{sect:results-sun})
\item They have mitigated or removed large systematic uncertainties in spectroscopic stellar parameter determination (Sect.\,\ref{sect:results-parameters})
\item They have removed the need for a cosmological origin of $^6$Li and helped us better define and interpret the Spite plataeu in the context of Big Bang nucleosynthesis and stellar depletion (Sect.\,\ref{sect:results-origin-li})
\item They have removed the need for C-production in excess of O in the early universe (Sect.\,\ref{sect:results-origin-calpha})
\item They have helped calibrate and validate core-collapse SNe and HNe yields of $\alpha$-elements (Sect.\,\ref{sect:results-origin-calpha})
\item They have uncovered previously unseen substructure in the [Mg/Fe]$-\feh$ plane in low-$\alpha$ halo (Sect.\,\ref{sect:results-origin-calpha})
\item They have highlighted he importance of the sub-Chandrasekhar mass channel of SNIa (Sect.\,\ref{sect:results-origin-fepeak})
\item They have helped us trace the r-process origin of Ba in very metal-poor stars (Sect.\,\ref{sect:results-origin-barium})
\item They have helped uncover a relationship between stellar C/O-ratios and likelihood to host Jupiter-mass exo-planets (Sect.\,\ref{sect:results-planets-abund})
\item By their ability to accurately predict the CLV of spectral lines, they are fundamental for the characterization of exo-planet atmospheres through transmission spectroscopy (Sect.\,\ref{sect:results-planets-transmission})
\end{enumerate}
We have every reason to believe that this list will continue to grow in the years to come, but the rate of progress will reflect on the investment of researchers' time and resources.

\section*{DISCLOSURE STATEMENT}
The authors are not aware of any affiliations, memberships, funding, or financial holdings that might be perceived as affecting the objectivity of this review. 

\section*{ACKNOWLEDGMENTS}
We are grateful to several colleagues for valuable contributions to this work in the form of discussions, detailed comments and figures, in particular Paul Barklem, Gloria Canocchi, Bengt Edvardsson, Nicolas Grevesse, Cis Lagae, Jonas Klevas, Hans Ludwig, Poul-Erik Nissen, Luisa Fernanda {Rodr{\'\i}guez D{\'\i}az}, Matthias Steffen, and Ella Xi Wang.
KL acknowledge funds from the Knut and Alice Wallenberg foundation, and the European Research Council (ERC) under the European Union’s Horizon 2020 research and innovation programme (Grant agreement No. 852977). AMA acknowledges support from the Swedish Research Council (VR 2020-03940).

%




\bibliographystyle{ar-style2}
\bibliography{AAR_review}

\newpage
\begin{appendix}

\begin{longtable}{@{}p{1.35in}|p{1.1in}|p{1.1in}|l@{}}
\caption{\label{table:Lit} Refereed 3D LTE and NLTE studies of metal abundances and/or hydrogen line profiles for stars other than the Sun.\\}\\
\hline\hline 
Reference & Species LTE$\rm^{a}$ & Species NLTE$\rm^{b}$ & Stars/models$\rm^{c}$  \\
\hline
\endfirsthead
\caption{continued.}\\
\hline\hline
Reference & Species LTE$\rm^{a}$ & Species NLTE$\rm^{b}$ & Stars/models$\rm^{c}$  \\
\hline
\endhead
\hline
\endfoot

{\citet{1999A&A...346L..17A,2000A&A...357L..49N}} & Li\,I & & \mbox{HD84937, HD140283}  \\ \hline
{\citet{2000A&A...364L..42P}} & Be\,II & &  \mbox{G64-12}  \\ \hline
{\citet{2001A&A...372..601A}} & OH  & &  \mbox{\#8 D $\rm-3<[Fe/H]<0$}  \\ \hline
{\citet{2002ApJ...567..544A}} & Fe\,I, Fe\,II & &  \mbox{Procyon}  \\ \hline
{\citet{2002A&A...390..235N}} & O\,I, Fe\,II & &  \mbox{\#7 D/SG $\rm-2.5<[Fe/H]<0$}  \\ \hline
{\citet{2003A&A...399L..31A,2003A&A...409L...1B}} & & Li\,I & \mbox{HD84937, HD140283}  \\ \hline
{\citet{2004A&A...414..931A}} & C\,I, O\,I  & &  \mbox{\#8 D $\rm-3<[Fe/H]<0$}  \\ \hline
{\citet{2004A&A...415..993N}} & S\,I, Fe\,II, Zn\,I  & &  \mbox{\#8 D $\rm-3<[Fe/H]<0$}  \\ \hline
{\citet{2005ApJ...618..939S}} &  & O\,I, Fe\,I, Fe\,II &  \mbox{HD140283}  \\ \hline
{\citet{2006ApJ...644L.121C,2006ApJ...638L..17F}} & 15 neutral, 7 singly ionized, 5 molecules & & \mbox{HE 0107-5240, HE 1327-2326}  \\ \hline
{\citet{2006ApJ...644..229A}} & Li\,I & & \mbox{\#10 D/SG $\rm-3<[Fe/H]<0$}  \\ \hline
{\citet{2007A&A...473L..37C}} & & Li\,I & HD74000 \\ \hline
{\citet{2007A&A...469..687C}} & Li\,I, O\,I, Na\,I, Mg\,I, Ca\,I, Fe\,I, Fe\,II, CH, NH, OH & & \mbox{\#8 G $\rm-3<[Fe/H]<0$} \\ \hline
{\citet{2007ApJ...671..402K}} & Mg\,I, Ti\,II, Fe\,I, Fe\,II &  & \mbox{\#2 D/G $\rm[Fe/H]=-2$}  \\ \hline
{\citet{2008A&A...480..233G}} & Li\,I, OH &  & CS 22876-032  \\ \hline
{\citet{2008ApJ...684..588F}} & 12 neutral, 8 singly ionized, 3 molecules &  &  HE 1327-2326 \\ \hline
{\citet{2009BaltA..18..193M}} & Na\,I, Mg\,I, Ba\,II &  & \mbox{\#1 G $\rm[Fe/H]=-1$}  \\ \hline
{\citet{2009A&A...501..519B}} & Si\,I, Sc\,II, Ti\,II, Cr\,I, Mn\,I, Fe\,I, Co\,I, Zn\,I, CH &  & \mbox{\#3 D/G $\rm[Fe/H]=-3$}  \\ \hline
{\citet{2009A&A...501.1087R}} & Fe\,I, Fe\,II &  & \mbox{\#1 D $\rm[Fe/H]=0$}  \\ \hline
{\citet{2009A&A...503..121S}} & S\,I & & \mbox{\#4 D/SG $\rm-1<[Fe/H]<0$} \\ \hline
{\citet{2009PASA...26..330C}} & Ba\,II &  & \mbox{HD140283}  \\ \hline
{\citet{2009A&A...505L..13G}} & & H\,I, Li\,I & \mbox{\#11 D/SG $\rm[Fe/H]=-2$} \\ \hline
{\citet{2009A&A...502L...1L}} & H\,I &  & \mbox{\#6 D/SG $\rm-3<[Fe/H]<0$}  \\ \hline 
{\citet{2009A&A...508.1429W}} & FeH &  & \mbox{\#8 D $\rm[Fe/H]=0$}  \\ \hline
{\citet{2010A&A...513A..72B}} & C\,I, O\,I, CH, NH, OH, $\rm C_2$ &  & \mbox{\#4 D} \mbox{$\rm-3<[Fe/H]<-2$}  \\ \hline 
{\citet{2010A&A...519A..46G}} & Fe\,I, OH &  & \mbox{\#52 D/SG $\rm-3<[Fe/H]<0$}  \\ \hline
{\citet{2010A&A...522A..26S}} &  & Li\,I & \mbox{\#17 D/SG $\rm-3<[Fe/H]<-2$}  \\  \hline
{\citet{2010A&A...524A..96B}} & Cu\,I & & \mbox{\#6 D/G $\rm-2<[Fe/H]<-1$}  \\  \hline
{\citet{2010AN....331..725C,2011A&A...528A...9S}} & S\,I & & \mbox{\#6 D/SG $\rm-3<[Fe/H]<-1$} \\ \hline
{\citet{2011A&A...529A.158H}} & Fe\,I, Fe\,II, CH, NH, OH & & \mbox{\#4 G $\rm-3<[Fe/H]<0$}  \\  \hline
{\citet{2011A&A...530A.144J}} & S\,I & & \mbox{\#7 G $\rm-3<[Fe/H]<-1$} \\ \hline
{\citet{2011A&A...534A...4C}} & Mg\,I, Ca\,I, Ca\,II, Fe\,I & & \mbox{\#1 D $\rm[Fe/H]=-3$} \\ \hline 
{\citet{2011Natur.477...67C,2012A&A...542A..51C}}$\rm^{c}$ & Mg\,I, Ca\,I, Ca\,II, Si\,I, Ti\,II, Fe\,I, Ni\,I, Sr\,II, CH, NH &  & \mbox{SDSS J102915.14+172927.9}  \\ \hline
{\citet{2012A&A...540A.128D}} & Ba\,II &  & \mbox{\#4 G $\rm-3<[Fe/H]<0$}  \\ \hline
{\citet{2012A&A...541A.143S}} & Ca\,I & & \mbox{\#3 D/G $\rm[Fe/H]=-3$} \\ \hline
{\citet{2012ApJ...753..150N}} & CH, OH &  & \mbox{\#1 G $\rm[Fe/H]=-3$}  \\ \hline
{\citet{2012A&A...544A.102B}} & S\,I &  & HE 1327-2326 \\ \hline
{\citet{2013A&A...549A..14K}} & 17 neutral, 8 singly ionized, 6 molecules  & & \mbox{\#1 G $\rm[Fe/H]=0$}\\ \hline
{\citet{2013A&A...550A.122S}} & Pb\,I  & & \mbox{\#1 G $\rm[Fe/H]=-3$}\\ \hline
{\citet{2013ApJ...765...51L}} & O\,I, HF & & \mbox{\#4 G $\rm[Fe/H]=-2$}\\ \hline
{\citet{2013A&A...552A.107S}} & CN, CH  & & \mbox{\#1 D $\rm[Fe/H]=-3$}  \\ \hline
{\citet{2013A&A...554A..96L}} &  & Li\,I, Na\,I, Ca\,I, Ca\,II & \mbox{\#4 D/SG $\rm-3.5<[Fe/H]<-2$}  \\ \hline
{\citet{2013A&A...559A.102D}} & 12 neutral, 8 singly ionized, 6 molecules &  & \mbox{\#4 G $\rm-3<[Fe/H]<0$}  \\ \hline
{\citet{2014A&A...564L...6M}} &  & Li\,I & \mbox{\#4 G $\rm-1<[Fe/H]<0$}  \\ \hline
{\citet{2014A&A...565A..93S}} & Mg\,I, Al\,I, Si\,I, Sc\,II, V\,I, V\,II, Mn\,I  &  & \mbox{\#2 G $\rm-3<[Fe/H]<-2$}  \\ \hline
{\citet{2014A&A...565A.121D,2014A&A...568L...4K}} & Na\,I, O\,I & Li\,I  & \mbox{\#4 D $\rm-1<[Fe/H]<0$}  \\ \hline
{\citet{2014A&A...568A..29C}} & S\,I & & \mbox{\#2 D $\rm-2<[Fe/H]<-1$} \\ \hline
{\citet{2015A&A...576A.128D}} & OH, O\,I &  & \mbox{\#4 G $\rm-3.5<[Fe/H]<-2.5$} \\ \hline
{\citet{2015ApJ...806L..16B}} & CH, NH, OH &   & SMSS0313-6708 \\ \hline 
{\citet{2015A&A...579A..94G}} & Ba\,II, Fe\,I, Fe\,II &   & HD140283  \\ \hline
{\citet{2015MNRAS.454L..11A,2016MNRAS.455.3735A}} &  & O\,I & \mbox{\#60 D/SG $\rm-3<[Fe/H]<0$}  \\ \hline
{\citet{2016A&A...586A.156K}} &  & Li\,I & \mbox{\#16 D/SG/G $\rm-2<[Fe/H]<0$} \\ \hline
{\citet{2016A&A...593A..48G,2018A&A...614A..68C,2018A&A...612A..65B}} & CH &  & \mbox{\#18 D $\rm-3<[Fe/H]<-1$} \\ \hline
{\citet{2016MNRAS.463.1518A}} &  & Fe\,I, Fe\,II  & \mbox{\#4 D/SG/G $\rm-3.5<[Fe/H]<-2$} \\ \hline
{\citet{2016A&A...588A..66T,2017ApJ...843..144T}} & MgH & & \mbox{\#22 D/G $\rm-3<[Fe/H]<-0.5$} \\ \hline
{\citet{2017A&A...597A...6N}} &  & Li\,I, Na\,I, Mg\,I, Al\,I Ca\,II, Si\,I, Fe\,I  & SMSS0313-6708  \\ \hline
{\citet{2017A&A...598L..10G}} & CH, OH & & \#1 D $\rm[Fe/H]=-3$ \\ \hline
{\citet{2017A&A...599A.128P}} & OH &   & HD122563 \\ \hline
{\citet{2017A&A...600A..26S}} & C\,I, CH, OH &   & HD84937 \\ \hline
{\citet{2017A&A...604A..35C}} & Na\,I, Mg\,I, K\,I & & \mbox{\#1 G} \mbox{$\rm[Fe/H]=-1$} \\ \hline
{\citet{2017A&A...604A..44M}} &  & Li\,I  & HD 123351 \\ \hline
{\citet{2017ApJ...847...15B}} & & Mg\,I & \mbox{\#2 D/G $\rm-2<[Fe/H]<-0.5$} \\ \hline
{\citet{2017A&A...606A..26W}} &  & H\,I, Ca\,II & Aldebaran \\ \hline  
{\citet{2018MNRAS.475.3369C}} & CH, NH, OH, CN &   & HD122563 \\ \hline
{\citet{2018A&A...615A.139A}} &  & H\,I  & \mbox{\#169 D/SG/G} \mbox{$\rm-4<[Fe/H]<0.5$} \\ \hline
{\citet{2018A&A...615A.173C}} & Mg\,I, K\,I & & \mbox{\#4 D $\rm-1<[Fe/H]<0$}   \\ \hline
{\citet{2018A&A...618A..16H}} &  & Li\,I  &  \mbox{\#24 D $\rm-1<[Fe/H]<0.5$} \\ \hline
{\citet{2019A&A...624A..10G}} & H\,I &  & \mbox{\#8 D $\rm-0.5<[Fe/H]<0$}  \\ \hline
{\citet{2019A&A...627A.177R}} &  & K\,I  & Procyon  \\ \hline
{\citet{2019A&A...628A.111G}} &  & Li\,I  & CS 22876-032  \\ \hline
{\citet{2019A&A...622L...4A,2019A&A...630A.104A}} & & C\,I, O\,I & \mbox{\#74 D/SG $\rm-4<[Fe/H]<0.5$}  \\ \hline
{\citet{2019A&A...622L...4A,2019A&A...630A.104A}} & Fe\,II & & \mbox{\#164 D/SG/G $\rm-4<[Fe/H]<0.5$}  \\ \hline
{\citet{2019A&A...631A..80B}} &  & Mn\,I  & \mbox{\#8 D/SG/G $\rm-2.5<[Fe/H]<0$}  \\ \hline
{\citet{2020A&A...635A.104H}} & Ti\,I, Ti\,II & & \mbox{\#1 G $\rm[Fe/H]=-2$} \\ \hline
{\citet{2020A&A...638A..58M}} &  & Li\,I  & \mbox{\#70 D/SG $\rm-3<[Fe/H]<0$}  \\ \hline
{\citet{2021MNRAS.500.2159W,2022MNRAS.509.1521W}} &  & Li\,I  & \mbox{\#195 D/SG/G $\rm-4<[Fe/H]<0.5$}  \\ \hline
{\citet{2021ApJ...908..245B}} & & Li\,I, Na\,I & \mbox{\#4 D/G $\rm-3<[Fe/H]<0$}  \\ \hline
{\citet{2021A&A...647A..24M}} & Mg\,I, Si\,I, Ca\,I, Fe\,I &   & \mbox{\#9 G $\rm-3<[Fe/H]<0$}  \\ \hline
{\citet{2022A&A...668A..68A}} &  & Fe\,I, Fe\,II  & \mbox{\#32 D/SG $\rm-4<[Fe/H]<0.5$}  \\ \hline
{\citet{2023A&A...672A..90L}} & CH, NH, Ti\,II, Ni\,I & Li\,I, Na\,I, Mg\,I, Al\,I, Ca\,I, Ca\,II, Si\,I, Fe\,I  & \mbox{SDSS J102915.14+172927.9}  \\ \hline
{\citet{2023A&A...675A.146D}}$^{\rm d}$ & CO, OH, CH, CN, $\rm C_2$ &  & \mbox{\#5 D $\rm-3<[Fe/H]<0$}  \\ \hline
\hline
\hline
\multicolumn{4}{l}{
$^{\rm a}$ includes 3D LTE + 1D (NLTE-LTE); $^{\rm b}$ includes LTE and NLTE; $\rm^{c}$ D=dwarf, SG=subgiant,}\\
\multicolumn{4}{l}{G=red giant, \# is the number of 3D models; 
$^{\rm d}$ Non-equilibrium chemistry }\\

\end{longtable}

\end{appendix}
\end{document}